\documentclass[11pt,aps,preprintnumbers,nofootinbib,superscriptaddress,notitlepage,longbibliography]{revtex4-2}

\usepackage[T1]{fontenc}
\usepackage[english]{babel}
\usepackage{stix}
\usepackage{epsfig}
\usepackage{amsmath,bm,bbm,tensor}
\usepackage{graphicx}
\usepackage{verbatim}
\usepackage{csquotes}
\usepackage{simplewick}
\usepackage{appendix}
\usepackage[usenames,dvipsnames,svgnames]{xcolor}
\usepackage[normalem]{ulem}
\usepackage{hyperref}
\usepackage{cleveref}
\usepackage{soul}

\usepackage[framemethod=default]{mdframed}
\newmdenv[skipabove=7pt,
skipbelow=7pt,
rightline=false,
leftline=false,
topline=false,
bottomline=false,
backgroundcolor=gray!10,
linecolor=gray,
innerleftmargin=5pt,
innerrightmargin=5pt,
innertopmargin=5pt,
innerbottommargin=5pt,
leftmargin=0cm,
rightmargin=0cm,
linewidth=4pt]{eBox}

\newcommand{\md}{\mathrm{d}}

\begin{document}

\title{Unique gravitational wave signatures of GLPV scalar-tensor theories}

\author{\textsc{Guillem Dom\`enech }}
    \email{{guillem.domenech}@{itp.uni-hannover.de}}
    \affiliation{ Institute for Theoretical Physics, Leibniz University Hannover, Appelstraße 2, 30167 Hannover, Germany}
    \affiliation{ Max-Planck-Institut für Gravitationsphysik, Albert-Einstein-Institut, 30167 Hannover, Germany}

\author{\textsc{Alexander Ganz}}
    \email{alexander.ganz@itp.uni-hannover.de}
    \affiliation{ Institute for Theoretical Physics, Leibniz University Hannover, Appelstraße 2, 30167 Hannover, Germany}

\author{\textsc{Mohammad Ali Gorji}}
    \email{{gorji@ibs.re.kr}}
    \affiliation{Cosmology, Gravity, and Astroparticle Physics Group, Center for Theoretical Physics of the Universe,
Institute for Basic Science (IBS), Daejeon, 34126, Korea}

\author{\textsc{Masahide Yamaguchi}}
    \email{{gucci@ibs.re.kr}}
    \affiliation{Cosmology, Gravity, and Astroparticle Physics Group, Center for Theoretical Physics of the Universe,
    Institute for Basic Science (IBS), Daejeon, 34126, Korea}
    \affiliation{Department of Physics, Institute of Science Tokyo, 2-12-1 Ookayama, Meguro-ku, Tokyo 152-8551, Japan}
    \affiliation{Department of Physics \& Institute of Physics and Applied Physics (IPAP), Yonsei University, 50 Yonsei-ro, Seodaemun-gu, Seoul 03722, Korea}

\begin{abstract}
We study gravitational waves induced by scalar primordial fluctuations in  Gleyzes-Langlois-Piazza-Vernizzi (GLPV), beyond Horndeski, scalar-tensor theories. We uncover, at the level of the action, a new scalar-scalar-tensor interaction, unique to GLPV models disconnected from Horndeski via disformal transformation. The new interaction, arising in the unitary-degenerate (U-DHOST) sector of GLPV, leads to third derivatives in the source for scalar-induced tensor modes, which are absent in Horndeski-related theories. Such new higher-derivative terms lead to a further enhanced production of induced gravitational waves. We predict that for a scale-invariant primordial spectrum, the induced gravitational wave spectral density has a characteristic frequency dependence proportional to $f^5$. Such a fast-rising spectrum offers a potential unique signature of modified gravity in the early universe. 

\end{abstract}

\maketitle

\section{Introduction}

While General Relativity (GR) describes gravity accurately on local scales, it does not fully answer some cosmological questions. The most notable one is the nature of dark energy. There are also some hints of non-minimal couplings in the primordial (e.g., in Starobinsky and Higgs inflation \cite{Planck:2018jri,Planck:2018vyg,Drees:2025ngb,ACT:2025fju,ACT:2025tim}) and late universes, see e.g. Ref.~\cite{Ye:2024ywg}. Thus, it is crucial to explore new physics beyond GR. Especially now that we can test them using Gravitational Waves (GWs).


 In recent years, there has been a global effort to establish a general framework for extensions of GR. Motivated by a cosmological set-up, the simplest extension is to add an additional scalar field, in the so-called scalar-tensor theories, potentially allowing for non-minimal and derivative couplings to the spacetime curvature. However, a theory with higher time derivatives, may contain the Ostrogradsky ghost \cite{Ostrogradsky:1850fid}, rendering the theory unstable (see, e.g., Refs.~\cite{Woodard:2015zca,Ganz:2020skf} and references therein). In this respect, there has been a lot of focus on determining the most general scalar-tensor theories without such a ghost \cite{Gao:2014fra,Gao:2014soa,Gleyzes:2014dya,Langlois:2015cwa,DeFelice:2018ewo,Kobayashi:2019hrl}, with later applications to dark energy, inflation and black holes (see Refs.~\cite{Kobayashi:2019hrl,Frusciante:2019xia} for a review). Similar ideas have also been applied to vector fields \cite{Heisenberg:2014rta,Heisenberg:2016eld,Kimura:2016rzw,deRham:2020yet}.

Horndeski proposed the first such theory in 1974 in Ref.~\cite{Horndeski:1974wa}, and later rediscovered in the context of Galileon in 2011 by Refs.~\cite{Charmousis:2011bf,Deffayet:2011gz}. The equivalence between Horndeski~\cite{Horndeski:1974wa} and Generalized Galileon theories~\cite{Deffayet:2011gz} was shown in Ref.~\cite{Kobayashi:2011nu}, and then we can safely use the action of Generalized Galileon as Horndeski action even though both of them apparently look quite different. Horndeski constructed the most general scalar-tensor theory by requiring second-order equations of motion. However, it was later realized that
such a condition can be relaxed without necessarily introducing an Ostrogradsky ghost.
This has led to the establishment of the beyond Horndeski, GLPV models \cite{Gleyzes:2014dya,Gleyzes:2014qga} and later on to the systematic classification of the Degenerate Higher Order Scalar Tensor (DHOST) theories \cite{Langlois:2015cwa,Langlois:2015skt,Crisostomi:2016czh,BenAchour:2016fzp}.

General metric transformations involving the scalar field, known as disformal transformations \cite{Bekenstein:1992pj}, play a crucial role in mapping and classifying scalar-tensor theories \cite{Bettoni:2013diz,Zumalacarregui:2013pma,Ezquiaga:2017ner,Crisostomi:2016czh,BenAchour:2016cay,deRham:2016wji}. For instance, the functional form of DHOST theories is invariant under a general disformal transformation \cite{BenAchour:2016cay,BenAchour:2016fzp}, say $ g_{\mu\nu} = \Omega^2(\phi,\tilde X) \tilde g_{\mu\nu} + D(\phi,\tilde X) \partial_\mu \phi \partial_\nu \phi$, where $\Omega$ and $D$ are arbitrary functions of the scalar field $\phi$ and its kinetic term, that is $\tilde X=\tilde g^{\mu\nu}\partial_\mu \phi \partial_\nu \phi$. It is also known that the functional form of (the DHOST subclass) GLPV and Horndeski theories are invariant only for a restricted set of disformal transformation, that is  $\Omega=\Omega(\phi)$ \cite{Crisostomi:2016tcp,Gleyzes:2014qga} and $\Omega=\Omega(\phi)$ and $D=D(\phi)$ \cite{Bettoni:2013diz}, respectively.

Interestingly, one can use disformal transformations to connect GR, or Horndeski theories, with GLPV or DHOST. In other words, only a subset of GLPV, or DHOST, is equivalent to GR, or Horndeski, up to field redefinitions \cite{Gleyzes:2014qga,Crisostomi:2016tcp}. It then follows that the equations of motion of such subset can be recast with only second-order derivatives \cite{Zumalacarregui:2013pma}. This is, in general, not possible for all DHOST theories. We will show that GLPV, or DHOST, models disformally disconnected from Horndeski theories, have a fundamentally different phenomenology, at least regarding scalar-scalar-tensor interactions. For cubic scalar and tensor interactions it has been shown in Refs.~\cite{Renaux-Petel:2011zgy,Fasiello:2014aqa,Akita:2015mho,Passaglia:2018afq} that the operators are the same as in Horndeski \cite{Gao:2012ib}, after field redefinitions. 

In the present universe, models of dark energy described by general DHOST theories are highly constrained, e.g., by GW observations \cite{Ezquiaga:2017ekz,Langlois:2017dyl,Sakstein:2017xjx,Crisostomi:2017lbg,Creminelli:2017sry}  (although see Refs.~\cite{Amendola:2018ltt,Bordin:2020fww,deRham:2018red}). We propose a way to test the presence of such theories in the very early Universe, after inflation but before Big-Bang Nucleosynthesis (BBN), using GWs. In particular, we focus on the GWs induced by scalar primordial fluctuations \cite{Tomita,Matarrese:1992rp,Matarrese:1993zf,Matarrese:1997ay}, the existence of which has been measured in Cosmic Microwave Background (CMB) Observations \cite{Planck:2018vyg}. These GWs are often referred to as induced GWs \cite{Ananda:2006af,Baumann:2007zm,Assadullahi:2009nf,Alabidi:2013lya}. For recent reviews see Refs.~\cite{Yuan:2021qgz,Domenech:2021ztg,Domenech:2024kmh,Roshan:2024qnv}. 

A non-minimally, derivatively coupled scalar field will significantly alter the spectrum of the induced GWs with respect to the GR case. For instance, two of us investigated in Ref.~\cite{Domenech:2024drm} the impact of a Horndeski scalar field. We found that the presence of higher-derivative terms (proportional to powers of second-order derivatives) may lead to a significant enhancement of the GW spectrum. For a scale-invariant primordial spectrum, the resulting induced GW spectrum scales as $f^3$, where $f$ is the frequency of the GWs. Unfortunately, such scaling coincides with the universal prediction for the low-frequency tail of GWs generated by finite-time sources \cite{Cai:2019cdl}. Thus, although primordial fluctuations with the CMB normalization can yield a detectable induced GW signal within Horndeski theories \cite{Domenech:2024drm}, it may be challenging to distinguish it from other sources.

It is important to note that the enhancement found in Ref.~\cite{Domenech:2024drm} is directly linked to higher-order spatial derivatives in scalar-scalar-tensor interactions. These terms are, e.g., not present in the case of a non-minimally coupled scalar field. Indeed, Ref.~\cite{Kugarajh:2025rbt} analyzes the impact of $f(R)$ gravity (which is equivalent to a non-minimally coupled scalar field \cite{Maeda:1988ab}) on the induced GWs, showing that the main effect on the GWs emerges from the modified curvature perturbation at linear order and not the source term itself (see also Ref.~\cite{Lopez:2025gfu} for a similar analysis in the Palatini formalism). In a complementary direction, Refs.~\cite{Hu:2024hzo,Feng:2024yic,Zhang:2024vfw} investigate the induced GW spectrum in parity-violating modifications of GR in scalar-tensor theories and teleparallel gravity. Note that the higher derivative interactions have also been discussed in detail in the context of the EFT of dark energy, where they can lead to a decay of GW into the scalar field \cite{Creminelli:2018xsv}. In our case, we consider the interaction to be only relevant in the very early Universe before BBN. 

In this work, we generalize the results of Ref.~\cite{Domenech:2024drm} by studying the generation of induced GWs in GLPV theories. We uncover a new higher-derivative scalar-scalar-tensor interaction, unique to GLPV theories disformally disconnected from Horndeski. The new interaction explicitly breaks the second-order nature of the equations of motion, further enhancing the generation of GWs from small scale primordial fluctuations. Note that such modification only appears at second order in perturbation theory for scalar-scalar-tensor interactions, as linear perturbation obeys a standard dispersion. Induced GWs are specially suited to test such unique interactions. As we will show, the new interaction leads to an enhancement of the GW power spectrum scaling as $f^5$ for a scale-invariant curvature power spectrum. The growth is insensitive to the details of the models as long as the corrections to GR at linear order are sufficiently small.

This paper is organized as follows. In Sec. \ref{sec:II}, we derive the new cubic interaction and show that it belongs to GLPV models disconnected from Horndeski. In Sec.~\ref{sec:New_Interaction_Toy_Model}, we discuss the scalar-induced GW spectrum for a toy model within GLPV following the construction in \cite{Domenech:2024drm}. The new interaction enhances the GW spectrum leading to a tail scaling as $f^5$ for a scale-invariant primordial spectrum. We then discuss the robustness of the results in Sec.~\ref{sec:Effective_Cubic_Interaction}, 
by considering an effective ansatz for the tensor and scalar perturbation which includes all of GLPV and show that the tail and the resonance are robust as long as the corrections to the linear curvature perturbation are sufficiently small. We end our paper with several discussion and future directions in Sec.~\ref{sec:conclusions}. Details of the calculations can be found in the appendices.

\section{Disformal Transformation in GLPV and Horndeski gravity \label{sec:II}}

We start by showing, using cosmological perturbation theory, that GLPV models, disformally disconnected from Horndeski, contain a new cubic scalar-scalar-tensor interaction that explicitly breaks the second-order nature found in Horndeski gravity. We argue that such a unique interaction may provide a new way to discriminate beyond Horndeski models. Readers interested in the induced GWs may skip to Sec.~\ref{sec:New_Interaction_Toy_Model}.


\subsection{Preliminaries}

Our starting point is the GLPV action \cite{Gleyzes:2014dya,Gleyzes:2014qga}. In the ADM decomposition and unitary gauge\footnote{The action in the covariant formulation and the matching to the unitary gauge are discussed in App.~\ref{subse:GLPV_Covariant}}, it reads
\begin{align}
S=\int d^4x \,{\cal L}\,,
\end{align}
where
\begin{align}\label{eq:unitarygaugeaction}
\begin{split}
    {\cal L} = \sqrt{h} N \Big[ & A_2(t,N) + A_3(t,N) K + A_4(t,N)  (K^2 - K_{ij} K^{ij} ) + B_4(t,N) R  \\
    & + A_5(t,N) ( K^3 - 3 K_{ij} K^{ij} K + 2 K_{ij} K^{jk} K_k^i)  + B_5(t,N) R_{ij} \left( K^{ij} - \frac{1}{2} h^{ij} K \right) \Big]\,,
\end{split}
\end{align}
where $t$ is cosmic time, $N$ is the lapse function, $h_{ij}$ is the metric on the 3 dimensional spatial hypersurface, $R_{ij}$ the corresponding intrinsic Ricci curvature and $K_{ij} =  \frac{1}{2N} \left( \dot h_{ij} - D_i N_j -D_j N_i \right)$ is the extrinsic curvature. In general GLPV, the functions $A_2$, $A_3$, $A_4$, $A_5$, $B_4$ and $B_5$ are arbitrary functions of $t$ and $N$. In contrast, Horndeski theory is characterized by the following relations (or constraints),
\begin{align}
    \label{eq:Horndeski_constraint_A_B}
   C_{4}^{\rm H}\equiv A_4+  B_4 + N  B_{4, N}=0~, \qquad C_{5}^{\rm H}\equiv  A_5 - \frac{N}{6}  B_{5, N}=0 \,,
\end{align}
where $ B_{4, N}=\partial B_4/\partial N$. Let us show that the new interaction term, which we focus on, is precisely proportional to the Horndeski constraints \eqref{eq:Horndeski_constraint_A_B} and therefore is absent in the Horndeski models.

Since we are interested in induced GWs, we focus on the scalar-scalar-tensor interactions, which lead to quadratic scalar source terms in the tensor mode equations of motion. We expand the metric around a spatially flat Friedmann–Lemaître–Robertson–Walker (FLRW) universe, in the unitary gauge (spatially uniform scalar field gauge), as\footnote{We used the scalar part of the spatial diffeomorphism freedom to eliminate the extra scalar perturbation in the spatial metric.}
\begin{align}
    N = 1+ \alpha~, \qquad N_i = \partial_i \beta~, \qquad h_{ij} = a^2 e^{2\zeta} (e^\gamma)_{ij}~,
\end{align}
where $a$ is the scale factor and we neglected vector perturbations. The tensor perturbations $\gamma_{ij}$ are transverse and traceless, i.e $\delta^{ij}\gamma_{ij} = \partial^i \gamma_{ij}=0$. From now on, we focus on the scalar-scalar-tensor interaction terms including the highest derivatives. Details on the linear order perturbations and the quadratic action can be found in App.~\ref{subsec:GLPV_Quadratic_Action}. 

We find that the highest derivative scalar-scalar-tensor interaction term, treating space and time derivatives equally, is given by
\begin{align}
    {\cal L}_{\zeta\zeta\gamma} \supset & \frac{1}{a} \Big[ 3  A_5 \dot  \gamma_{ij} ( \partial_i \partial_k \beta \partial_j \partial_k \beta - \partial_i \partial_j \beta \partial^2 \beta ) - B_{5,N} \gamma_{ij} ( \frac{1}{2} \partial_i \partial_j \alpha \partial^2 \beta + \frac{1}{2} \partial^2 \alpha \partial_i \partial_j \beta - \partial_i \partial_k \alpha \partial_j \partial_k \beta ) \Big] \nonumber \\
    \supset & \frac{1}{a} \Big[ 2 \partial^2 \dot \gamma_{ij} \partial_i \zeta \partial_j  \zeta   \frac{ (6 A_5 (B_4 + B_{4,N} )  + A_4 B_{5,N} ) (2 (B_4 + B_{4,N}) - H B_{5,N})}{(- 4 A_4 H + 12 A_5 H^2 + A_{3,N} + 4 H A_{4,N} + 6 H^2 A_{5,N})^2} \Big]\,,
    \label{eq:Cubic_General_Leading_Term}
\end{align}
where $H=\dot{a}/a$ is the Hubble parameter and, in the second step, we used the linear solution for $\alpha$ and $\beta$ given by Eq.~\eqref{eq:Solution_Hamiltonian_Constraint}. Interestingly, we see that the first term in the numerator of the coefficient in Eq.~\eqref{eq:Cubic_General_Leading_Term} vanishes in Horndeski theories, as it is proportional to the constraints \eqref{eq:Horndeski_constraint_A_B} (after setting $N=1$). Explicitly, we have that
\begin{align}\label{eq:coefficientnumerator}
6 A_5 (B_4 + B_{4,N} )  + A_4 B_{5,N} =6\left(A_5 C_4^{\rm H}-A_4C_5^{\rm H}\right)\,,
\end{align}
where $C_4^{\rm H}=C_5^{\rm H}=0$ in Horndeski. 
However, this term is in general non-vanishing in GLPV theories. This means that, while in Horndeski gravity the highest derivative cubic scalar-scalar-tensor interaction scales as $\partial^4 \sim k^4$, in GLPV it scales as $\partial^5 \sim k^5$. Note that the interaction term \eqref{eq:Cubic_General_Leading_Term} also vanishes if $2 (B_4 + B_{4,N}) - H B_{5,N}=0$. However, such a cancellation depends on the background solution and will not occur in general.

 At this point, let us clarify that GLPV theories are constructed in the uniform field slicing. This means that a subset of this theories may contain an instantaneous mode in the covariant formulation. In general, starting from a covariant formulation and requiring three propagating degrees of freedom, that is the absence of instantaneous modes, imposes additional degeneracy conditions (see \cite{BenAchour:2016fzp,Langlois:2015cwa} and appendix \ref{app:Degeneracy_conditions} for more details).\footnote{We would like to thank David Langlois and Karim Noui for pointing it out.} We find that it is precisely this degeneracy condition, given in Eq.~\eqref{eq:GLPV_degeneracy_condition}, which leads to a vanishing leading interaction \eqref{eq:coefficientnumerator}.
In other words, the GLPV theory under consideration belongs to U-DHOST theories, or more appropriately to U-beyond Horndeski theories \cite{DeFelice:2018ewo,DeFelice:2021hps,DeFelice:2022xvq}.

It is important to note that Eq.~\eqref{eq:Cubic_General_Leading_Term} explicitly leads to third derivatives of $\zeta$ in the equations of motion of $\gamma_{ij}$. This is in contrast to  Horndeski gravity, where one can always find a field redefinition to remove such terms \cite{Gao:2012ib}, due to the (imposed) second-order nature of the equations of motion. From this, it follows that in Horndeski the maximum scaling in derivatives at the n-th order in perturbation theory must be an even power of derivatives, that is $\partial^{2n} \sim k^{2n}$. Instead, in GLPV (in the uniform slicing), additional constraints prevent third, or higher, order time derivatives but allow for higher-order spatial derivatives. This is the reason for the odd derivative power scaling, namely $\partial^5 \sim k^5$.

In passing, let us point out that, for cubic interactions around a flat FLRW, scalar-scalar-tensor interactions are the only ones which explicitly introduce higher than second-order derivatives in the equation of motions. Cubic scalar and tensor interactions scale as $\partial^4 \sim k^4$, as explicitly shown in Refs.~\cite{Renaux-Petel:2011zgy,Fasiello:2014aqa,Akita:2015mho,Passaglia:2018afq}. Similarly, the tensor-tensor-scalar interaction only scales at maximum as $\partial^4 \sim k^4$ as demonstrated in the App.~\ref{subsec:Tensor_Tensor_Scalar_GLPV}, since higher-order derivative terms require terms quadratic in the intrinsic curvature which is beyond GLPV. This might be possible in the context of scordatura DHOST theories \cite{Motohashi:2019ymr,Gorji:2020bfl,Gorji:2021isn}, where the degeneracy conditions are slightly broken, or in spatially covariant theories \cite{Gao:2014soa,Gao:2014fra}, where full spacetime diffeomorphism invariance is reduced to time-dependent spatial diffeomorphisms.

\subsection{GLPV theories disformally connected to Horndeski}

One may wonder if the interaction term \eqref{eq:Cubic_General_Leading_Term} may be removed via a disformal transformation. Or, in other words, if one can obtain such an interaction term starting from Horndeski action and performing a disformal transformation. Let us show that this is not the case. Consider a disformal transformation given by
\begin{align}\label{eq:metricdisformal}
    g_{\mu\nu} = \Omega^2(\phi,\tilde X) \tilde g_{\mu\nu} + D(\phi,\tilde X) \tilde\nabla_\mu \phi \tilde\nabla_\nu\phi\,,
\end{align}
where $\tilde X = \tilde\nabla_\mu \phi \tilde\nabla^\mu \phi$. The transformation \eqref{eq:metricdisformal} corresponds, in the unitary gauge, to
\begin{align}\label{eq:Nchangeandh}
    N =  \Xi \tilde N, \qquad N^k = \tilde N^k, \qquad h_{ij} = \Omega^2 \tilde h_{ij}\,,
\end{align}
where, for compactness, we introduced
\begin{align}
    \Xi^2 = \Omega^2(t,\tilde N)  -  \frac{D(t,\tilde N)}{\tilde N^2} \,.
\end{align}
For detailed studies of disformal transformations in the context of GLPV and DHOST theories, we refer the reader to Refs.~\cite{Gleyzes:2014dya,Gleyzes:2014qga,Crisostomi:2016tcp,BenAchour:2016cay,BenAchour:2016fzp,Crisostomi:2016czh,BenAchour:2024hbg} (see also \cite{Bettoni:2013diz,Zumalacarregui:2013pma} for earlier works). There, it is shown that  Horndeski theories are closed under disformal transformation if $\Omega(\phi)$ and $D(\phi)$, while GLPV theories are invariant if $\Omega(\phi)$ and $D(\phi,\tilde X)$. One ends up in DHOST when $\Omega=\Omega(\phi,\tilde X)$. In what follows, we start with a Horndeski action and consider first a pure disformal transformation and then move on to the general case.

\subsubsection{Purely Disformal Transformation}

Take a pure disformal transformation, namely $\Omega=1$ and $D = D(\phi,X)$. From Eq.~\eqref{eq:Nchangeandh}, this corresponds to a rescaling of the lapse function.
In that case, borrowing the results of Ref.~\cite{Gleyzes:2014qga}, we have that
\begin{align}
    \tilde A_4 = \frac{ 1}{\Xi  } A_4(t, \Xi \tilde N)~, \qquad
    \tilde B_4 = \Xi  B_4(t, \Xi \tilde N)~, \qquad
    \tilde A_5 = \frac{ 1}{\Xi^2} A_5(t,\Xi \tilde N)~, \qquad
    \tilde B_5 = B_5(t, \Xi \tilde N)~.
\end{align}
If we start from Horndeski gravity, the functions $A_4$, $A_5$, $B_4$ and $B_5$ are related via Eq.~\eqref{eq:Horndeski_constraint_A_B}, namely they satisfy $C_4^{\rm H}=C_5^{\rm H}=0$. Given this, it follows that the new tilded functions, after the disformal transformation, are now related by
\begin{align}
    \tilde A_4 = - \frac{1}{\Xi} \frac{1}{\Xi + \Xi_{\tilde N} \tilde N} \left(  \tilde B_4 + \tilde N {\tilde B}_{4, \tilde N} \right)  \quad,\quad
    \tilde A_5 =  \frac{ 1}{ \Xi}  \frac{1}{\Xi + \Xi_{,\tilde N} \tilde N} \frac{\tilde N}{6} \tilde B_{5,\tilde N}\,.
\end{align}
This means that we arrived at a GLPV theory, since the combinations given by
\begin{align}
\begin{split}
{\tilde C}_{4}^{\rm H}
&\equiv {\tilde A}_4+  {\tilde B}_4 + {\tilde N}  {\tilde B}_{4, {\tilde N}}
=\tilde A_4 (1-\Xi(\Xi + \Xi_{\tilde N} \tilde N)) \neq 0 \,,
\\
{\tilde C}_{5}^{\rm H}
&\equiv {\tilde A}_5 - \frac{{\tilde N}}{6}  {\tilde B}_{5, {\tilde N}}
=\tilde A_5 (1-\Xi(\Xi + \Xi_{\tilde N} \tilde N)) \neq 0 \,,
\end{split}
\end{align}
are non-vanishing. However, their relation is such that the coefficient of the highest derivative term in Eq.~\eqref{eq:Cubic_General_Leading_Term}, proportional to Eq.~\eqref{eq:coefficientnumerator}, now trivially vanishes in the tilded frame
\begin{align}
\tilde A_5 \tilde  C_4^{\rm H}-\tilde  A_4 \tilde C_5^{\rm H}=(\tilde A_5 \tilde A_4-\tilde  A_4 \tilde A_5)(1-\Xi(\Xi + \Xi_{\tilde N} \tilde N)=0\,.
\end{align}
Indeed, $\tilde A_5 \tilde  C_4^{\rm H}-\tilde  A_4 \tilde C_5^{\rm H}= \left(A_5 C_4^{\rm H} - A_4 C_5^{\rm H} \right) (\Xi+\Xi_{\tilde N}{\tilde N})/\Xi^2$ which explicitly shows that the condition for the vanishing of the highest-derivative interaction coefficient, Eq.~\eqref{eq:Cubic_General_Leading_Term}, is preserved under a pure disformal transformation. We thus conclude that a pure disformal transformation cannot generate the new higher-derivative terms not present in the original Horndeski theory.


\subsubsection{Full Disformal Transformation}
The general disformal transformation case is more involved, but the final result is similar to that in the pure disformal case. Consider a general disformal transformation with $\Omega= \Omega(t, \tilde N)$ and $D=D(t,\tilde N)$, which in general connects Horndeski gravity to DHOST. For simplicity, we will only study the transformation perturbatively as we are concerned with the cubic interactions. 

At the background level, the transformation corresponds to a rescaling of the lapse function and the scale factor, that is
\begin{align}
    N_0 = \Xi_0 \tilde N_0~, \qquad a = \Omega \tilde a~,
\end{align}
where the subscript $0$ denotes $0$th order. Up to second order, we find that
\begin{align}\label{eq:relationsgeneraldisformal}
\begin{split}
    \alpha & =  \frac{\partial ( \Xi \tilde N)}{\partial \tilde N} \tilde \alpha + \frac{\partial^2 ( \Xi \tilde N)}{2 \partial \tilde N^2} \tilde \alpha^2, \\
    \zeta & =  \tilde \zeta + \frac{\Omega_{,\tilde N}}{\Omega} \tilde \alpha +  \frac{\Omega_{,\tilde N \tilde N } \Omega - 3 \Omega_{,\tilde N}^2}{2 \Omega^2} \tilde \alpha^2, \\
    \partial_i \beta &= \Omega^2 \left(1  + \frac{2 \Omega_{,\tilde N}}{ \Omega} \tilde \alpha \right)  \partial_i \tilde \beta\,.
 \end{split}
\end{align}
The tensor perturbations $\gamma_{ij}$ are not impacted up to this order. Note that in Eq.~\eqref{eq:relationsgeneraldisformal}, and in what follows, the functions $\Xi$, $\Omega$ and their derivatives are to be evaluated at the background. We omit the subscript ``0'' to avoid cluttered notation.

We now substitute the relations \eqref{eq:relationsgeneraldisformal} into the (untilded) Horndeski quadratic action (see Eq. \eqref{eq:Quadratic_Action_GLPV_General} for details). The resulting action in the tilded frame depends on the time derivative of the lapse function. This seemingly leads to an extra degree of freedom. However, the term can be absorbed by redefining the curvature perturbation $\tilde \zeta \rightarrow \hat \zeta$ which follows directly from the disformal transformation $\hat \zeta \equiv \zeta(\tilde \zeta, \tilde \alpha)$. The new solution to the momentum and Hamiltonian constraints, at linear order and in Fourier space, in the new tilded frame yields 
\begin{align}
& \tilde \alpha = -\frac{\tilde t_6 }{\tilde t_4 \Xi (\Xi \tilde N)_{,\tilde N}} \dot{\hat \zeta} ~, \qquad \tilde \beta = - \frac{\tilde t_3}{\tilde t_4 \Omega^2} \hat \zeta - \frac{1}{(\Xi \tilde N)_{,\tilde N} } \left( \frac{\tilde t_5}{\tilde t_4} - \frac{2 \tilde t_6 \tilde t_0}{\tilde t_4^2} \right) \frac{a^2}{k^2}\dot{\hat \zeta} ~,
\end{align}
 where the explicit form of $t_i$ are provided in App.~\ref{subsec:GLPV_Quadratic_Action}. However, it is important to note that $\tilde t_3 = - \tilde t_6$ if we start from Horndeski gravity. After reducing the third-order action, the highest-derivative term for scalar-scalar-tensor interactions, Eq.~\eqref{eq:Cubic_General_Leading_Term}, reads
 \begin{align}\label{eq:newactiongeneraldisformal}\begin{split}
     {\cal L}_{\zeta\zeta\gamma} \supset & \frac{1}{\Omega \tilde a} \Big[ \frac{1}{2} \tilde B_{5,\tilde N} \frac{\tilde t_3^2}{\tilde t_4^2 } \frac{\dot{\tilde \gamma}_{ij}}{\Xi} ( \partial_i \partial_k \hat \zeta \partial_j \partial_k \hat \zeta - \partial_i \partial_j \hat \zeta \partial^2 \hat \zeta ) \\
     & - \tilde B_{5,\tilde N} \frac{\tilde t_6^2}{\tilde t_4^2} \frac{1}{\Xi} \tilde{\gamma}_{ij} \left( \frac{1}{2 }   \partial_i \partial_j \dot{\hat \zeta} \partial^2 \hat \zeta + \frac{1}{2} \partial^2 \dot{\hat \zeta} \partial_i \partial_j \hat \zeta - \partial_i \partial_k \dot{\hat \zeta} \partial_j \partial_k \hat \zeta  \right)  \Big]\,.
\end{split}
 \end{align}
One can check that, after several integration by parts, the cubic interaction \eqref{eq:newactiongeneraldisformal} vanishes exactly, since $\tilde t_3 = - \tilde t_6$. This means that the highest-derivative interaction scales as ${\cal L}_{\zeta\zeta\gamma} \sim {\cal O}(k^4)$. 

Therefore, general disformal transformations, with non-linear redefinitions, do not introduce any higher derivative of the order $\partial^5 \sim k^5$ starting from Horndeski theories. We conclude that the new interaction \eqref{eq:Cubic_General_Leading_Term} belongs to GLPV models disformally disconnected from Horndeski. 

It is worth mentioning that, unlike a conformal transformation, a pure disformal transformation can lead to observable effects in the GW spectrum. However, such an effect entirely depends on the frame to which the GW detector is minimally coupled \cite{BenAchour:2024tqt}. On the other hand, the effect we are considering here is entirely universal and, in principle, should be observable regardless of the frame to which the GW detector is minimally coupled.

\section{Induced Gravitational Waves for GLPV Toy Model}
\label{sec:New_Interaction_Toy_Model}

We proceed to study the generation of induced GWs in GLPV. For simplicity, we focus on a toy model that contains mainly the new cubic interaction. For the effect of scalar-scalar-tensor interactions included in Horndeski see Ref.~\cite{Domenech:2024drm}. The action density of the toy model we consider is given by
\begin{align}
    \label{eq:GLPV_Toy_Model}
    {\cal L} = \sqrt{h} N \left( P(t,N) + \frac{1}{2} \left( R + K_{ij} K^{ij} - K^2 \right) + B_5(t,N) R_{ij} \left( K^{ij} - \frac{1}{2} h^{ij} K \right) \right)~.
\end{align}
The new cubic interaction term stems from a non-vanishing $B_5(t,N)$, see also Eq.~\eqref{eq:Cubic_General_Leading_Term}. For the covariant formulation of the toy model see App.~\ref{subse:GLPV_Covariant}. 

To further simplify calculations at second order, we will consider that the background effectively behaves as a radiation fluid and GLPV term has minimal impact to linear perturbations
\begin{align}
\label{eq:Ansatz_B_5-ADM}
P(t,N) = \frac{1}{4t^2N^4} \,,
\qquad
B_5(t,N) =  b_5\, t \left( N^{-2} -1 \right) \,, 
\end{align}
where $b_5$ is an arbitrary constant quantifying the amplitude of the modification of gravity. Because of time reparametrization, we are free to choose $N=1$ at the background level. Note that $B_{,N}, B_{,NN}, \cdots$ are non-vanishing. The background solution for the Hubble parameter is $H=\sqrt{-(P+NP_{,N})/3}|_{N=1}=1/(2t)$ which gives $a\propto t^{1/2}$. There are different covariant realizations of the ADM configuration \eqref{eq:Ansatz_B_5-ADM}. For example, the following covariant realization is consistent with the power-law solutions found in Ref.~\cite{Domenech:2024drm}, namely
\begin{align}
\label{eq:Ansatz_B_5}
P(\phi,X) = \frac{\phi_\star^2}{4 \phi^2 t_\star^2} \frac{X^2}{X_\star^2}\,,
\qquad
B_5(\phi,X) =  b_5 \frac{\phi t_\star}{\phi_\star} \left( \frac{X}{X_\star} -1 \right) \,,
\end{align}
where $X=\nabla_\mu \phi \nabla^\mu \phi$ and $\phi = \phi_\star (t/t_\star)$ on the power-law solution. Quantities with subscript ``$\star$'' are evaluated at an arbitrary pivot scale. 

Below, we first compute the quadratic and cubic actions and present the explicit form for the source of induced GWs. Later we focus on the Kernel integral and discuss possible gauge ambiguities. Lastly, we compute the induced GW spectrum for two examples: a log-normal and a scale-invariant spectra.

\subsection{Quadratic and cubic actions}
The second-order action densities for tensor and scalar modes for the ansatz \eqref{eq:Ansatz_B_5} are given by
\begin{align}
    {\cal L}_{\gamma\gamma} = \frac{a^3}{8} \left(  \dot \gamma_{ij}^2 - \frac{(\partial_k \gamma_{ij})^2}{a^2} \right)\,,
    \quad
    {\cal L}_{\zeta\zeta} = a^3{\cal G}_s \left(  \dot \zeta^2 - c_s^2 \frac{(\partial_k \zeta)^2}{a^2} \right)\,,
\end{align}
where we defined
\begin{align}
    {\cal G}_s = \frac{2 P_{,N} + P_{,NN}}{2 H^2}=6~, \qquad
    c_s^2  = \frac{1}{{\cal G}_s} \left( \epsilon - H B_{5,N} - \dot B_{5} - \dot B_{5,N} \right) = \frac{1}{3}+\frac{b_5}{2}\label{eq:cs}\,,
\end{align}
where $\epsilon=-\dot{H}/H^2$ is the slow-roll parameter. For details of the derivation see App.~\ref{subsec:GLPV_Quadratic_Action}.
Note that tensor modes satisfy the same linear equations of motion as in GR. For the curvature perturbation, the equation of motion in conformal time ($d\tau=dt/a$) and Fourier space reads
\begin{align}
    \zeta^{\prime\prime} + 2 {\cal H} \zeta^\prime + c_s^2 k^2 \zeta = 0\,,
\end{align}
where $\zeta^{\prime}=\partial\zeta/\partial\tau$ and ${\cal H}=a'/a$.
Thus, we obtain standard GR + K-essence scalar field \cite{Armendariz-Picon:1999hyi} up to a constant shift in the sound speed $c_s$. We parameterize the solution for the curvature perturbation as $\zeta = T_\zeta(k\tau) \zeta_{\bf k}^p$, where $\zeta_{\bf k}^p$ are the initial conditions set by inflation and $T_\zeta$ is the transfer function  given by
\begin{align}\label{eq:Tzeta}
    T_\zeta(x) = \frac{\sin(c_s x)}{c_s x}\quad{\rm with}\quad x=k\tau\,.
\end{align}
Note that $T_\zeta$ \eqref{eq:Tzeta} only depends on $b_5$ through $c_s$ \eqref{eq:cs}, otherwise $T_\zeta$ takes the same form as in GR.
Tensor perturbations obey similar solutions at linear order but with $c_s=1$, as in GR.

Moving on to the cubic action for the scalar-scalar-tensor interactions, we find that it is given by
\begin{align}\label{eq:actiontoymodel}
\begin{split}
    {\cal L}_{\gamma\zeta\zeta} =& a \Big[ \frac{1}{4} \gamma_{ij} \partial^2 (\partial_i \bar\beta \partial_j \bar\beta) - \frac{a}{2 H} \dot \gamma_{ij} \partial_i \bar\beta \partial_j \dot \zeta - \frac{1}{H}  \gamma_{ij} \partial_t (\partial_i \zeta \partial_j \zeta) + \frac{3 a}{2} \dot \gamma_{ij} \partial_i \bar\beta \partial_j \zeta - \gamma_{ij} \partial_i \zeta \partial_j \zeta \\
    & - b_5 \left( \gamma_{ij} \frac{\partial_t (\partial_i \zeta \partial_j \zeta)}{H} + \dot \gamma_{ij} \frac{\partial_t (\partial_i \zeta \partial_j \zeta)}{4 H^2} - \gamma_{ij} \frac{\partial^2(\partial_i \dot \zeta \partial_j \bar\beta) }{2 H^2 a} \right) \Big]~,
\end{split}
\end{align}
where we introduced a rescaled shift $\bar\beta = \beta/ a$ for convenience. The term proportional to $b_5$ is the unique GLPV interaction. From Eq.~\eqref{eq:actiontoymodel}, it follows that the equations of motion of the induced tensor modes in Fourier space\footnote{Our Fourier convention is given by
\begin{align}
    \gamma_{ij} = \int \frac{\md^3 k}{(2\pi)^3} \gamma_{\bf k}^\lambda e^\lambda_{ij} e^{i \bf k\cdot x}~\,,\nonumber
\end{align} 
and similarly for the curvature perturbation.} with the quadratic scalar source is expressed as
\begin{align}
    \gamma_{{\bf k}, \lambda}^{\prime\prime} + 2 {\cal H} \gamma_{{\bf k}, \lambda}^\prime +  k^2 \gamma_{{\bf k},\lambda} = - 4 \int \frac{\md^3 q}{(2 \pi)^3} \epsilon_{\lambda}^{i j}( {\bf k} ) q_i q_j f(\tau, \vert {\bf k} - {\bf q} \vert, q) \zeta_{\bf q} \zeta_{{\bf k} - {\bf q}}~,
\end{align}
where $e^\lambda_{ij}$ are the polarization tensors and the source function is given by
\begin{align}\label{eq:ftoymodel}
    f(\tau, u, v) = &   \frac{ k^2 }{4} T_\beta(vx) T_\beta(ux) - \frac{1}{2 a^2 }  \left( \frac{a^2}{{\cal H}} T_\beta(vx) T_\zeta^\prime(ux) \right)^\prime + \frac{1}{{\cal H}} \left( T_\zeta(vx) T_\zeta(ux) \right)^\prime + \frac{3}{2 a^2} (a^2 T_\beta(vx) T_\zeta(ux) )^\prime \nonumber \\ &+  T_\zeta(vx) T_\zeta(ux)
    - \frac{b_5}{4 {\cal H}^2} \left( T_\zeta(vx) T_\zeta(ux) \right)^{\prime\prime} - \frac{b_5 x^2}{2}  T^\prime_\zeta(vx)T_\beta(ux) + ( u \leftrightarrow v )\,,
\end{align}
and we introduced $v=q/k$ and $u=\vert {\bf k} - {\bf q} \vert/k$.
In Eq.~\eqref{eq:ftoymodel}, we have introduced for compactness a transfer-function for the rescaled shift, that is
\begin{align}
    T_{\beta}(x) = - \frac{1 + b_5 }{{\cal H}} T_\zeta(x) - \frac{{\cal G}_s}{k^2} T_\zeta^\prime(x)\,,
\end{align}
which also depends on $b_5$. Note that, in contrast to $T_\zeta$ \eqref{eq:Tzeta}, $T_{\beta}(x)$ contains an explicit dependence on $b_5$, besides the one coming from a shifted sound speed $c_s$ \eqref{eq:cs} inside $T_\zeta$. 

As we will later argue, the constant shift in $c_s$ due to $b_5$ does not lead to new spectral features of the induced GW spectrum, for obvious reasons. It may slightly shift the position of the resonant peak and the overall GW spectrum amplitude. But as we will be mostly concerned in the case where $b_5\ll1$, these effects are minimal. Thus, it is convenient to treat separately terms in the induced GW source which contain prefactors proportional to $b_5$ from those which do not. The latter terms remain even in the limit $b_5\to 0$. In this way, we split $T_\beta$ into two contributions given by
\begin{align}\label{TF-beta}
      T_{\beta}(x) = T_{\beta}^{0}(x) - \frac{b_5}{{\cal H}} T_\zeta(x) \,,
\end{align}
where
\begin{align}\label{TF-beta-GR}
T_{\beta}^{0}(x) 
\equiv - \frac{1}{{\cal H}} \left( T_\zeta(x) + {\cal G}_s \frac{1}{x} \frac{\md{T}_\zeta(x)}{\md{x}}
\right) \,.
\end{align}
We proceed similarly with the source $f(\tau, u,v)$, and separate it into $f(\tau, u,v) = f_{0}(\tau,u,v) + \delta f(\tau,u,v)$, where $\delta f$ is the term with coefficients proportional to $b_5$. Both terms are explicitly are given by
\begin{align}\label{eq:fGR}
\begin{split}
    f_{0} =&  \frac{k^2}{4} T_\beta^{0}(vx) T_\beta^{0}(ux) - \frac{1}{2 {\cal H}} \left( T_\beta^{0}(vx) T_\zeta^\prime(ux) \right)^\prime + T_\zeta(vx) T_\zeta(ux) + \frac{1}{{\cal H}} (T_\zeta(vx) T_\zeta(ux))^\prime \\ &+ 3 {\cal H} T_\beta^{0}(vx) T_\zeta(ux) + \frac{3}{2} T_\beta^{0\prime}(vx) T_\zeta(ux)+(u\leftrightarrow v)~\,,
\end{split}
\end{align}
and
\begin{align}\label{eq:delta_f}
\begin{split}
    \delta f = & - \frac{b_5 x^2}{2}  T_\zeta^\prime(vx) \left( T_\beta^{\rm GR}(ux) - \frac{b_5}{{\cal H}} T_\zeta(ux) \right) 
    - \frac{ b_5 k x}{2} T_\beta^{\rm GR}(vx) T_\zeta(ux) - \frac{b_5^2 x^2}{4} T_\zeta(vx) T_\zeta(ux)  \\ &- \frac{ b_5}{ 2 {\cal H} } \left( T_\zeta(vx) T_\zeta(ux) \right)^\prime - \frac{9 b_5}{2 }  T_\zeta(vx) T_\zeta(ux)+(u\leftrightarrow v) \,.
\end{split}
\end{align}
 Note that $f_0$ contains the same interaction as in standard GR, however, with a shifted sound speed given by Eq.~\eqref{eq:cs}. In the limit $b_5\to 0$, this term exactly coincides with the GR case. On the other hand, $\delta f$ contain the new source term coming from the new contributions from GLPV which leads to the enhancement of the GW production. This term vanishes when $b_5\to 0$.
 
It is interesting to note that the first term in Eq. \eqref{eq:delta_f}, originating from the last cubic interaction in Eq.\eqref{eq:ftoymodel}, is proportional to $x^2$ and thus dominates on subhorizon scales. As can be seen in Eq.~\eqref{TF-beta-GR}, the second term is also proportional to $x^2$, but arises from the first term in Eq.~\eqref{eq:ftoymodel}. The former $x^2$ contribution comes directly from the new GLPV cubic interaction, scaling as $k^5$, while the latter appears via the substitution of the non-dynamical shift variable $\beta$ into the standard GR cubic interaction. In the next subsection, we show that the contribution from the new GLPV interaction indeed dominates when performing the time integration in Eq.~\eqref{eq:I}, compared to the other term.

\subsection{Modified induced GW Kernel}

We now compute the two-point function of induced tensor modes and analytically integrate the Kernel. Following the standard approach for induced GWs (see, e.g., Ref.~\cite{Domenech:2021ztg}), the dimensionless power-spectrum\footnote{The dimensionless power-spectrum is defined by
\begin{align}
    \langle h_{\mathbf{k},\lambda}h_{\mathbf{k}',\lambda'}\rangle=\frac{2\pi^2}{k^3}{\cal P}_{h,\lambda}(k)\times (2\pi)^3\delta_{\lambda\lambda'}\delta^{(3)}(\mathbf{k}+\mathbf{k'})\,.
\end{align}} for the scalar-induced tensor modes can then be written as
\begin{align}\label{eq:Phktau}
    {\cal P}_{h}(k,\tau)&=\sum_\lambda {\cal P}_{h,\lambda}(k,\tau)\nonumber\\&=8\int_0^\infty dv\int_{|1-v|}^{1+v}du\left(\frac{4v^2-(1-u^2+v^2)^2}{4uv}\right)^2 I^2(v,u,x){\cal P}_\zeta(vk){\cal P}_\zeta(uk)\,,
\end{align}
in which we introduced the kernel function $I(v,u,x)$ explicitly given by
\begin{align}\label{eq:I}
    I(u,v,x)=\int_0^x dx_1 G_k(x,x_1)( f_{0}(u,v,x_1) + \delta f(u,v,x_1) ) \equiv I_{0} + \delta I\,,
\end{align}
where in the last step we separated the kernel for the different contributions from $f_0$ and $\delta f$. For clarity, let us remind the reader that $\delta I$ is directly proportional to powers of $b_5$, while $I_0$ depends on $b_5$ only through $c_s$.
In Eq.~\eqref{eq:I}, $G_k(x,x_1)$ is the Green's function of tensor modes in radiation domination, that is 
\begin{align}
G_k(x,x_1) = \frac{x_1}{x} \left( \cos x_1 \sin x - \cos x \sin x_1 \right)~.
\end{align}

We focus now on the analytical integral for the Kernel. We first plug in the linear solution to the curvature perturbation, Eq.~\eqref{eq:Tzeta}, into $f_{0}$ \eqref{eq:fGR} and $\delta f$ \eqref{eq:delta_f}. The form of $f_{0}$ is that of the standard GR case with a shifted sound speed, which can be found in the same notation as in the present paper in Ref.~\cite{Domenech:2024drm}. The new term, $\delta f(u,v,x)$, explicitly reads 
\begin{align}\label{eq:fsimplifiedincis}
\begin{split}
\delta f(v,u,x)=&\left(c^-_{0}+\frac{c^-_{2}}{x^2}\right)\cos(c_s(u-v)x)+\left(c^+_{0}+\frac{c^+_{2}}{x^2}\right)\cos(c_s(u+v)x)
\\&
+\left(c_{-1}^- x +\frac{c^-_{1}}{x}\right)\sin(c_s(u-v)x)+\left(c_{-1}^+ x + \frac{c^+_{1}}{x}\right)\sin(c_s(u+v)x)\,,
\end{split}
\end{align}
where the coefficients $c_i$ are constant in time and their explicit expressions in terms of $u$, $v$ and $b_5$ is listed in App.~\ref{subsec:Coefficients_c_i}. Note that $c_i$ vanish for $b_5=0$ giving $\delta{f}=0$ as expected. Most importantly, note that the terms proportional to $c_{-1}^{\pm} $ grow linearly in $x$. These terms arise from the new GLPV cubic interaction and will dominate for $x \gg 1$, that is on subhorizon scales. On superhorizon scales, the contributions from standard GR dominate. 

The time integral in \eqref{eq:I} can be performed analytically. Substituting \eqref{eq:fsimplifiedincis} into the kernel \eqref{eq:I}, and focusing on the subhorizon limit $x \gg 1$ (see App.~\ref{subsec:Modified_Kernel} for the full result), we obtain that 
\begin{align}\label{eq:deltaIexplicit}
\begin{split}
   &  \delta I(u,v,x\gg 1) \approx  - \frac{\sin x}{x}  \delta I_{c,0} - \frac{\cos x}{x}  \delta I_{s,0} + \left( \frac{c_{-1}^-}{y_4^2} - \frac{c_{-1}^-}{y_1^2} + \frac{c_0^- }{2  c_s^2  u  v (1 - y_t) } \right) \cos(c_s (u-v) x) \\
    & + \left(  \frac{c_{-1}^+ }{y_2^2} - \frac{c_{-1}^+ }{y_3^2} -\frac{c_0^+ }{2  c_s^2  u  v (1 + y_t)} \right) \cos (c_s (u+v) x) + d_1 \frac{\sin(c_s (u-v) x)}{x} + d_2 \frac{\sin(c_s (u+v) x)}{x}\,.
\end{split}
\end{align}
In Eq.~\eqref{eq:deltaIexplicit}, the standard wave mode functions have time-independent coefficients given by
\begin{align}\nonumber
\begin{split}
    \delta I_{c,0}(v,u)=&-   \frac{c_{-1}^-}{y_1^3} + \frac{c_{-1}^-}{y_4^3} - \frac{c_{-1}^+}{y_3^3} + \frac{c_{-1}^+}{y_2^3} + \frac{c_0^- \left(1+c_s^2
   (u-v)^2\right)}{\left(1-c_s^2
   (u-v)^2\right)^2}+\frac{c_0^+ \left(1+c_s^2
   (u+v)^2\right)}{\left(1-c_s^2
   (u+v)^2\right)^2}\\& +\frac{c_1^- c_s (u-v)}{1-c_s^2
   (u-v)^2}+\frac{c_1^+ c_s (u+v)}{1-c_s^2 (u+v)^2} +\frac{c_2^-}{2}\ln\left|\frac{1-c_s^2(u-v)^2}{1-c_s^2(u+v)^2}\right|\,,
\end{split}
\\
   \delta I_{s,0}(v,u) =&-\frac{\pi c_2^-}{2}{\rm sign}(1+y_t)\Theta(1-y_t^2)\,,
\end{align}
where in the last equality we used that $c_2^+=-c_2^-$ (see App.~\ref{subsec:Coefficients_c_i})
and we defined for compactness
\begin{align}
   y_1 &=1 -c_s(u-v)~,\quad y_2= 1 +c_s(u+v)~,\quad y_3= 1 -c_s(u+v)\,,\\ y_4 &=1 +c_s(u-v)\,,\quad y_t= \frac{-y_2 y_3}{2c_s^2uv}\,.
\end{align}
The coefficients of the remaining terms in Eq.~\eqref{eq:deltaIexplicit} explicitly read
\begin{align}\label{coef-d-i}
\begin{split}
    d_1 =& c_{-1}^-  \left(   \frac{-2 + y_4^2 x^2}{2 y_4^3 }  + \frac{-2 + y_1^2 x^2}{2y_1^3 }  \right)+  \frac{1}{2c_s (u-v)  (c_s^2 uv(1-y_t))^2}
\left(c_0^-+\frac{c_1^-}{c_s(u-v)}
   c_s^2 u v (1-y_t)\right)\,, \\
   d_2 =& c_{-1}^+ \left( \frac{-2 + y_2^2 x^2}{2 y_2^3} + \frac{-2 + y_3^2 x^2}{2 y_3^3 } \right) +\frac{1}{2c_s (u+v) (c_s^2uv(1+y_t))^2}
\left(c_0^+-\frac{c_1^+}{c_s(u+v)}
   c_s^2 u v (1+y_t)\right)\,.
\end{split}
\end{align}

Let us now discuss the structure of the Kernel. Overall, we see that the Kernel has the same structure as in Horndeski gravity \cite{Domenech:2024drm}. Namely, due to the higher-derivative interactions, the induced GWs grow on subhorizon scales and mix with the scalar modes. For instance, we get GW mode propagation with the scalar sound speed.\footnote{Note that, while these terms are also present in standard GR, they decay rapidly deep inside the horizon.} However, in contrast to Horndeski gravity, the new higher-order interaction leads to an even faster growth. In the resonance band, that is $y_3=0$ or, equivalently, $1=c_s(u+v)$, the amplitude of induced tensor modes grow as fast as $x^2$ (note the presence of $x^2$ in the coefficients $d_{1,2}$ defined in \eqref{coef-d-i}). Outside the resonance band, the growth is proportional to $c_{-1}^\pm$ and linear in $x$. Therefore, the new interaction term with third-order derivatives leads to an additional factor $x$ enhancement relative to Horndeski (and its disformally connected theories).


\subsection{Gauge transformation and possible gauge ambiguities}

 It is known that the scalar-induced GW spectrum is gauge-dependent \cite{Hwang:2017oxa}, and that the unitary (or uniform density) gauge is not appropriate for such calculations \cite{Hwang:2017oxa,Gong:2019mui,Tomikawa:2019tvi,Domenech:2020xin}.  Nevertheless, it has been shown in GR that the Newton, flat or synchronous yield the same prediction deep inside the horizon \cite{DeLuca:2019ufz,Inomata:2019yww,Yuan:2019fwv,Lu:2020diy,Domenech:2020xin}.\footnote{See \cite{Yuan:2024qfz} for a related discussion in the case of isocurvature perturbations.} Although the gauge issue is more subtle in modified gravity \cite{Domenech:2024drm}, we expect similar conclusions once GR is recovered. Given this, following Ref.~\cite{Domenech:2024drm}, we perform a gauge transformation into Newton gauge to avoid possible gauge artifacts. 

 The tensor modes in the Newton gauge are related to those in the unitary gauge at second order via 
\begin{align}
h_{ij}^N= h^u_{ij}- T^{ab}_{ij} \partial_a \bar\beta\partial_b\bar\beta\,,
\end{align}
where $T^{ab}_{ij}$ is the projector onto the traceless-transverse components. After performing the gauge transformation, the kernel in the Newton gauge becomes
\begin{align}
    I^N = & I^u - \frac{1}{4} k^2 \left( T_\beta^{0} - \frac{b_5}{{\cal H}} T_\zeta \right) \left( T_\beta^{0} - \frac{b_5}{{\cal H}} T_\zeta \right) \nonumber \\
\begin{split}
    \approx & I^u  - \frac{ (1  +  b_5)^2 }{8 c_s^2 u v } \cos (c_s (u-v) x) + \frac{(1  +  b_5)^2 }{8 c_s^2 u v } \cos (c_s (u+v) x) 
    \\
    & - \frac{ {\cal G}_s  (1+b_5) (u -v)  }{8 c_s u^2 v^2  x} \sin(c_s (u-v) x) - \frac{ {\cal G}_s (1+b_5) (u +v)  }{8 c_s u^2 v^2 x} \sin (c_s (u+v) x)\,,
\end{split}
    \label{eq:Inewtongauge}
\end{align}
where the superscripts ``$N$'' and ``$u$'' refer to Newton and unitary gauges, respectively. Interestingly, the additional terms from the gauge transformation are subleading with respect to the new GLPV term in $I^\mu$, which grows with $x$, in contrast to Horndeski and GR. In other words, the leading contribution from the new term is not affected by the transformation to the Newton gauge. 

We can generalize the above statement to an arbitrary gauge transformation. Consider a gauge transformation $x^\mu\to x^\mu+\xi^\mu$, where $\xi^\mu = (T,\partial^i L)$, whereby we neglect the vector components. The metric perturbations $\gamma_{ij}$ and $\zeta$ transform up to second order as \cite{Domenech:2017ems}
\begin{align}\label{eq:transformationgeneral}
    \gamma_{ij}^G = & \gamma_{ij}^u - T_{ij}^{ab} \left( \partial_a T \partial_b T + \partial_a \partial_k L  \partial_b \partial_k L + ...\right)~, \\
    \zeta^G = & \zeta^u - \frac{\Delta^{-1}}{4} \left( \dot \gamma_{ij}  \partial_i \partial_j T + \partial_i \partial_j  \partial_k L \partial_k \gamma_{ij} + ... \right)~, 
\end{align}
where we only consider the terms relevant for the scalar-scalar-tensor interaction. The general argument goes as follows. First, notice that the transformation \eqref{eq:transformationgeneral} of the tensor perturbation can only lead to even powers of derivatives in the scalar-scalar-tensor interaction, since it depends quadratically in derivatives of $T$ and $L$. Therefore, it is not possible to modify the new source term in GLPV by gauge transformation of the tensor modes as it contains on an odd power of derivatives. Second, a gauge transformation of $\zeta$ \eqref{eq:transformationgeneral} can a priori modify the new source term by considering, for instance, $T= \partial^2 \zeta$. However, if we consider such transformation, namely $T= \partial^2 \zeta$, the transformation of $\gamma_{ij}$ \eqref{eq:transformationgeneral} would artificially introduce higher derivatives in the action.
Therefore, we conclude that it is not possible to gauge away the new source term in GLPV scaling as $\partial^5 \sim k^5$ without introducing an even higher-order spatial derivative term in the cubic action.
The fundamental reason of such ``gauge invariance'' follows from the explicit breaking the second-order nature of the equations of motion. 

\subsection{Induced GW spectral density}

\begin{figure}
	\centering
	\includegraphics[width=0.7\linewidth]{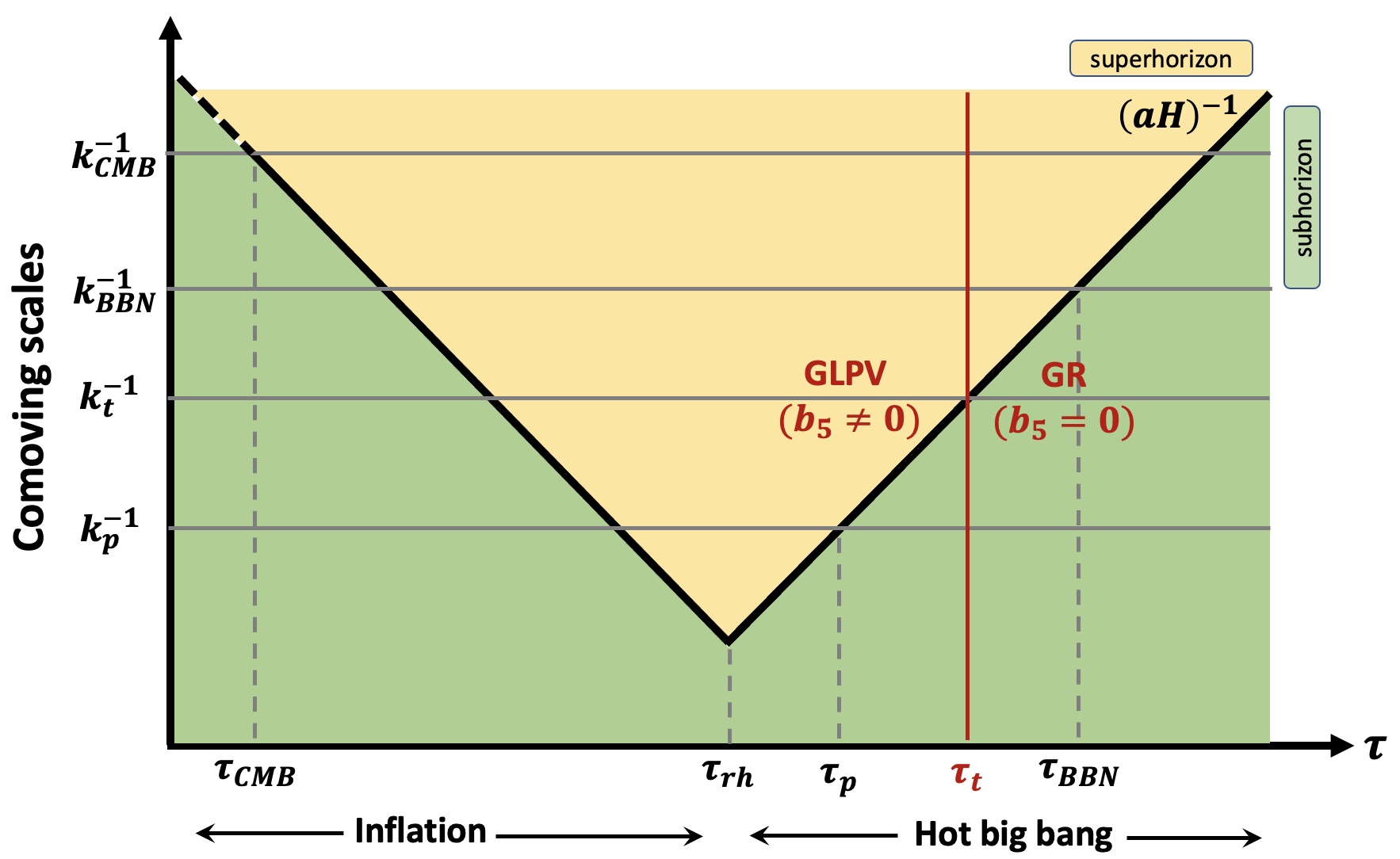}
	\caption{GLPV effects dominate for modes that become subhorizon $k\tau>1$ ($k > aH$) before the transition to GR ($\tau<\tau_t$), with $\tau_t$ occurring before BBN ($\tau_{BBN}$). Assuming the initial power spectrum peaks at $k_p$, the dominant GLPV impact is on the subhorizon modes with $k_p>k_t$, where $k_t=1/\tau_t$.}
	\label{fig:modes}
\end{figure}

To be consistent with observations, we require that we recover GR at some point well before BBN. As the modified gravity corrections to the induced GWs grow in time and, therefore, the enhancement depends on the duration of the modified gravity epoch. For simplicity, we assume that GR is recovered instantaneously at an arbitrary conformal time $\tau=\tau_t$ (see Fig.~\ref{fig:modes}). In other words, we take $b_5=0$ for $\tau>\tau_t$. The kernel derived above is valid as long as the scales of interested entered the horizon during the GLPV era. If the primordial power spectrum has a peak at $k=k_p$, then above requirement translates into $k_p\tau_t=k_p/k_t\gg 1$.
In that case, we cut-off the non-GR source term at an arbitrary time $\tau=\tau_t$, i.e. 
\begin{align}\label{eq:generalI}
    I =  \int^{x_t}_0 \md x^\prime G_k(x,x^\prime) f  + \int^x _{x_t}\md x^\prime G_k(x,x^\prime) f_{\rm GR} ~\,,
\end{align}
where the first term denotes the contribution arising during the modified gravity epoch (given by eq.~\eqref{eq:I}) while the second term denotes the contributions after the transition to standard GR (given by eq.~\eqref{eq:I} in the limit $b_5\to 0$).

Let us show that for our purposes we can neglect the second term in Eq.~\eqref{eq:generalI}, which corresponds to the induced GWs sourced after the GLPV phase. To do that, let us assume that there are no jumps in the background solutions, or in the curvature perturbation $\zeta$, or its derivatives at $\tau=\tau_t$. Although this might not be in general the case, we will show later that for a smooth transition the effect only introduces ${\cal O}(1)$ factors. With no jumps, the matching conditions for the curvature perturbation at $\tau_t$, i.e., at the end of the GLPV era and the beginning of the GR era, are the continuity of $\zeta$ and its time derivative. Doing so, we find that
\begin{align}
    T_\zeta(\tau>\tau_t) = A\frac{\sin\left(x/\sqrt{3}\right)}{ x/\sqrt{3}}+B\frac{\cos\left(x/\sqrt{3}\right)}{ x/\sqrt{3}}\,,
\end{align}
where $A$ and $B$ are ${\cal O}(1)$ coefficients.\footnote{Explicitly they are given by
\begin{align}
    A&=\frac{1}{4c_s}\left(2\left(c_s+\tfrac{1}{\sqrt{3}}\right) \cos \left(\tfrac{b_5
   x_t}{2}\right)+b_5 \cos \left[\left(c_s+\tfrac{1}{\sqrt{3}}\right) x_t\right]\right)\nonumber\,,\\
   B&=\frac{1}{4c_s}\left(2\left(c_s+\tfrac{1}{\sqrt{3}}\right) \sin \left(\tfrac{b_5
   x_t}{2}\right)-b_5 \sin \left[
   \left(c_s+\tfrac{1}{\sqrt{3}}\right) x_t\right]\right)\nonumber
  \,.
\end{align}
For $b_5\ll1$, we see that scales for which $b_5x_t\ll1$, the coefficient of the decaying mode is suppressed and do not feel any effect from the transition. For scales which $b_5x_t\gg1$ both coefficients are ${\cal O}(1)$.
}
Thus, the main effect of the transition is to excite the ``decaying'' mode of the curvature perturbation regardless of the value of $b_5$. Nevertheless, their coefficients are at most ${\cal O}(1)$. This shows that there are no enhancements due to the transition, since the source terms after the transition is exactly that of GR in a radiation dominated universe. Furthermore, since our kernel during the GLPV (first term in Eq.~\eqref{eq:generalI}) era is exact, that is it includes the GR contributions as well, we can use well-known results (e.g. from Ref.~\cite{Kohri:2018awv}) to conclude that, for $x_t\gg1$, the integral from $x_t$ to $\infty$ (second term in Eq.~\eqref{eq:generalI}) is suppressed by an additional factor $1/x_t$ (see also Ref.~\cite{Domenech:2020kqm}). Thus, we can safely neglect this contribution for $k\gg k_t$.


In general, one also should be careful to account for additional contributions at the matching hypersurface due to the instantaneous transition. However, as we discussed in Ref.~\cite{Domenech:2024drm} in the case of Horndeski gravity, we expect that the corrections from the matching are of ${\cal O}(1)$ and, therefore, not relevant for our general conclusions. To confirm this, we will consider a more realistic scenario in Sec.~\ref{sec:Effective_Cubic_Interaction} with a decaying coefficient for the GLPV interaction. We will see the results are consistent up to ${\cal O}(1)$ coefficients, as we anticipated. For this reason, let us focus for now on the contributions arising during the modified gravity epoch, corresponding to the first term in Eq.~\eqref{eq:generalI}.

The induced GW spectral density, once GR is recovered, is then given by \cite{Domenech:2021ztg}
\begin{align}\label{eq:inducedGWspectrum}
    \Omega_{\rm GW}(\tau> \tau_t) &=\frac{k^2}{12{\cal H}^2}\overline{{\cal P}_h(k\tau)} \nonumber\\&=\frac{1}{3} \int_0^\infty  \md v \int_{\vert 1-v \vert}^{1+v} \md u\, \left( \frac{4 v^2 - (1-u^2+ v^2)^2}{4 u v} \right)^2 {\cal P}_\zeta(v k) {\cal P}_\zeta(u k) \nonumber \\
    & \qquad\qquad\times \left[ ( I_{c}^{0,N}(u,v) + \delta I_c^N(u,v,x_t) )^2 + \left( I_s^{0,N}(u,v) + \delta I_s^N(u,v,x_t) \right)^2  \right]\,,
\end{align}
where an overline denotes oscillation average, $\delta I_{c/s}$ are given by Eqs. \eqref{eq:generalIc} and $\eqref{eq:generalIs}$ evaluated at $x_t$, and $I_{c/s}^{0}$ are the standard expressions of the kernel in GR in the Newton gauge up to the shift in the sound speed, namely
\begin{align}
   I^{0}_{c}(u,v)=-\frac{y_t}{c_s^2uv}\left(1-\frac{1}{2}y_t\,\ln\left|\frac{1+y_t}{1-y_t}\right|\right)\quad,\quad
I^{0}_{s}(u,v)=\frac{\pi y_t^2}{2 c_s^2 u v}\,{\rm sign}(1+y_t)\Theta(1-y_t^2)\,.\label{eq:IsGR}
\end{align}
 
We discuss the results for two different primordial spectra, log-normal and scale-invariant, in more details below.

\subsubsection{Log-Normal Spectrum}
First, we consider a log-normal primordial spectrum given by
\begin{align}
    \label{eq:Log_Normal_Spectrum}
    {\cal P}_\zeta = \frac{A_\zeta}{\sqrt{2 \pi \Delta^2}} e^{- \frac{\log^2(k/k_p)}{2 \Delta^2}}\,,
\end{align}
where $A_\zeta$ is the amplitude, $\Delta$ the dimensionless width, and $k_p$ the peak position. We calculate the spectrum numerically. In the left panel of Fig.~\ref{fig:Omega_GW_toy_model}, we show the induced GW spectrum for $\Delta =0.2$, $b_5 =10^{-4}$ and $c_s^2=1/3$. We see that we recover GR with the universal slope $k^3 \log^2 k$ in the infrared tail \cite{Cai:2018dig,Yuan:2019wwo} (though it becomes $k^3$ in the presence of dissipation \cite{Domenech:2025bvr}). At an intermediate scale the spectrum gets enhanced and grows as $k^7$ while the resonance peak is enhanced by a factor $b_5^2 x_t^5$. Therefore, the new resonance leads to an enhancement factor proportional to $x_t^2$ compared to Horndeski \cite{Domenech:2024drm}.

\subsubsection{Scale-Invariant Spectrum}
\label{subsubsec:Scale_Invariant}

\begin{figure}
    \centering
     \includegraphics[width=0.45\linewidth]{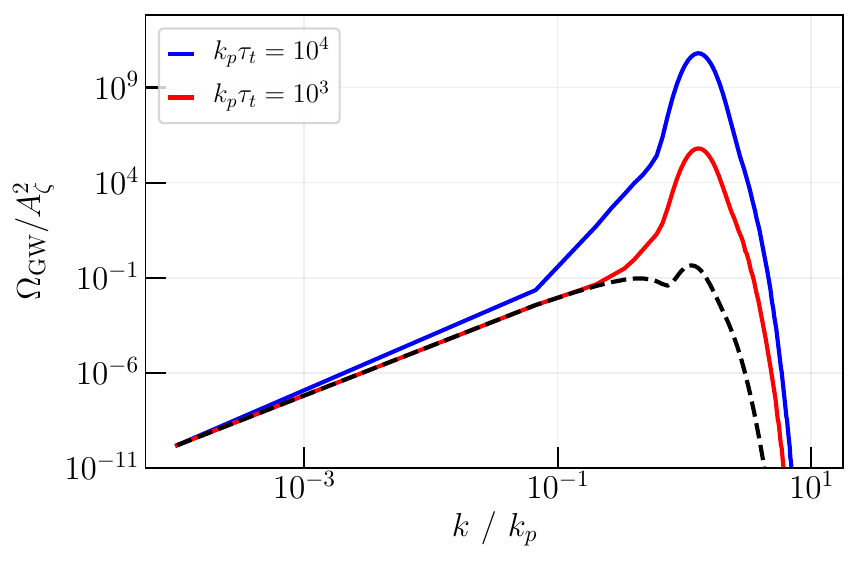}
    \includegraphics[width=0.45\linewidth]{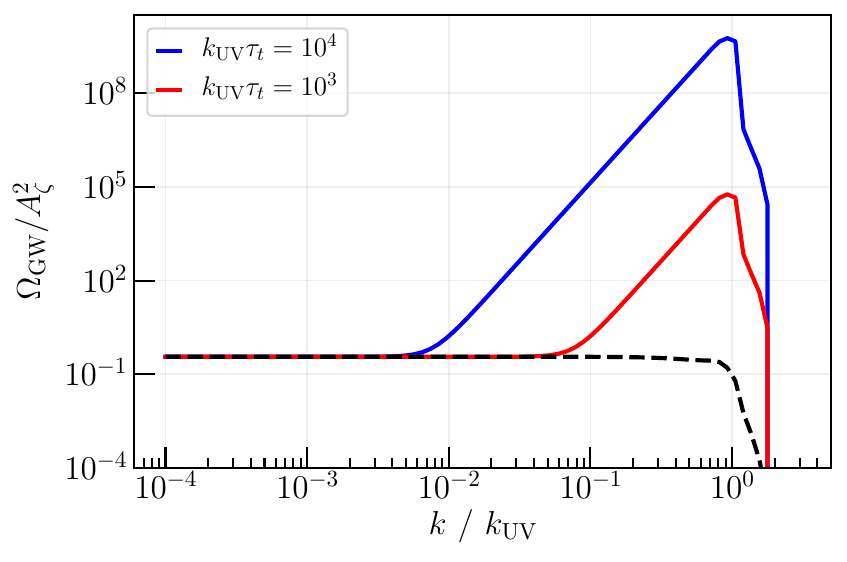}
    \caption{Spectral density of the induced GWs normalized by $A_\zeta^2$ for $b_5=10^ {-4}$ for two different values of $\tau_t$. The left hand side is a log-normal primordial spectrum and the right hand side is a scale-invariant one. The dashed lines show the GR limit.}
    \label{fig:Omega_GW_toy_model}
\end{figure}

Second, we take a scale-invariant primordial spectrum, that is
\begin{align}
    {\cal P}_\zeta = A_\zeta \Theta( k_{\rm UV} - k)\,,
\end{align}
where $k_{\rm UV}$ is a small-scale cut-off\footnote{
Note that $k_{\rm UV}$ is a cutoff we introduced to later avoid non-linearities. This is not to be confused with the UV cutoff of from an EFT point of view, which is customarily  called $\Lambda_{\rm UV}$. See discussion in Sec.~\ref{sec:backreaction}.} to avoid overproduction of induced GWs. For analytical considerations it is sufficient to focus solely on the resonance $y_3 = 1 - c_s (u+v) =0$ band, as it dominates over other contributions. Following \cite{Domenech:2024drm}, a good approximation is given by integration around the resonance scales. In this way, the induced GW peak is approximately given by
\begin{align}
    \Omega_{\rm GW}^{\rm res}  & \approx\frac{A_\zeta^2}{3}  \int_0^\infty  \md v \int_{\vert 1-v \vert}^{1+v} \md u\,   \left( \frac{4 v^2 - (1-u^2+ v^2)^2}{4 u v} \right)^2  \nonumber\\& \qquad\qquad\qquad\qquad\times\left[ \frac{b_5^2 (1+b_5)^2 (u+v)^2}{64 c_s^2 u^2 v^2} \frac{\sin(y_3 x_t)^2 x_t^6}{4 y_3^2 x_t^2} \Theta(u_{\rm uv} - u) \Theta( v_{\rm uv} - v ) \right] \nonumber \\
     & \approx\frac{A_\zeta^2 (1-c_s^2)^2}{96} \frac{b_5^2(1+b_5)^2 \pi}{c_s}   x_t^5 \int_{-d_0(k)}^{d_0(k)} \md d\,  \frac{(1-d^2)^2 }{(1- c_s^2 d^2)^4} \,,\label{eq:inducedgwscaleinv}
\end{align}
where we introduced $d=u-v$, $v_{\rm UV}=k_{\rm UV}/k$ and
\begin{align}\label{eq:d0}
d_0(k)=\left\{
\begin{aligned}
&1\qquad &v_{UV}>\tfrac{1+c_s}{2c_s}\\
&2v_{\rm UV}-\frac{1}{c_s}\qquad & \tfrac{1}{2c_s}<v_{\rm UV}<\tfrac{1+c_s}{2c_s}\\
&0\qquad &v_{\rm UV}<\tfrac{1}{2c_s}
\end{aligned}
\right.\,,
\end{align}
which stems from momentum conservation. 
For $b_5^2 x_t^5 > 1$ the contribution from the resonance dominates over the standard GR contribution and leads to a growth scaling as $k^5$, as anticipated from the $x_t^5$ factor in Eq.~\eqref{eq:inducedgwscaleinv}. In the right panel of Fig.~\ref{fig:Omega_GW_toy_model}, we plot the resulting induced GW spectrum for a scale-invariant primordial spectrum and $b_5 =10^{-4}$. 

\begin{figure}
    \centering
    \includegraphics[width=0.8\linewidth]{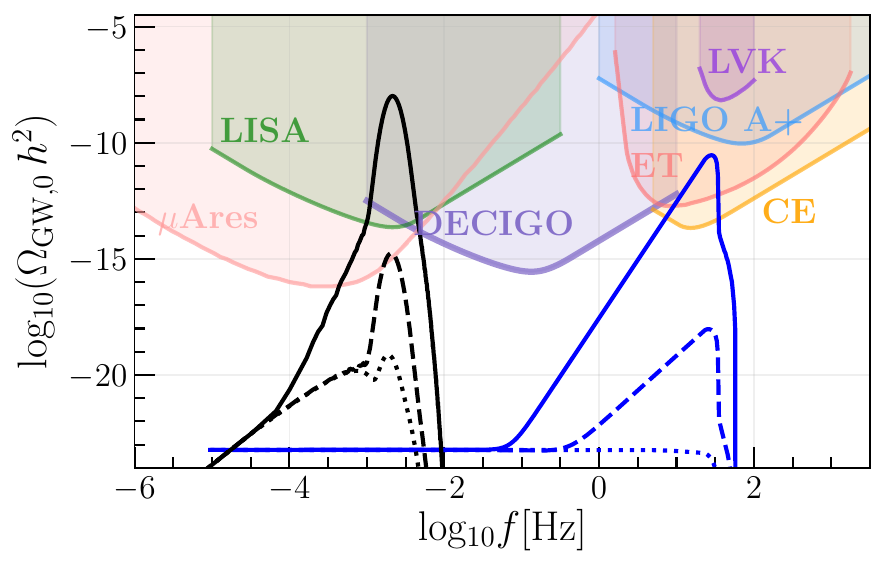}
    \caption{We show the induced GW spectrum today for a log-normal (peaked at $k_p$) and a scale-invariant (with a UV cut-off at $k_{\rm UV}$) spectrum of curvature fluctuations, respectively shown in black and blue lines. Solid lines indicate the induced GWs from the GLPV model, while dashes lines indicate the Horndeski model in \cite{Domenech:2024drm} (we also provide details in App.~ \ref{subsubsec:Horndeski_Toy_Model}). The dotted lines indicate the GR limit. We fix the amplitude of the scalar power spectrum to $A_\zeta = 10^{-7}$ and $A_\zeta = 10^{-9}$ for the log-normal and scale-invariant case, respectively. Further, we set $\tilde \Gamma = b_5= 10^{-4}$ and $k_p/k_t= 10^4$ and $k_{\rm UV}=k_t=10^{4.5}$. For the log-normal distribution we fixed $f_p=k_p/(2\pi) \simeq 1.7 \times 10^{-3}\,{\rm Hz}$ and for the scale-invariant spectrum $f_{\rm UV} \simeq 30\,{\rm Hz}$. For illustration purposes, we show the power-law integrated sensitivity curves as described in Ref.~\cite{Thrane:2013oya,Schmitz:2020syl}
    for LIGO A+ \cite{A+}, Einstein Telescope (ET) (15km arms triangular configuration \cite{Branchesi:2023mws}), Cosmic Explorer (CE) \cite{ce}, DECIGO \cite{Yagi:2011wg,Kawamura:2020pcg}, LISA \cite{Barke:2014lsa} and $\mu$-Ares \cite{Sesana:2019vho} experiments. The purple-shaded region shows the upper bounds from the LVK collaboration \cite{KAGRA:2021kbb}. }
    \label{fig:Omega_Sensitivity_Curves}
\end{figure}

In Fig.~\ref{fig:Omega_Sensitivity_Curves} we show the induced GW spectral density today for the log-normal and the scale-invariant spectra for $b_5=10^{-4}$. We also include the sensitivity curves of the current and upcoming GW experiments. In addition, we compare the new induced GW spectrum in GLPV with the standard GR prediction and the results from Horndeski gravity from the model constructed in Ref.~\cite{Domenech:2024drm} (see also in App.~\ref{subsubsec:Horndeski_Toy_Model} for the explicit model).

\subsection{Non-linear backreaction }
\label{sec:backreaction}

Higher derivative interactions also affect curvature perturbations. Since, in our model, their strength grows with time, we require that the terms with higher derivatives of the third-order action never dominate over the second-order action for consistency of the perturbative expansion. Otherwise, one would have to consider all higher derivative terms at higher orders in perturbation theory, like in the Vainshtein screening \cite{Vainshtein:1972sx,Kimura:2011dc,Koyama:2013paa,Babichev:2013usa}. This is why we refer to this requirement as the non-linear backreaction bound. As we shall see later, the backreaction bound is stronger than the UV cut-off if we treat GLPV as an EFT-like expansion.

To estimate the backreaction of the non-linear interaction on the evolution of the curvature perturbation we follow the approach in \cite{Domenech:2024drm}.
The cubic interaction of the curvature perturbation in GLPV has been analyzed in \cite{Renaux-Petel:2011zgy,Passaglia:2018afq}. Focusing on the terms with the highest number of derivatives we obtain terms scaling as
\begin{align}
    \label{eq:sss_interaction}
    {\cal L}_{\zeta\zeta\zeta} \sim b_5 \frac{\partial^2 \zeta}{a H^2} (\partial_k \zeta)^2~.
\end{align}
Note that even in standard GR one obtains the same operator, however, these terms can be removed by a gauge-transformation and, therefore, we do not consider them. 

Schematically, the equation of motion for the curvature perturbation up to quadratic order focusing only the terms with the highest number of derivatives can be expressed as
\begin{align}
    \zeta^{\prime\prime } + 2 {\cal H} \zeta^\prime + c_s^2 \left( 1 + \frac{b_5}{{\cal H}^2} \partial^2 \zeta + \frac{b_5}{{\cal H}^3} \partial^2 \gamma^\prime \right) \partial^2 \zeta \sim 0\,,
\end{align}
where we neglected order one coefficients and the index structure as we are only interested in the scale at which the higher order terms will get important.  Thus, in order to avoid the backreaction of the higher derivatives on the curvature perturbation we need to require that
\begin{align}\label{eq:nonlinearbound}
  \Big \vert  b_5 \frac{\partial^2 \zeta}{{\cal H}^2} \Big \vert \lesssim 1~, \qquad  \Big \vert b_5 \frac{\partial^2 \gamma^\prime}{{\cal H}^3} \Big \vert \lesssim 1~.
\end{align}
Using the linear solution of the curvature and tensor perturbation in Fourier space, that is
\begin{align}
    \zeta \sim \sqrt{P_\zeta} \frac{\sin c_s x}{c_s x}~, \qquad \gamma \sim \sqrt{P_h} \frac{\sin x}{x}\,,
\end{align}
we interpret \eqref{eq:nonlinearbound} as an upper bound on $x$, which yields
\begin{align}\label{eq:xNLbound}
    x\lesssim x_{\rm NL} \equiv {\rm min} \left( b_5^{-1} P_\zeta^{-1/2}, b_5^{-1/2} P_h^{-1/4} \right)~.
\end{align}
Considering that the GLPV phase terminates at $\tau=\tau_t$, $x_{\rm NL}$ determines the maximum linear scale below which we can trust the perturbative expansion. Namely, evaluated at the transition we have $x_{\rm NL}=k_{\rm NL}\tau_t$ and only consider  modes with $k\lesssim k_{\rm NL}$.

In our current setup, the strongest bound comes from the second term inside the parenthesis in Eq.~\eqref{eq:xNLbound}. Using that the induced GW power spectrum in the resonant regime scales as (see Eqs. \eqref{eq:inducedgwscaleinv} and \eqref{eq:inducedGWspectrum})
\begin{align}
    P_h \sim b_5^2 P_\zeta^2 x_{\rm NL}^3~,
\end{align}
the latter bound dominates leading to
\begin{align}\label{eq:boundbound}
    x\lesssim x_{\rm NL} \equiv {b_5^{-4/7} P_\zeta^{-2/7}}~.
\end{align}
For example, for $b_5 \sim 10^{-4}$ and a scale-invariant power-spectra with $P_\zeta \sim 10^{-9}$ or a peaked spectrum with $P_\zeta \sim 10^{-7}$  the bound is given by $x_{\rm NL} \lesssim 10^{34/7}$ and $x_{\rm NL} \lesssim 10^{30/7}$, respectively.

Before concluding this section, let us clarify that our non-linear backreaction bound is a classical effect. Namely, we want to avoid the system to dynamically enter a higher derivative dominated regime. One may wonder whether the non-linear backreaction bound is similar to the UV cutoff scale if we consider GLPV as an EFT expansion \cite{Gubitosi:2012hu}. We find that, in the EFT derivative expansion, the UV cutoff  scale $\Lambda_{\rm UV}$ for the tensor-scalar-scalar interaction is given by $\Lambda_{\rm UV}^4 \sim {H}^3M_{\rm pl}/b_5 $. The UV cutoff is lowest at the transition, that is at $H=H_t$, leading to
\begin{align}
    \Lambda_{\rm UV} \sim 6 \times 10^{21} \,{\rm Hz}\left( \frac{b_5}{10^{-4}} \right)^{-1/4}  \left( \frac{H_t}{10^{-3}\,{\rm eV}} \right)^{3/4} \left( \frac{M_{\rm pl}}{2.4\times 10^{18}\,{\rm GeV}} \right)^{1/4}~.
\end{align}
We then require the cutoff to be much higher that the GW frequencies of interest, that is $k_{\rm GW}/a(\tau) \ll \Lambda_{\rm UV}$. We have checked that this is easily satisfied in our model because $b_5\ll 1$. It should be noted that our situation is different than in the case of the EFT of dark energy. There one finds that $k_{\rm LIGO} \sim \Lambda_{\rm UV}^{\rm DE} = (H_0^2 M_{\rm pl})^{1/3}$, which limits the validity of the EFT of dark energy for describing the gravitational waves at the LIGO frequency band \cite{deRham:2018red}.

It is worth mentioning that, one finds a cutoff
$\tilde \Lambda_{\rm UV} \sim \big(H^2 M_{\rm pl}/b_5\big)^{1/3}$  from the scalar-scalar-scalar interaction in Eq.~\eqref{eq:sss_interaction}, which can be higher than $\Lambda_{\rm UV} \sim \big(H^3 M_{\rm pl}/b_5\big)^{1/4}$ that is read off from tensor-scalar-scalar interaction. The origin of this mismatch is the fact that the $k^5$ terms survive only in the tensor-scalar-scalar sector, while they cancel in the reduced scalar-scalar-scalar cubic action in unitary gauge after solving the constraints. This behavior is consistent with the fact that our model lies in the U-DHOST (unitary-degenerate) class \cite{DeFelice:2018ewo,DeFelice:2021hps,DeFelice:2022xvq}, where the degeneracy conditions hold only in unitary/uniform gauge. In a generic gauge one generally finds an additional non-propagating shadowy mode, governed by an elliptic equation rather than a wave equation. Integrating out this mode generically regenerates higher-derivative scalar self-interactions suppressed by the same scale as the tensor-scalar-scalar vertex. Therefore, the apparently higher scale $\tilde\Lambda_{\rm UV}$ inferred from the unitary-gauge scalar cubic action should not be interpreted as the physical EFT cutoff; instead, the physical UV cutoff is expected to be controlled by the lower scale $\Lambda_{\rm UV}$. It is interesting that the tensor-scalar-scalar interaction already knows about this cutoff even in unitary gauge.

\section{Effective Model}
\label{sec:Effective_Cubic_Interaction}

In the previous section we considered a simple toy model within GLPV where the coefficients are constant. Here we consider the effect of a general time-dependence in such coefficients. For simplicity, we will focus only on the impact on the new interaction term. This in turn allows us to check if the induced GW spectrum prediction is robust in more realistic scenarios. In particular, we will see that the characteristic $k^5$-growth of the induced GW spectrum, in the resonance regime, is mostly insensitive to time-dependent coupling functions and propagation speeds. 

To simplify our task, we take an effective action containing only the new cubic leading interaction in addition to the the standard GR source term, say ${\cal L}_{\rm GR}^{(3)}$. Namely, our starting effective action up to cubic order is given by
\begin{align}
    \label{eq:Effective_Ansatz_GLPV}
    {\cal L}_{\rm eff} =  \Big[ \frac{z^2_{\zeta}}{2} \left(  \zeta^{\prime 2}  - c_s^2 (\partial_k \zeta)^2\right) + \frac{z^2_\gamma}{8} \left(  \gamma_{ij}^{\prime 2} - c_T^2 (\partial_k \gamma_{ij})^2 \right) - \frac{C(\tau)}{H^3 a}  \gamma_{ij}^\prime \partial^2(\partial_i  \zeta \partial_j \zeta)  \Big] + {\cal L}_{\rm GR}^{(3)}\,,
\end{align}
where $z_\zeta$, $z_\gamma$, $c_s$, $c_T$ and $C(\tau)$ are arbitrary functions of time.
The above ansatz covers any model within GLPV up to leading order in derivatives at cubic order. 
As before, we assume that the background is effectively equivalent to that given by a radiation fluid, and so $a \propto \tau$. At the linear level, we take a perturbative ansatz for the free functions for simplicity, that is
\begin{align}
    z_\zeta = a \sqrt{ \frac{2 \epsilon}{c_r}}  \left( 1 + \delta z_\zeta(\tau) \right), \qquad z_\gamma = a ( 1+ \delta z_\gamma(\tau))\,, \qquad c_s = c_r (1  +  \delta c_s(\tau) ) , \qquad c_T = 1 + \delta c_T(\tau) , 
\end{align}
where $c_r^2 =1/3$. We also require that at late times GR is recovered. Therefore, we impose that $\delta z_\zeta(\tau \rightarrow \infty) =0$ and similarly for the other variables. In this section, we consider that the perturbations to the GR solution are small, i.e. $\delta z_\zeta \ll 1$. In the appendix \ref{sec:Time_dependent_z_zeta} we consider the general case $\delta z_\zeta \gg 1$ and demonstrate that it is possible to mimic the $k^5$ slope by tuning the time-dependence of $\delta z_\zeta$ even if the cubic interaction is the same as GR. Nevertheless, let us emphasize that in order to reproduce the $k^5$ one needs a very specific, rapidly varing function for $z_\zeta(\tau)$. In contrast, the $k^5$ scaling is a general prediction of GLPV.

\subsection{Linear solution}
In general, there are no closed analytic solutions. However, as long as the modifications can be considered as small corrections, we can solve the system perturbatively. Further, as we are interested in the new cubic interaction, which dominates deep inside the horizon, the evolution on superhorizon scales and around horizon crossing can be effectively absorbed by shifting the initial conditions for the curvature perturbation. We discuss the analytic solution for $\zeta$ separately for superhorizon and subhorizon scales. The solution for the tensor modes are constructed in the same way. 

\subsubsection{Superhorizon Modes}
Outside the horizon the scalar mode behaves as
\begin{align}
    \partial_\tau \left( z^2_\zeta \zeta^\prime \right) \simeq 0~.
\end{align}
Therefore, if $\delta z_\zeta \ll 1$ the mode still freezes outside the horizon, namely
\begin{align}
    \zeta( x \ll 1) \simeq \zeta_0 = \mathrm{const.}
\end{align}
Note that in our previous discussion we have seen that the higher-derivative terms are only important deep inside the horizon. Therefore, even a subhorizon growth of $\zeta$ will not impact the new interaction up to a rescaling of the initial conditions. However, it may be important for the standard interaction in GR.

\subsubsection{Subhorizon Modes}
In the subhorizon regime $x \gg 1$ and for $\delta z_\zeta \ll 1$, we can approximate the equation of motion for $\zeta$ as 
\begin{align}\label{sol-zeta-sub}
    \frac{\md^2 u_\zeta}{\md x^2} + c_s^2(x) u_\zeta \simeq 0~,
\end{align}
where we have introduced the canonical normalized curvature perturbation $u_\zeta = \zeta z_\zeta$. The solution at leading order in the WKB approximation reads
\begin{align}\label{eq:zetacsgeneral}
    \zeta \simeq \frac{1}{z_\zeta \sqrt{c_s }} e^{\pm i \int^x \md \tilde x c_s(\tilde x)}\,.
\end{align}
Matching both solutions at horizon crossing leads to an approximate transfer function given by
\begin{align}\label{eq:tzetaapprox}
    T_\zeta \simeq \frac{\sqrt{1+ \delta z_\zeta(0)} \sin ( \int^x_0 \md \tilde x c_s(\tilde x) ) }{\sqrt{c_s(x) c_s(0) (1+\delta z_\zeta(x))}  x } \,.
\end{align}

From Eq.~\eqref{eq:tzetaapprox}, we foresee two possible effects. First, the time dependence of $\delta z_\zeta$ may alter the growth rate of the curvature perturbation. Nevertheless, as long as we are only interested in the impact on the new cubic interaction we can regard it as a subleading effect for $\delta z_\zeta\ll1$. 
Second, the time-dependent propagation speed may spoil the resonance in the Kernel of induced GWs.

\subsection{Cubic coupling function}
We investigate first the impact of a time-dependent coupling function, $C(\tau)$ in Eq.~\eqref{eq:Effective_Ansatz_GLPV}, and keep constant propagation speeds. In particular, we fix $c_T=1$. We also neglect the effect of $\delta z_\gamma,\delta z_\zeta\ll1$ and require that GR is recovered at late times, that is $C(\tau)\to 0$. The latter imposes that $C(\tau)$ should decay at least as $1/x^2$ for $x\gg1$. For analytical viability, we take that the transition occurs exponentially fast, namely that
\begin{align}
    \label{eq:Ansatz_Coupling_function}
    C(\tau) = C_0 (k_D \tau)^m e^{-k_D \tau}=C_0 (\kappa x)^m e^{-\kappa x}\quad {\rm with}\quad \kappa=k_D/k\,,
\end{align}
and where $k_D$ is an effective cutoff scale. As we shall see, $k_D$  essentially plays the role of the transition scale $k_t$ of the previous section. Similar exponential damping occur when there are dissipative effects, see, e.g., Ref.~\cite{Domenech:2025bvr} for applications to induced GWs.

After integration, the leading contributions to the kernel from the new source term are given by
\begin{align}\label{eq:Iexponential}
    I = \frac{\sin x}{x}  \delta I_c - \frac{\cos x}{x} \delta I_s\,,
\end{align}
where
\begin{align}
    \delta I_c \simeq & \frac{2 C_0 \kappa^{m} }{c_s u v} \int \md x^\prime x^{\prime 2 + m} \cos( x^\prime) \left( (u+v) \sin( c_s ( u +v) x^\prime ) + (u-v) \sin( c_s (u-v) x^\prime) \right) e^{- \kappa x^\prime} \nonumber \\
    \simeq & \frac{ C_0 \kappa^{m}  }{c_s u v} \Big[ - (u-v) T_1 + (u-v) T_4 - (u+v) T_3 + (u+v) T_2 \Big]\,,
\end{align}
and 
\begin{align}
    \delta I_s \simeq & \frac{2 C_0 \kappa^m}{c_s u v} \int \md x^\prime x^{\prime 2 + m} \sin(x^\prime) \left( (u+v) \sin( c_s ( u +v) x^\prime ) + (u-v) \sin( c_s (u-v) x^\prime) \right) e^{- \kappa x^\prime} \nonumber \\
    \simeq & \frac{ C_0 \kappa^{m}  }{c_s u v} \Big[  (u-v) Z_1 - (u-v) Z_4 + (u+v) Z_3 - (u+v) Z_2 \Big]\,,
\end{align}
and we have introduced the time integrals
\begin{align}
   T_i=& \int_0^\infty \md x^\prime  x^{\prime 2+m} e^{-\kappa x^\prime} \sin(y_i x^\prime)  = \Gamma(3+m) \left( \kappa^2 + y_i^2  \right)^{\frac{-3-m}{2}} \sin\left( (3+m) \arctan(y_i/\kappa)  \right), \\
   Z_i =&   \int_0^\infty \md x^\prime  x^{\prime 2+m} e^{-\kappa x^\prime} \cos(y_i x^\prime)  = \Gamma(3+m) \left( \kappa^2 + y_i^2  \right)^{\frac{-3-m}{2}}  \cos\left( (3+m) \arctan(y_i/\kappa)  \right)\,,
\end{align}
for $i=\{1,2,3,4\}$. It is interesting to note that in Eq.~\eqref{eq:Iexponential} there are no oscillations proportional to the scalar momenta as in Eqs.~\eqref{eq:deltaIexplicit} and \eqref{eq:Inewtongauge}. This is because the coefficients of such terms are directly proportional to $C(\tau)$ and decay away exponentially.

In the resonance regime, that is $y_3=0$, We find that the leading contribution to the Kernel is given by
\begin{align}
    \delta I_s(y_3 =0) \simeq \frac{\Gamma(3+m) C_0 (u+v)}{c_s u v} \left( \frac{k}{k_D} \right)^3
\end{align}
Therefore, we still recover the same growth as for constant coefficients with a sudden transition where the transition scale $k_t$ is replaced by $k_D$ (see Eq. \eqref{eq:generalIs} in the limit $y_3=0$). 

In passing, it is interesting to note that the new growth in the resonance regime is robust for any ansatz of the form
\begin{align}
    C(\tau) = C_0 (\kappa x)^m e^{-(\kappa x)^\alpha}\quad{\rm with}\quad \alpha >0\,.
\end{align}
For instance, the relevant integral yields
\begin{align}
    Z_3^{\alpha}(y_3=0)= \int_0^\infty \md x\, x^{2+m} e^{-(\kappa x)^\alpha} = \frac{\Gamma\left( \frac{3+m}{\alpha} \right)}{\alpha \kappa^{-3-m}}
\end{align}
which still leads to $\delta I_s(y_3=0) \sim (k/k_D)^3$. In other words, the different damping coefficient only lead to order ${\cal O}(\alpha,m)$ correction factors. 
This confirms our expectations that the simplified ansatz of a sudden transition is a good approximation as long as the transition is exponentially fast. Further, the characteristic growth $\kappa^3$ for the kernel in the resonance regime is not impacted by the time-dependent coupling function. 

\subsection{Time-dependent propagation speeds \label{sec:timedependentcs}}
So far we have assumed constant propagation speeds for scalar and tensor modes. However, in modified gravity one generically expects a time dependence. In that case, oscillations of the curvature fluctuations do not have a constant frequency and may hinder possible resonances. For simplicity, we still consider that $C(\tau)$ decays exponentially is given by Eq.~\eqref{eq:Ansatz_Coupling_function}.


Let us explicitly estimate in which cases the resonance in the induced GW spectrum is still effective. For that purpose, it is convenient to work directly with the oscillation average of the Kernel \eqref{eq:I}. For $ \kappa x=k_D\tau \gg 1$, we note that the Kernel has a similar form to that of Eq.~\eqref{eq:Iexponential}, namely the oscillations proportional to the scalar momenta decays way and we are left with the standard free wave solutions. In that limit, the averaged Kernel is well-approximated by
\begin{align}
    \overline{I^2(u,v,\kappa x\gg 1)}= \frac{1}{2 x^2} \left  \vert \int_0^\infty \md x\, x \, e^{i \int^x \md \tilde x c_T(\tilde x)} f(x,u,v) \right \vert^2 \,.  
\end{align}
At this point, we are only interested in possible resonances. Thus, we only consider the oscillations in $f(x,u,v)$ which are responsible for the original resonance. Namely, we focus on terms such that $f(x,u,v)\sim e^{(u+v)\int^x d\tilde x \,c_s(\tilde x) }$ (see Eq.~\eqref{eq:zetacsgeneral}). This means that in practice we have an integral of the type
\begin{align}
    \bar{I}_{\rm res}(u,v,x)^2 \simeq  \frac{1}{2 x^2} \frac{ C_0^2 \kappa^{2m}}{u^2 v^2} \left  \vert \int_0^\infty \md x\,   e^{i \omega_3(x) - \kappa x} (u+v)    \frac{x^{2+m}}{c_s(x)} \right \vert^2      \,,
\end{align}
where we only considered the leading term in $x$ in the Kernel and we have defined
\begin{align}\label{eq:w3}
    \omega_3(x) =\int^{x}  (c_T(\tilde x) - c_s(\tilde x) (u+v) )  \md \tilde x \,.
\end{align}
The resonance momenta are determined by $\omega_3(x) =0$. For constant $c_s$ and $c_T$ we recover $c_T=c_s(u+v)$. However, for general time dependent functions the resonant momenta configuration depends on time. 


Let us look at the resonance directly in the induced GW spectrum, dealing with the whole momentum integration plane. Introducing $d = u-v$ and $s = u +v$, the induced GW spectrum \eqref{eq:inducedGWspectrum} is expressed as 
\begin{align}\label{eq:OmegaGWresonancegeneral}
    \Omega_{\rm GW}^{\rm res} \simeq & \frac{1}{3}  \int_1^\infty \md s\, \int_0^1 \md d\, \frac{(s^2-1)^2 (d^2-1)^2}{(s^2-d^2)^2} P_\zeta \left( \frac{(s+d)k}{2} \right) P_\zeta\left( \frac{(s-d)k}{2} \right) \bar I(x,s,d)^2 \nonumber \\
    \simeq & \frac{2 C_0^2 \kappa^{2m}}{3 } \int_1^\infty \md s\, \int_0^1 \md d\,  \frac{  s^2 (s^2-1)^2 (d^2-1)^2}{(s^2-d^2)^4}  P_\zeta \left( \frac{(s+d)k}{2} \right) P_\zeta\left( \frac{(s-d)k}{2} \right) \nonumber \\
    & \times \int_0^\infty \md x_1 \int_0^\infty \md x_2 \frac{(x_1 x_2)^{2+m}}{c_r^2 ( 1 + \delta c_s(x_1)) (1 + \delta c_{s}(x_2) )} e^{i \frac{x_1 - x_2}{2} z - \kappa (x_1 + x_2) }\,,
\end{align} 
where we defined
\begin{align}
    2 z = & 1 - s c_r \nonumber\\& + \frac{1}{x_1 - x_2} \Big[ \int^{x_1} \delta c_T(\tilde x) \md \tilde x - \int^{x_2} \md \tilde x \delta c_T(\tilde x) - s c_r \left( \int^{x_1} \md \tilde x \delta c_s(\tilde x) - \int^{x_2} \md \tilde x \delta c_s(\tilde x) \right)  \Big]\,.
\end{align} 
The resonant configuration is determined by the condition $z(s_{\rm res})=0$, which only depends on $s$. Solving the resonant conditions yields
\begin{align}\label{eq:sres}
    s_{\rm res}(x_1,x_2) = & \frac{1}{c_r } \frac{ 1 + \frac{1}{x_1 -x_2} \int^{x_1}_{x_2} \md \tilde x \delta c_T(\tilde x)   }{1 + \frac{1}{x_1 - x_2} \int^{x_1}_{x_2} \md \tilde x \delta c_s(\tilde x) } 
    \simeq \frac{1}{c_r } \left( 1 - \frac{1}{x_1 - x_2} \int_{x_2}^{x_1} \md \tilde x\,  (\delta c_T(\tilde x) - \delta c_s(\tilde x)  ) \right)\,,
\end{align}
where in the last step we Taylor expanded for small modifications of the sound speed, $\delta c_s\ll1$, as well as its integral over time.

In Eq.~\eqref{eq:OmegaGWresonancegeneral}, the main resonant contributions come from the $s=s_{\rm res}$ momentum surface. For constant coefficients, this corresponds to integrating over a constant slice, i.e. $s=s_{\rm res}=\mathrm{constant}$ (see section \ref{subsubsec:Scale_Invariant}). However, for time-dependent propagation speeds $s_{\rm res}$ is time-dependent. Now, instead of integrating along a constant $s$ slice, we need to choose a time-dependent slice in the four-dimensional momenta integration space, that is $s=s_{\rm res}(x_1,x_2)$ given by \eqref{eq:sres}. This is always possible as long as $c_s < c_T$.

In the limit where $\delta c_s \ll 1$ and $\delta c_t \ll 1$, the resonance scale $s_{\rm res}$ remains in a very narrow band, namely around
\begin{align}
   \left \vert s_{\rm res} - \frac{1}{c_r} \right \vert \simeq \left \vert \frac{1}{x_1 - x_2} \int_{x_2}^{x_1} \md \tilde x (\delta c_T(\tilde x) - \delta c_s(\tilde x)) \right \vert \ll 1\,,
\end{align}
over the relevant time integration. Therefore, the resonance remains all the time inside the momentum integration plane allowed by momentum conservation, if the primordial spectrum is sufficiently broad. Namely, if the dimensionless width of the primordial spectrum $\Delta$ (see, e.g., Eq. \eqref{eq:Log_Normal_Spectrum}) satisfies 
\begin{align}\label{eq:deltamin}
    \Delta > \Delta_{\rm min} \equiv \left \vert s_{\rm res} - \frac{1}{c_r} \right \vert \,,
\end{align}
we can integrate along the resonance slice $z(s_{\rm res})=0$ and pick up only the resonant contributions. In this case, we recover the same scaling as for constant propagation speeds. We demonstrate this explicitly using a toy model in App.~\ref{subsec:Resonance_Time_dependent_Propagation}.  As an interesting application, we conclude that the resonance, and the resulting $k^5$ scaling, is always present for a scale-invariant primordial spectrum, independent of the time dependence of $c_s$ and $c_T$, at least in the perturbative regime.

However, in the opposite case, namely when the primordial spectrum is sharply peaked and $\Delta<\Delta_{\rm min}$ \eqref{eq:deltamin}, the resonant slice may at some point fall outside the integration plane restricted by momentum conservation. Then, the resonance effectively stops. For instance, for a delta-peak in the spectrum, the resonant condition is only valid at one instant in time and, therefore, the resonance is highly suppressed. In other words, there is only one scalar momenta contributing to the integral, whose oscillations have a time-dependent frequency. Then, there is little time for a resonant configuration to build up.


It is important to note that our method to estimate the contributions from the resonance is also applicable to standard GR, for example for a general K-essence scalar field or in phase transitions. As one interesting example, the approach could be useful for the primordial black hole dominated universe \cite{Inomata:2020lmk,Papanikolaou:2020qtd} with an initial broad mass function, where the dominant contribution to the GW spectrum arises from the resonances after evaporation \cite{Domenech:2020ssp}.

\section{Conclusion\label{sec:conclusions}}

GWs offer a way of testing gravity in the very early Universe. We studied the impact of a GLPV scalar field on the spectrum of the induced GWs. Generally speaking, GLPV models do not have manifestly second-order equations of motion. However, around FLRW the linear equations of motion have a standard dispersion. We showed that there is a new source term at the nonlinear level for the GLPV models that are disformally disconnected from Horndeski. Note that these models are only degenerate in the uniform slicing and belong to the U-DHOST class. It is only the scalar-scalar-tensor interaction that breaks the second-order form of the equation of motion at cubic order and provides a fundamental distinction between Horndeski gravity and GLPV. 

The new source term dominates deep inside the horizon and further enhances the production of GWs. For a scale-invariant curvature power spectrum, the resonant contributions to the GWs spectrum scale as $\Omega_{\rm GW} \propto (k/k_D)^5 P_\zeta^2 $ where $k_D$ is the transition scale from GLPV to standard GR. The new scaling is different from the previous results in Horndeski gravity \cite{Domenech:2024drm} where the spectrum growths as $k^3$, which coincidentally is the same as the universal causal infrared tail \cite{Cai:2019cdl}. Therefore, our work shows that induced GWs provide a powerful tool not only to test modifications of gravity but also to test between conventional Horndeski models and GLPV. Equivalently, the presence (or absence) of this $k^5$ scaling provides a sharp discriminator between the U-DHOST sector of GLPV and Horndeski models.

It should be noted that, as we show in the App.~\ref{sec:Time_dependent_z_zeta}, it is in principle possible to construct a model that leads to the same scaling as in GLPV by modifying the linear curvature perturbation while keeping the same cubic interaction as in GR. However, it requires a large degree of fine-tuning in addition to a very different evolution of linear perturbations compared to standard GR. In contrast, the $k^5$ scaling from GLPV is a general feature of the U-DHOST subset of the theory. Furthermore, we demonstrated that the scaling of the GWs spectrum in GLPV is robust as long as the deviations in the linear curvature perturbation are small. We also showed that time-dependent propagation speeds for the curvature or tensor perturbation will not spoil the resonance as long as the width of the curvature spectrum is broader than the variation in the sound speeds. The latter result may be also relevant for transitions in standard GR when the scalar sound speed changes gradually.  

In our current set-up we have assumed that the background behaves as radiation and is dominated by the scalar field. In future works, it would be interesting to relax this assumption and consider an additional radiation component and a subdominant scalar field and study how the amplitude of the induced GWs depends on the relative energy density of the scalar field and the standard radiation.
\bigskip

\textsc{\textbf{Note:}} During the final stages of our work, Ref.~\cite{Jiang:2025ysb} appeared on arXiv discussing induced GWs in spatially covariant gravity. Our results agree when there is overlap. However, their analysis of the induced GW spectrum differs from ours, as they do not consider modes of the induced GWs which oscillate with the scalar sound speed and which grow in time.

\begin{acknowledgments}
   We would like to thank David Langlois and Karim Noui for the helpful comments and discussions.
 G.D. and A.G. are supported by the DFG under the Emmy-Noether program, project number 496592360. GD is also supported by the JSPS KAKENHI grant No. JP24K00624. M.A.G. and M.Y. were supported by the Institute for Basic Science under the project code IBS-R018-D3. M.Y. is supported in part by JSPS Grant-in-Aid for Scientific Research Number JP23K20843.
\end{acknowledgments}

\appendix

\section{GLPV model}
\label{sec:GLPV_Model}
Here we list the detailed calculations and notations for the GLPV model. We first show the covariant formulation. Then we present the quadratic action and, lastly, we discuss the scalar-tensor-tensor interaction.

\subsection{Covariant formulation}
\label{subse:GLPV_Covariant}
In the covariant formulation the GLPV action is given by \cite{Gleyzes:2014dya,Gleyzes:2014qga}
\begin{align}
    \mathcal{L} = \sqrt{-g} & \Big[G_2 + G_3 \Box \phi + G_4 \,^{(4)}R -2 G_{4,X} \left( (\Box \phi)^2 - \phi^{\mu\nu} \phi_{\mu\nu}  \right) + F_4 \tensor{\epsilon}{^\mu^\nu^\rho_\sigma} \tensor{\epsilon}{^{\mu^\prime}^{\nu^\prime}^{\rho^\prime}^\sigma } \phi_\mu \phi_{\mu^\prime} \phi_{\nu{\nu^\prime}} \phi_{\rho{\rho^\prime}} \nonumber \\
    & + G_{5}\,^{(4)}G_{\mu\nu} \phi^{\mu\nu} + \frac{1}{3} G_{5,X} (  (\Box \phi)^2 -3 \Box \phi \phi^{\mu\nu}\phi_{\mu\nu} +2 \phi^{\mu\nu} \phi_{\nu\alpha} \phi^{\alpha}_\mu) \nonumber \\
    & + F_5 \tensor{\epsilon}{^\mu^\nu^\rho^\sigma} \tensor{\epsilon}{^{\mu^\prime}^{\nu^\prime}^{\rho^\prime}^{\sigma^\prime} } \phi_\mu \phi_{\mu^\prime} \phi_{\nu{\nu^\prime}} \phi_{\rho{\rho^\prime}} \phi_{\sigma{\sigma^\prime}} \Big]
\end{align}
where $\,^{(4)}R$ and $\,^{(4)}G_{\mu\nu} $ are the 4-dim Ricci scalar and Einstein tensor and  we introduced the notation $\nabla_\mu \phi = \phi_\mu$ and $\nabla_\mu\nabla_\nu \phi = \phi_{\nu\mu}$. The parameters $G_i$ and $F_i$ are free functions of $\phi$ and $X=\phi_\mu \phi^\mu$.

The parameters in the action in the unitary gauge \eqref{eq:unitarygaugeaction}, that is $A_i$ and $B_i$, are given in terms of $G_i$ and $F_i$ via 
\begin{align}
    A_2 =& G_2 - \sqrt{-X} \int \frac{G_{3,\phi}}{2 \sqrt{-X}} \md X~, \\
    A_3 =& - \int G_{3,X} \sqrt{-X} \md X - 2 \sqrt{-X} G_{4,\phi}~, \\
    A_4=& - G_4 + 2 X G_{4,X} + \frac{X}{2} G_{5,\phi} {\color{red} +} X^2 F_4~, \\
    B_4 =& G_4 + \sqrt{-X} \int \frac{G_{5,\phi}}{4 \sqrt{-X}} \md X, \\
    A_5 =& - \frac{(-X)^{3/2}}{3} G_{5,X} + (-X)^{5/2} F_5 ~, \\
    B_5 =& - \int G_{5,X} \sqrt{-X}\md X
\end{align}

{\subsubsection{Degeneracy conditions in beyond Horndeski}
\label{app:Degeneracy_conditions}
In the covariant formulation one has to impose an additional degeneracy condition to avoid the presence of instantaneous modes \cite{Langlois:2015cwa,BenAchour:2016fzp}, which reads as
\begin{align}
    \label{eq:GLPV_degeneracy_condition}
    F_4 = - \frac{3 F_5 ( G_4 - 2  X G_{4,X} )}{X G_5}~.
\end{align}
Therefore, in the covariant formulation the form of the GLPV terms is constrained except for the specific case where $G_4(X) = \sqrt{X}$. For general $G_4$, $F_4$, $G_5$ and $F_5$  the GLPV action belongs to the U-DHOST class and propagates an instantaneous modes for a non uniform slicing of the scalar field \cite{DeFelice:2018ewo}. 

We can note that imposing the degeneracy condition Eq. \eqref{eq:GLPV_degeneracy_condition} the condition Eq. \eqref{eq:coefficientnumerator} is fulfilled such that the higher derivative interaction Eq. \eqref{eq:Cubic_General_Leading_Term} precisely vanishes.
}

\subsubsection{GLPV Toy Model}
The GLPV toy model, Eq.~\eqref{eq:GLPV_Toy_Model}, introduced in Sec.~\ref{sec:New_Interaction_Toy_Model}, is given in the covariant formulation by
\begin{align}
    & G_2 = P~,   \\ & G_4= B_4 - \sqrt{-X} \int \frac{G_{5,\phi}}{4 \sqrt{-X}} \md X = \frac{1}{2} + \frac{b_5 t_\star (-X)^{3/2}  }{2 X_\star \phi_\star}~,  \\
    & F_4= -\frac{G_4}{X^2} + \frac{2 }{X} G_{4,X} + \frac{1}{2 X} G_{5,\phi} = - \frac{1}{2 X^2} ~, \\
    & G_5 = - \int \frac{B_{5,X}}{\sqrt{-X}} \md X = 2 b_5 \frac{\phi t_\star}{\phi_\star X_\star}  \sqrt{-X}~, \\
    & F_5 = - \frac{1}{3 X} G_{5,X} = - \frac{ b_5 \phi t_\star}{3 \phi_\star X_\star} \frac{1}{(-X)^{3/2}}\,.
\end{align}

\subsubsection{Horndeski Toy Model}
\label{subsubsec:Horndeski_Toy_Model}
We shortly review the model parameters of the Horndeski toy model discussed in \cite{Domenech:2024drm}. The model is constructed such that the linear perturbations behave exactly like in standard GR, and the modifications to GR are only present at cubic order parameterized by a single free parameter $\tilde \Gamma$. 
The precise form of the free function is given by
\begin{align}
    G_2 =& P(\phi,X) - \frac{\tilde \Gamma  \phi_\star^2  }{16 \phi^2 t_\star^2} \left( \frac{X}{X_\star} -1 \right) \Theta(\phi_t - \phi) \nonumber \\
    G_3 =& - \frac{3 \tilde \Gamma}{32 \phi }  \left( \frac{X}{X_\star} -1 \right)^2 \Theta(\phi_t -\phi), \\
    G_4 =& \frac{1}{2} + \frac{1}{32} \tilde \Gamma \left( \frac{X}{X_\star} -1 \right)^2 \Theta(\phi_t - \phi), \\
    G_5 = & -  \frac{\tilde \Gamma \phi t_\star^2 }{4  \phi_\star^2} \left( \frac{X}{X_\star} -1 \right)^2 \Theta(\phi_t - \phi)
\end{align}
where the $\star$ denotes an arbitrary pivot scale and $P(\phi,X)$ is given by Eq. \eqref{eq:Ansatz_B_5}. 

\subsection{Quadratic action}
\label{subsec:GLPV_Quadratic_Action}

The quadratic action for tensor modes is given by
\begin{align}
    {\cal L}_{\gamma\gamma} = \frac{a^3}{8} \Big[ (-2 A_4 - 6 A_5 H) \dot \gamma_{ij}^2 - \frac{2 B_4 + \dot B_5}{a^2} (\partial_k \gamma_{ij})^2 \Big]\,.
\end{align}
On the other hand, the action for the scalar modes reads
\begin{align}
    \label{eq:Quadratic_Action_GLPV_General}
    {\cal L}_{\zeta\zeta} =& a^3 \Big[ t_0 \alpha^2 + t_1 \frac{k^2}{a^2}  \zeta^2 + t_2 \dot \zeta^2 + t_3 \frac{k^2}{a^2} \zeta \alpha + t_4 \frac{k^2}{a^2} \alpha \beta + t_5 \alpha \dot \zeta + t_6 \frac{k^2}{a^2} \dot \zeta \beta   \Big]\,,
\end{align}
where we defined
\begin{align}
    t_0 = & \frac{1}{2} \Big(6 H^3 A_{5,N^2}-24 H^3 A_{5,N}+6 H^2 A_{4,N^2}-12 H^2 A_{4,N}+3 H A_{3,N^2}+A_{2,N^2}+2 A_{2,N} \nonumber \\
    & +36 A_5 H^3+12 A_4 H^2\Big)~, \\
    t_1 = &   \left(2 B_4 + \dot B_5 \right) \,, \\
    t_2 =& 6 (A_4 + 3 A_5 H) \,,\\
    t_3 =& 2 \left(-H B_{5,N}+2 B_{4,N}+2 B_4\right)\,, \\
    t_4 =& 6 H^2 A_{5,N}+4 H A_{4,N}+A_{3,N}-12 A_5 H^2-4 A_4 H\,, \\
    t_5 =& 3 \left(6 H^2 A_{5,N}+4 H A_{4,N}+A_{3,N}-12 A_5 H^2-4 A_4 H\right) \,,\\
    t_6 =& 4 \left(3 A_5 H+A_4\right)\,.
\end{align}
The solutions to the momentum and Hamiltonian constraint yield
\begin{align}
    \alpha = -\frac{t_6 \dot \zeta}{t_4}~, \qquad \beta = - \frac{t_3}{t_4} \zeta - \left( \frac{t_5}{t_4} - \frac{2 t_6 t_0}{t_4^2} \right) \frac{a^2}{k^2} \dot \zeta ~. 
    \label{eq:Solution_Hamiltonian_Constraint}
\end{align}
Substituting it back into the quadratic action for the curvature perturbation we obtain
\begin{align}
    {\cal L}_{\zeta\zeta} = & a^3  \Big[ {\cal G}_S\dot \zeta^2 - {\cal F}_S  \frac{k^2}{a^2} \zeta^2 \Big]\,,
\end{align}
where
\begin{align}
    {\cal G}_S = t_0 \left( \frac{t_6}{t_4} \right)^2 + t_2 - t_5 \frac{t_6}{t_4}, \qquad {\cal F}_S = -t_1 - \frac{1}{ a} \frac{\md}{\md t} \left( \frac{a t_3 t_6}{2t_4} \right)\,.
\end{align}

\subsection{Cubic action for tensor-tensor-scalar interaction}
\label{subsec:Tensor_Tensor_Scalar_GLPV}

In this appendix we discuss the scaling of the tensor-tensor-scalar interactions and argue that they scale up to $\sim {\cal O}(k^4)$. First, we note that, focusing on the tensor modes, one has $K_{ij} \sim \dot \gamma_{ij}$ and $R_{ij} \sim \partial^2 \gamma_{ij}$. Namely, $K_{ij}$ has at most one derivative and $R_{ij}$ two derivatives even when one goes to the second order expansion. It then follows from the GLPV action \eqref{eq:unitarygaugeaction}, that the tensor-tensor-scalar interaction in GLPV can only contain terms up to the order $\partial^4 \sim k^4$. Any higher-derivative terms scaling as $\partial^5 \sim k^5$ require terms which are quadratic in the intrinsic curvature which is beyond both GLPV and DHOST.

For completeness, we show the highest derivative order terms in the tensor-tensor-scalar interaction in GLPV, which are given by
\begin{align}
    {\cal L}_{\zeta\gamma\gamma} \simeq  & a \Big[   \frac{3}{2} A_5 \Big( \frac{1}{2} \dot \gamma_{jl}^2 \partial^2 \beta -  \dot \gamma_{il} \dot \gamma_{jl} \partial_i \partial_j \beta  \Big) + B_{5,N} \Big( - \frac{1}{4} \dot \gamma_{ij} \partial^2 \gamma_{ij} \alpha  \Big) \Big]\,.
\end{align}
Thus, we confirm by explicit calculation the scaling of $\partial^4 \sim k^4$ and demonstrate that the leading order terms are the same as in Horndeski gravity \cite{Gao:2012ib}. 

\section{Full expression of the Kernel}
\label{sec:Full_Kernel}

In this appendix we provide the explicit expressions of the coefficients in Eq.~\eqref{eq:fsimplifiedincis}. and the exact full Kernel. 
\subsection{Coefficients $c_{i}^{\pm}$}
\label{subsec:Coefficients_c_i}
The coefficients $c_i^{\pm}$ in Eq. \eqref{eq:fsimplifiedincis} are given by
\begin{align}
    & c_{-1}^- = \frac{b_5 (1+b_5) (u- v)}{8 c_s u v}  ~, \\
    & c_{-1}^+ = - \frac{b_5 (1+b_5) (u+v)}{8 c_s u v}  ~,
\end{align}
\begin{align}
    & c_0^- = - \frac{- (4 b_5 + b_5^2) u v + 6 b_5 c_s^2 (u^2+v^2)  }{8 c_s^2 u^2 v^2}  ~, \\
    & c_0^+ = - \frac{ (4 b_5 + b_5^2) u v + 6 b_5 c_s^2 (u^2+v^2) }{8 c_s^2 u^2 v^2} ~, \\
   & c_1^- = \frac{(u-v) b_5 (3 u v - 3 (v^2 + u^2) + u^2 v^2 ) }{4 c_s u^3 v^3}  ~, \\
   & c_1^+ = \frac{(u+v) b_5 (3 ( u v + v^2 + u^2) - u^2 v^2 )}{4 c_s u^3 v^3}~, \\
   & c_2^- =- \frac{b_5 (6 (v^2 + u^2) + 7 u^2 v^2 ) }{4 c_s^2 u^3 v^3}~, \\
   & c_2^+ = \frac{b_5 (6 (v^2 + u^2) + 7 u^2 v^2 ) }{4 c_s^2 u^3 v^3}~.
\end{align}

\subsection{Modified Kernel}
\label{subsec:Modified_Kernel}
Here we give the exact expression of the Kernel \eqref{eq:I} after integration. For that,
it is convenient to split the calculation of the kernel as
\begin{align}
   \delta I = \frac{\sin x}{ x} \delta I_c(v,u,x)-\frac{\cos x}{ x} \delta I_s(v,u,x)\,.
\end{align}
where
\begin{align}\label{eq:Idefinition}
\delta I_{c/s}(v,u,x)=\int_0^x \md\tilde x \,\tilde x \left\{\begin{aligned}
\cos(\tilde x)\\
\sin(\tilde x)
\end{aligned}\right\}\delta f(v,u,\tilde x)\,.
\end{align}
After integration, we find that Eq.~\eqref{eq:Idefinition}  reads
\begin{align}\label{eq:generalIc}
\delta I_c(v,u,x)&=\frac{1}{4}\Bigg\{\cos (x y_1) \left(\frac{(-4 + 2 y_1^2 x^2) c_{-1}^- }{y_1^3}  +
   \frac{2c_0^-}{y_1^2}+\frac{2c_1^-}{y_1}\right)+\sin (x y_1) \left(- \frac{4 c_{-1}^- x}{y_1^2} +\frac{2c_0^- x}{y_1}\right)\nonumber\\&+\cos (x y_2) \left(\frac{(-4 + 2 y_2^2 x^2) c_{-1}^+ }{y_2^3}  +\frac{2c_0^+}{
   y_2^2}-\frac{2c_1^+}{
   y_2}\right)+\sin (x y_2) \left( \frac{4c_{-1}^+ x}{y_2^2} + \frac{2c_0^+
   x}{y_2}\right)\nonumber\\&+\cos (x
   y_3) 
   \left(\frac{(-4 + 2 y_3^2 x^2) c_{-1}^+ }{y_3^3}  + \frac{2c_0^+}{y_3^2}+\frac{2c_1^+}{y_3}\right)+\sin (x
   y_3) \left( -\frac{4c_{-1}^+ x}{y_3^2} + \frac{2c_0^+ x}{ y_3}\right)\nonumber\\&+\cos (x y_4) \left(\frac{(-4 + 2 y_4^2 x^2) c_{-1}^- }{y_4^3}  +\frac{2c_0^-}{
   y_4^2}-\frac{2c_1^- }{
   y_4}\right)+\sin (x y_4) \left( \frac{4c_{-1}^- x}{y_4^2} + \frac{2c_0^-
   x}{ y_4}\right)\nonumber\\&+2 c_2^- \text{Ci}(x |y_1|)+ 2 c_2^+ \text{Ci}(x
  |y_2|) + 2 c_2^+
   \text{Ci}(x |y_3|) +2 c_2^- \text{Ci}(x |y_4|)\Bigg\}   - \delta I_{c,0}(v,u)\,,
\end{align}
where we defined
\begin{align}\label{eq:Ic0appendix}
\delta I_{c,0}(v,u)=&- c_{-1}^- \left( \frac{1}{y_1^3} - \frac{1}{y_4^3} \right) - c_{-1}^+ \left( \frac{1}{y_3^3} - \frac{1}{y_2^3} \right) + \frac{c_0^- \left(1+c_s^2
   (u-v)^2\right)}{\left(1-c_s^2
   (u-v)^2\right)^2}+\frac{c_0^+ \left(1+c_s^2
   (u+v)^2\right)}{\left(1-c_s^2
   (u+v)^2\right)^2}\nonumber\\& +\frac{c_1^- c_s (u-v)}{1-c_s^2
   (u-v)^2}+\frac{c_1^+ c_s (u+v)}{1-c_s^2 (u+v)^2} +\frac{c_2^-}{2}\ln\left|\frac{1-c_s^2(u-v)^2}{1-c_s^2(u+v)^2}\right|\,,
\end{align}
and
\begin{align}
\delta I_s(v,u,x)&=\frac{1}{4}\Big\{-\cos (x y_1) \left(- \frac{4 c_{-1}^- x}{y_1^2} +\frac{2c_0^-
   x}{ y_1}\right)+\sin (x  y_1)
   \left( c_{-1}^- \frac{-4 + 2 y_1^2 x^2}{ y_1^3}  + \frac{2c_0^-}{y_1^2}+\frac{2c_1^-}{y_1}\right)\nonumber\\&
  -\cos (x y_2)\left( \frac{4 c_{-1}^+ x}{y_2^2} + \frac{2c_0^+ x}{ y_2}\right)+\sin (x y_2) \left(  c_{-1}^+ \frac{-4 + 2 y_2^2 x^2}{ y_2^3}  + \frac{ 2c_0^+ }{  y_2^2}-\frac{2c_1^+}{  y_2}\right)\nonumber\\&
   -\cos (x y_3) \left( - \frac{4 c_{-1}^+ x}{y_3^2} + \frac{2 c_0^+ x}{y_3}\right)+ \sin (x y_3) \left( c_{-1}^+ \frac{-4 + 2 y_3^2 x^2}{ y_3^3}  +\frac{2 c_0^+}{y_3^2}+\frac{2
   c_1^+}{y_3}\right)\nonumber\\&
   -\cos (x y_4)\left(  \frac{4 c_{-1}^- x}{y_4^2} + \frac{2 c_0^- x}{ y_4}\right)+\sin (x y_4)\left( c_{-1}^- \frac{-4 + 2 y_4^2 x^2}{ y_4^3}  +\frac{2 c_0^-}{ y_4^2}-\frac{2 c_1^-}{ y_4}\right)\nonumber\\&
   +2 c_2^- \text{Si}(x y_1) + 2 c_2^+ \text{Si}(x y_2) + 2 c_2^+ \text{Si}(x y_3) +2 c_2^- \text{Si}(x y_4) \Big\}\,.
   \label{eq:generalIs}
\end{align}

\section{Resonance for time-dependent propagation speed}
\label{subsec:Resonance_Time_dependent_Propagation}
In this appendix we discuss in more detail the time dependence of the propagation speed of Sec.~\ref{sec:timedependentcs}. In order to get a better understanding of the resonance, we use a toy model with $c_T=1$ and
\begin{align}
    c_s(x) = c_r (1 + \delta c_{s,0} \kappa x )\,.
\end{align}
As the growth of $c_s$ must stop to have a sensible example, we set a cutoff at $x_{\rm cutoff}=1/\kappa$. The oscillation frequency that determines the resonance, that is $\omega_3$ \eqref{eq:w3}, is given by
\begin{align}
    \omega_3(x) = \int^x \md \tilde x \left( 1 - s  c_s(\tilde x) \right) = \left( 1 - s c_r \left( 1 + \frac{\delta c_{s,0}}{2} \kappa x \right) \right) x\,.
\end{align}
Therefore, the time-dependent resonant momentum configuration is given by
\begin{align}
   \frac{2}{c_r \left( 2 + \delta c_{s,0} \right)} \leq s_{\rm res}(x) = \frac{2}{c_r \left( 2 + \delta c_{s,0} \kappa  x \right)} \leq \frac{1}{c_r}
\end{align}
Note that for $\delta c_{s,0} / \kappa \ll 1$ the change in the resonance scale becomes irrelevant. 

Regarding the kernel, it is more convenient to directly evaluate the average squared kernel. This yields 
\begin{align}
     \overline{ I^2(u,v) }\simeq & \frac{C_0^2 s^2}{2 x^2 c_r^2 u^2 v^2}  \int_0^{1/\kappa}  \md x_1 \int_0^{1/\kappa} \md x_2   x_1^2 x_2^2 e^{i \frac{x_1 -x_2}{2} ( 2 - s c_r (2+ (x_1+ x_2 ) \delta c_{s,0} \kappa )}\,,
\end{align}
where we took the exponential ansatz for the coupling function \eqref{eq:Ansatz_Coupling_function} but set $m=0$ for simplicity.
Then, we introduced $x_-=(x_1 - x_2)/2$ and $x_+ = (x_1 + x_2)/2$, which leads to
\begin{align}\label{eq:I2example}
     \overline{ I^2(u,v) }\simeq & \frac{C_0^2 s^2}{2 x^2 c_r^2 u^2 v^2}  \int_0^{1/\kappa}  \md x_+ \int_{-1/(2\kappa)}^{1/(2\kappa)} \md x_-\, 8 (x_+^2 -x_-^2)^2 e^{i x_- (2 - 2 c_r s (1+ \delta c_{s,0 } \kappa x_+))} \nonumber \\
     = & \frac{4 C_0^2 s^2}{ x^2 c_r^2 u^2 v^2} \int_0^{1/\kappa} \md x_+ \Bigg[ \frac{x_+^4}{\kappa} \pi \delta(y(s,x_+) ) - 2 x_+^2 \frac{ 2 y  \cos y +   (y^2- 2 ) \sin y}{2 y^3 \kappa^3} \nonumber \\
     & \qquad \qquad \qquad \qquad \qquad+ \frac{4 y   ( y^2 - 6 ) \cos y +  ( y^4 -12  y^2 + 24) \sin y }{ 16 y^5 \kappa^5}\Bigg]\,,
\end{align}
where we introduced 
\begin{align}
    y \kappa = 1 -  c_r s (1+ \delta c_{s,0} x_+ \kappa) \,.
\end{align}
The resonant regime is then given by $y(s,x_+) =0$. It is important to note that the last two terms in Eq.~\eqref{eq:I2example} do not diverge in the limit $y=0$ but instead scale as $\sim x_+^2 /\kappa^3$ and $\sim 1/\kappa^5$. Furthermore, the first term in Eq.~\eqref{eq:I2example} already picks only contributions in the resonance regime. 

To find the contribution from the resonance to the gravitational wave energy density, we must compute the integral in
\begin{align}
    \Omega_{\rm GR}^{\rm res} \simeq & \frac{1}{3} \int_1^{\infty} \md s\, \int_0^1 \md d\, \frac{(s^2-1)^2(d^2-1)^2}{(s^2-d^2)^2} P_\zeta\left( \frac{(s+d) k}{2} \right) P_\zeta \left( \frac{(s-d) k}{2} \right) 2 x^2 \overline{I^2(x,s,d)}~.
\end{align}
To do that, we first assume that the primordial spectrum is broad enough such that the whole resonance regime is inside the allowed momentum integration plane, namely
\begin{align}
  \left  \vert s_{\rm res}(x_+) - \frac{1}{c_r} \right \vert = \left \vert \frac{1}{c_r (1+ \delta c_{s,0} x_+ \kappa)} - \frac{1}{c_r} \right \vert \leq \Delta\,,
\end{align}
where $\Delta$ is the dimensionless width of the primordial spectrum. For further simplicity, we assume that the scalar spectrum is flat inside the resonance band with an amplitude of $A_{\rm peak}$. This would lead at most to order $O(1)$ errors. We then perform a coordinate change from $(s,x_+)$ to $(y,x_+)$ and then evaluate the integrand along $y \simeq 0$ which leads to
\begin{align}
    \Omega_{\rm GW}^{\rm res} 
    \simeq & \frac{8 C_0^2 A_{\rm peak}^2 }{3 c_r^2} \int_0^1 \md d\, \frac{(1-c_r^2)^2 (d^2-1)}{(1-c_r^2 d^2)^4 } \int_0^{1/\kappa} \md x_+\, \frac{1}{c_r } \left( \pi x_+^4 -   \frac{x_+^2}{3 \kappa^2} + \frac{1}{80 \kappa^4} \right) \sim   \frac{C_0^2 A_{\rm peak}^2 }{c_r^3 } \kappa^{-5}\,,
\end{align}
which recovers the scaling $\Omega_{\rm GW}^{\rm res} \sim \kappa^{-5}$ for the resonance. This calculation shows that the resonance is not affected by time dependence of the propagation speed if the spectrum broad enough.

\section{Mimicking $k^5$ growth }
\label{sec:Time_dependent_z_zeta}

Here we present an explicit example in which one can obtain a $k^5$ scaling by only modifying the linear solutions while keep the same cubic interactions as in GR. For simplicity, we assume that the only modification to GR in Eq. \eqref{eq:Effective_Ansatz_GLPV} is given by $ \delta z_\zeta$.  Then, we parameterize the full $z_\zeta$ as
\begin{align}\label{eq:zetaexample}
    z_\zeta = \tilde z a  = a \begin{cases}
        z_0  &  0 \leq \tau \leq \tau_1 \\
        z_0 \left( \frac{\tau}{\tau_1} \right)^{-n} & \tau_1 \leq \tau \leq \tau_2 \\
        z_{\rm GR} & \tau\geq \tau_2
    \end{cases} \,,
\end{align}
where $z_0 (\tau_2/\tau_1)^{-n} = z_{\rm GR}$ and we assume $n>0$. Eq.~\eqref{eq:zetaexample} is built such that it is well-behaved at early and late times.
This parameterization essentially corresponds to a transition of the effective Planck mass for the scalar modes. 

To estimate the effect it is helpful to consider the different time intervals separately, namely $ \tau \leq \tau_1$, $\tau_1 \leq \tau \leq \tau_2$ and $\tau \geq \tau_2$. 
In the intermediary region $\tau_1 < \tau < \tau_2$ we get
\begin{align}
    T_\zeta = c_1 \tau^{-1/2+ n} J_{-1/2+n}(c_s k \tau) + c_2 \tau^{-1/2+n} Y_{-1/2+n}(c_s k \tau)\,,
\end{align}
while in the other two periods we have the standard solution as in GR. 
Matching the solutions at $\tau=\tau_1$ yields
\begin{align}
    c_1 = &  \frac{- \pi \tau_1^{-1/2 - n }}{2 c_s k } \Big[  c_s x_1 Y_{1/2+n}(c_s x_1) \sin(c_s x_1) +  J_{-1/2+n}(c_s x_1) (c_s x_1 \cos (c_s x_1) - 2 n \sin ( c_s x_1) ) \Big] \,,\\
    c_2 =& \frac{\pi \tau^{-1/2 -n }}{2 c_s k } \Big[ c_s x_1 J_{1/2+n}(c_s x_1) \sin (c_s x_1) + J_{-1/2+n}(c_s x_1) (c_s x_1 \cos (c_s x_1) - 2 n \sin (c_s x_1) ) \Big]\,.
\end{align}
Similarly, we can match the solutions at $\tau=\tau_2$. 
Later, we will only be interested in two cases: $n=1$ and $n=3/2$.

To calculate the kernel we split the time integral into three different regions. First, we use the kernel in GR in Newton Gauge, which  is given by
\begin{align}
    I_c = & \int_0^\infty \md x^\prime \cos (x^\prime) x^\prime f_{\rm GR} \nonumber \\
    =& \int_0^{x_1} \md x^\prime \cos(x^\prime) x^\prime f_{\rm GR} + \int_{x_1}^{x_2} \md x^\prime \cos(x^\prime) x^\prime f_{\rm GR} + \int_{x_2}^{\infty} \md x^\prime \cos(x^\prime) x^\prime f_{\rm GR} \,,
\end{align}
where
\begin{align}
    f_{\rm GR} = & T_\Phi T_\Phi + \frac{1}{2} \left( T_\Phi + x \frac{\md }{\md x} T_\Phi \right) \left( T_\Phi + x \frac{\md }{\md x} T_\Phi \right)\,.
\end{align}
Note that, strictly speaking, we should start from the unitary gauge and perform a gauge transformation to the Newton gauge, namely
\begin{align}
    \Phi = \zeta + {\cal H} \bar\beta\,.
\end{align}
Therefore, generically we need to know the whole action, and the solution to $\bar\beta$, to perform the transformation properly. Nevertheless, as proof of concept, we simply take the standard GR case
\begin{align}
    \Phi \simeq {\cal H} \partial^{-2} \zeta^\prime\,,
\end{align}
and neglect potential corrections from modified gravity. Alternatively, we could directly take an effective ansatz for $\Phi$ in the Newton gauge. 

Let us now consider the three main regimes separately: (I)  $k \tau_2 \ll 1$, (II) $k \tau_1 \gg 1$ and $k \tau_2 \gg 1$ and (III) $k\tau_1 \gg 1$. In the regime (I), the tensor modes are still far superhorizon. Therefore, we can approximate that the main contributions are coming from the last integral and we can neglect the other two terms. This is essentially leads to the standard GR result, namely
\begin{align}
     I^{\rm (I)}(u,v) \vert_{k \tau_2 \ll 1} \simeq & \int_{k \tau_2 \ll 1}^{\infty} \md x^\prime f_{\rm GR} G_k(x,\tau^\prime) 
     \simeq  I_{\rm GR}\,.
\end{align}

Next, we consider the case (II), that is $k \tau_1  \ll 1$ but $k \tau_2 \gg 1$. As $k \tau_1 \ll 1$ we can neglect the first integral and we only need to consider the other two integrals, that is
\begin{align}
    I^{\rm (II)}_c = \int_{x_1}^{x_2} \md x^\prime \cos( x^\prime) x^\prime f_{\rm GR} + \int_{x_2}^{x} \md x^\prime \cos( x^\prime) x^\prime f_{\rm GR} \,.
\end{align}
Note that the integrand in the last integral decays as $1/x^\prime$. Therefore, if the transition period $\Delta x = x_2 - x_1$ is sufficiently long we can neglect the last integral and only focus on the second term. This becomes apparent for $n=1$ as the modes just oscillate for $\tau_1 \leq \tau \leq \tau_2$ and therefore the second integral will dominate if $\Delta x \gg 1$. 
The other interesting case is given by $n=3/2$ in which case the scalar modes growth as $T_\zeta \sim \sqrt{x^\prime} \cos (c_s u x^\prime)$ as soon as they enter the horizon. Again, the contribution during the transition is dominant. 

The other integral in case (II) is given by
\begin{align}
    I_s^{\rm (II)} (u,v) \vert_{k \tau_2 \gg 1, k \tau_1 \ll 1} \simeq \int_{x_1}^{x_2} \md x^\prime\, \sin(x^\prime) x^\prime f_{\rm GR} \,.
\end{align}
As the integrand is growing in time and $x_2 \gg 1$, the dominant contributions are coming from the resonance. As done in the previous sections we can estimate the relative growth by considering the integrals in the resonance regime. For $n=1$, this yields
\begin{align}
    I_s^{\rm (II), res}\vert_{n=1} \sim & \int_{x_1}^{x_2} \md x^\prime x^\prime \sin (x^\prime) \frac{1}{2 u^2 v^2 } \frac{\md^2 T_\zeta }{\md x^2} \frac{\md^2 T_\zeta}{\md x^2} \nonumber \\
     \sim & \frac{c_s^4}{2 }  \int_{0}^{x_2} \md x^\prime x^\prime  \sin x^\prime \cos(c_s u x^\prime) \cos(c_s v x^\prime) \sim  \frac{c_s^4}{4} x_2^2\,,
\end{align}
where in the first step we only considered the dominant contribution for the resonance and in the second step we took the limit $x_1 \rightarrow 0$. The resulting induced GW spectrum will growth as 
\begin{align}
    \Omega_{\rm GW} \sim x_2^3\,,
\end{align}
for $k \tau_1 \ll k \tau \ll k \tau_2$. Therefore, we get the same cubic scaling as we did for the resonance in Horndeski gravity. 

Similarly, we can estimate the growth factor for $n=3/2$. We find that it scales as
\begin{align}
    I_s^{\rm (II),res}\vert_{n=3/2} \sim & \int_{x_1}^{x_2} \md x^\prime x^\prime \sin (x^\prime) \frac{1}{2 u^2 v^2 } \frac{\md^2 T_\zeta }{\md x^2} \frac{\md^2 T_\zeta}{\md x^2} 
    \sim  c_s^4 x^{3}_2
\end{align}
which leads to a gravitational wave energy spectrum scaling as $\Omega_{\rm GW} \sim x_2^5$ as we found in GLPV.

Calculations in regime (III), corresponding to $k \tau_1 \gg 1$, become more involved. Nevertheless, deep in the UV we can neglect the last two integrals and only consider the first integral which leads again to the standard GR result. Thus, there will be an intermediate scaling as $k^3$ ($n=1$) or $k^5$ ($n=3/2$) in the induced GW spectrum. We conclude that it is possible to get the same growth factor for a scale-invariant power spectrum from a change in the normalization factor as we got from the higher-order cubic interactions. Note, however, that it is not feasible to get such a behavior for $z$ in standard GR.

\bibliography{bibliography.bib} 

\begin{thebibliography}{111}%
\makeatletter
\providecommand \@ifxundefined [1]{%
 \@ifx{#1\undefined}
}%
\providecommand \@ifnum [1]{%
 \ifnum #1\expandafter \@firstoftwo
 \else \expandafter \@secondoftwo
 \fi
}%
\providecommand \@ifx [1]{%
 \ifx #1\expandafter \@firstoftwo
 \else \expandafter \@secondoftwo
 \fi
}%
\providecommand \natexlab [1]{#1}%
\providecommand \enquote  [1]{``#1''}%
\providecommand \bibnamefont  [1]{#1}%
\providecommand \bibfnamefont [1]{#1}%
\providecommand \citenamefont [1]{#1}%
\providecommand \href@noop [0]{\@secondoftwo}%
\providecommand \href [0]{\begingroup \@sanitize@url \@href}%
\providecommand \@href[1]{\@@startlink{#1}\@@href}%
\providecommand \@@href[1]{\endgroup#1\@@endlink}%
\providecommand \@sanitize@url [0]{\catcode `\\12\catcode `\$12\catcode
  `\&12\catcode `\#12\catcode `\^12\catcode `\_12\catcode `\%12\relax}%
\providecommand \@@startlink[1]{}%
\providecommand \@@endlink[0]{}%
\providecommand \url  [0]{\begingroup\@sanitize@url \@url }%
\providecommand \@url [1]{\endgroup\@href {#1}{\urlprefix }}%
\providecommand \urlprefix  [0]{URL }%
\providecommand \Eprint [0]{\href }%
\providecommand \doibase [0]{https://doi.org/}%
\providecommand \selectlanguage [0]{\@gobble}%
\providecommand \bibinfo  [0]{\@secondoftwo}%
\providecommand \bibfield  [0]{\@secondoftwo}%
\providecommand \translation [1]{[#1]}%
\providecommand \BibitemOpen [0]{}%
\providecommand \bibitemStop [0]{}%
\providecommand \bibitemNoStop [0]{.\EOS\space}%
\providecommand \EOS [0]{\spacefactor3000\relax}%
\providecommand \BibitemShut  [1]{\csname bibitem#1\endcsname}%
\let\auto@bib@innerbib\@empty
\bibitem [{\citenamefont {Akrami}\ \emph {et~al.}(2020)\citenamefont {Akrami}
  \emph {et~al.}}]{Planck:2018jri}%
  \BibitemOpen
  \bibfield  {author} {\bibinfo {author} {\bibfnamefont {Y.}~\bibnamefont
  {Akrami}} \emph {et~al.} (\bibinfo {collaboration} {Planck}),\ }\href
  {https://doi.org/10.1051/0004-6361/201833887} {\bibfield  {journal} {\bibinfo
   {journal} {Astron. Astrophys.}\ }\textbf {\bibinfo {volume} {641}},\
  \bibinfo {pages} {A10} (\bibinfo {year} {2020})},\ \Eprint
  {https://arxiv.org/abs/1807.06211} {arXiv:1807.06211 [astro-ph.CO]}
  \BibitemShut {NoStop}%
\bibitem [{\citenamefont {Aghanim}\ \emph {et~al.}(2020)\citenamefont {Aghanim}
  \emph {et~al.}}]{Planck:2018vyg}%
  \BibitemOpen
  \bibfield  {author} {\bibinfo {author} {\bibfnamefont {N.}~\bibnamefont
  {Aghanim}} \emph {et~al.} (\bibinfo {collaboration} {Planck}),\ }\href
  {https://doi.org/10.1051/0004-6361/201833910} {\bibfield  {journal} {\bibinfo
   {journal} {Astron. Astrophys.}\ }\textbf {\bibinfo {volume} {641}},\
  \bibinfo {pages} {A6} (\bibinfo {year} {2020})},\ \bibinfo {note} {[Erratum:
  Astron.Astrophys. 652, C4 (2021)]},\ \Eprint
  {https://arxiv.org/abs/1807.06209} {arXiv:1807.06209 [astro-ph.CO]}
  \BibitemShut {NoStop}%
\bibitem [{\citenamefont {Drees}\ and\ \citenamefont
  {Xu}(2025)}]{Drees:2025ngb}%
  \BibitemOpen
  \bibfield  {author} {\bibinfo {author} {\bibfnamefont {M.}~\bibnamefont
  {Drees}}\ and\ \bibinfo {author} {\bibfnamefont {Y.}~\bibnamefont {Xu}},\
  }\href {https://doi.org/10.1016/j.physletb.2025.139612} {\bibfield  {journal}
  {\bibinfo  {journal} {Phys. Lett. B}\ }\textbf {\bibinfo {volume} {867}},\
  \bibinfo {pages} {139612} (\bibinfo {year} {2025})},\ \Eprint
  {https://arxiv.org/abs/2504.20757} {arXiv:2504.20757 [astro-ph.CO]}
  \BibitemShut {NoStop}%
\bibitem [{\citenamefont {Louis}\ \emph {et~al.}(2025)\citenamefont {Louis}
  \emph {et~al.}}]{ACT:2025fju}%
  \BibitemOpen
  \bibfield  {author} {\bibinfo {author} {\bibfnamefont {T.}~\bibnamefont
  {Louis}} \emph {et~al.} (\bibinfo {collaboration} {ACT}),\ }\href@noop {} {\
  (\bibinfo {year} {2025})},\ \Eprint {https://arxiv.org/abs/2503.14452}
  {arXiv:2503.14452 [astro-ph.CO]} \BibitemShut {NoStop}%
\bibitem [{\citenamefont {Calabrese}\ \emph {et~al.}(2025)\citenamefont
  {Calabrese} \emph {et~al.}}]{ACT:2025tim}%
  \BibitemOpen
  \bibfield  {author} {\bibinfo {author} {\bibfnamefont {E.}~\bibnamefont
  {Calabrese}} \emph {et~al.} (\bibinfo {collaboration} {ACT}),\ }\href@noop {}
  {\  (\bibinfo {year} {2025})},\ \Eprint {https://arxiv.org/abs/2503.14454}
  {arXiv:2503.14454 [astro-ph.CO]} \BibitemShut {NoStop}%
\bibitem [{\citenamefont {Ye}\ \emph {et~al.}(2025)\citenamefont {Ye},
  \citenamefont {Martinelli}, \citenamefont {Hu},\ and\ \citenamefont
  {Silvestri}}]{Ye:2024ywg}%
  \BibitemOpen
  \bibfield  {author} {\bibinfo {author} {\bibfnamefont {G.}~\bibnamefont
  {Ye}}, \bibinfo {author} {\bibfnamefont {M.}~\bibnamefont {Martinelli}},
  \bibinfo {author} {\bibfnamefont {B.}~\bibnamefont {Hu}},\ and\ \bibinfo
  {author} {\bibfnamefont {A.}~\bibnamefont {Silvestri}},\ }\href
  {https://doi.org/10.1103/PhysRevLett.134.181002} {\bibfield  {journal}
  {\bibinfo  {journal} {Phys. Rev. Lett.}\ }\textbf {\bibinfo {volume} {134}},\
  \bibinfo {pages} {181002} (\bibinfo {year} {2025})},\ \Eprint
  {https://arxiv.org/abs/2407.15832} {arXiv:2407.15832 [astro-ph.CO]}
  \BibitemShut {NoStop}%
\bibitem [{\citenamefont {Ostrogradsky}(1850)}]{Ostrogradsky:1850fid}%
  \BibitemOpen
  \bibfield  {author} {\bibinfo {author} {\bibfnamefont {M.}~\bibnamefont
  {Ostrogradsky}},\ }\href@noop {} {\bibfield  {journal} {\bibinfo  {journal}
  {Mem. Acad. St. Petersbourg}\ }\textbf {\bibinfo {volume} {6}},\ \bibinfo
  {pages} {385} (\bibinfo {year} {1850})}\BibitemShut {NoStop}%
\bibitem [{\citenamefont {Woodard}(2015)}]{Woodard:2015zca}%
  \BibitemOpen
  \bibfield  {author} {\bibinfo {author} {\bibfnamefont {R.~P.}\ \bibnamefont
  {Woodard}},\ }\href {https://doi.org/10.4249/scholarpedia.32243} {\bibfield
  {journal} {\bibinfo  {journal} {Scholarpedia}\ }\textbf {\bibinfo {volume}
  {10}},\ \bibinfo {pages} {32243} (\bibinfo {year} {2015})},\ \Eprint
  {https://arxiv.org/abs/1506.02210} {arXiv:1506.02210 [hep-th]} \BibitemShut
  {NoStop}%
\bibitem [{\citenamefont {Ganz}\ and\ \citenamefont
  {Noui}(2021)}]{Ganz:2020skf}%
  \BibitemOpen
  \bibfield  {author} {\bibinfo {author} {\bibfnamefont {A.}~\bibnamefont
  {Ganz}}\ and\ \bibinfo {author} {\bibfnamefont {K.}~\bibnamefont {Noui}},\
  }\href {https://doi.org/10.1088/1361-6382/abe31d} {\bibfield  {journal}
  {\bibinfo  {journal} {Class. Quant. Grav.}\ }\textbf {\bibinfo {volume}
  {38}},\ \bibinfo {pages} {075005} (\bibinfo {year} {2021})},\ \Eprint
  {https://arxiv.org/abs/2007.01063} {arXiv:2007.01063 [hep-th]} \BibitemShut
  {NoStop}%
\bibitem [{\citenamefont {Gao}(2014{\natexlab{a}})}]{Gao:2014fra}%
  \BibitemOpen
  \bibfield  {author} {\bibinfo {author} {\bibfnamefont {X.}~\bibnamefont
  {Gao}},\ }\href {https://doi.org/10.1103/PhysRevD.90.104033} {\bibfield
  {journal} {\bibinfo  {journal} {Phys. Rev. D}\ }\textbf {\bibinfo {volume}
  {90}},\ \bibinfo {pages} {104033} (\bibinfo {year} {2014}{\natexlab{a}})},\
  \Eprint {https://arxiv.org/abs/1409.6708} {arXiv:1409.6708 [gr-qc]}
  \BibitemShut {NoStop}%
\bibitem [{\citenamefont {Gao}(2014{\natexlab{b}})}]{Gao:2014soa}%
  \BibitemOpen
  \bibfield  {author} {\bibinfo {author} {\bibfnamefont {X.}~\bibnamefont
  {Gao}},\ }\href {https://doi.org/10.1103/PhysRevD.90.081501} {\bibfield
  {journal} {\bibinfo  {journal} {Phys. Rev. D}\ }\textbf {\bibinfo {volume}
  {90}},\ \bibinfo {pages} {081501} (\bibinfo {year} {2014}{\natexlab{b}})},\
  \Eprint {https://arxiv.org/abs/1406.0822} {arXiv:1406.0822 [gr-qc]}
  \BibitemShut {NoStop}%
\bibitem [{\citenamefont {Gleyzes}\ \emph
  {et~al.}(2015{\natexlab{a}})\citenamefont {Gleyzes}, \citenamefont
  {Langlois}, \citenamefont {Piazza},\ and\ \citenamefont
  {Vernizzi}}]{Gleyzes:2014dya}%
  \BibitemOpen
  \bibfield  {author} {\bibinfo {author} {\bibfnamefont {J.}~\bibnamefont
  {Gleyzes}}, \bibinfo {author} {\bibfnamefont {D.}~\bibnamefont {Langlois}},
  \bibinfo {author} {\bibfnamefont {F.}~\bibnamefont {Piazza}},\ and\ \bibinfo
  {author} {\bibfnamefont {F.}~\bibnamefont {Vernizzi}},\ }\href
  {https://doi.org/10.1103/PhysRevLett.114.211101} {\bibfield  {journal}
  {\bibinfo  {journal} {Phys. Rev. Lett.}\ }\textbf {\bibinfo {volume} {114}},\
  \bibinfo {pages} {211101} (\bibinfo {year} {2015}{\natexlab{a}})},\ \Eprint
  {https://arxiv.org/abs/1404.6495} {arXiv:1404.6495 [hep-th]} \BibitemShut
  {NoStop}%
\bibitem [{\citenamefont {Langlois}\ and\ \citenamefont
  {Noui}(2016{\natexlab{a}})}]{Langlois:2015cwa}%
  \BibitemOpen
  \bibfield  {author} {\bibinfo {author} {\bibfnamefont {D.}~\bibnamefont
  {Langlois}}\ and\ \bibinfo {author} {\bibfnamefont {K.}~\bibnamefont
  {Noui}},\ }\href {https://doi.org/10.1088/1475-7516/2016/02/034} {\bibfield
  {journal} {\bibinfo  {journal} {JCAP}\ }\textbf {\bibinfo {volume} {02}},\
  \bibinfo {pages} {034}},\ \Eprint {https://arxiv.org/abs/1510.06930}
  {arXiv:1510.06930 [gr-qc]} \BibitemShut {NoStop}%
\bibitem [{\citenamefont {De~Felice}\ \emph {et~al.}(2018)\citenamefont
  {De~Felice}, \citenamefont {Langlois}, \citenamefont {Mukohyama},
  \citenamefont {Noui},\ and\ \citenamefont {Wang}}]{DeFelice:2018ewo}%
  \BibitemOpen
  \bibfield  {author} {\bibinfo {author} {\bibfnamefont {A.}~\bibnamefont
  {De~Felice}}, \bibinfo {author} {\bibfnamefont {D.}~\bibnamefont {Langlois}},
  \bibinfo {author} {\bibfnamefont {S.}~\bibnamefont {Mukohyama}}, \bibinfo
  {author} {\bibfnamefont {K.}~\bibnamefont {Noui}},\ and\ \bibinfo {author}
  {\bibfnamefont {A.}~\bibnamefont {Wang}},\ }\href
  {https://doi.org/10.1103/PhysRevD.98.084024} {\bibfield  {journal} {\bibinfo
  {journal} {Phys. Rev. D}\ }\textbf {\bibinfo {volume} {98}},\ \bibinfo
  {pages} {084024} (\bibinfo {year} {2018})},\ \Eprint
  {https://arxiv.org/abs/1803.06241} {arXiv:1803.06241 [hep-th]} \BibitemShut
  {NoStop}%
\bibitem [{\citenamefont {Kobayashi}(2019)}]{Kobayashi:2019hrl}%
  \BibitemOpen
  \bibfield  {author} {\bibinfo {author} {\bibfnamefont {T.}~\bibnamefont
  {Kobayashi}},\ }\href {https://doi.org/10.1088/1361-6633/ab2429} {\bibfield
  {journal} {\bibinfo  {journal} {Rept. Prog. Phys.}\ }\textbf {\bibinfo
  {volume} {82}},\ \bibinfo {pages} {086901} (\bibinfo {year} {2019})},\
  \Eprint {https://arxiv.org/abs/1901.07183} {arXiv:1901.07183 [gr-qc]}
  \BibitemShut {NoStop}%
\bibitem [{\citenamefont {Frusciante}\ and\ \citenamefont
  {Perenon}(2020)}]{Frusciante:2019xia}%
  \BibitemOpen
  \bibfield  {author} {\bibinfo {author} {\bibfnamefont {N.}~\bibnamefont
  {Frusciante}}\ and\ \bibinfo {author} {\bibfnamefont {L.}~\bibnamefont
  {Perenon}},\ }\href {https://doi.org/10.1016/j.physrep.2020.02.004}
  {\bibfield  {journal} {\bibinfo  {journal} {Phys. Rept.}\ }\textbf {\bibinfo
  {volume} {857}},\ \bibinfo {pages} {1} (\bibinfo {year} {2020})},\ \Eprint
  {https://arxiv.org/abs/1907.03150} {arXiv:1907.03150 [astro-ph.CO]}
  \BibitemShut {NoStop}%
\bibitem [{\citenamefont {Heisenberg}(2014)}]{Heisenberg:2014rta}%
  \BibitemOpen
  \bibfield  {author} {\bibinfo {author} {\bibfnamefont {L.}~\bibnamefont
  {Heisenberg}},\ }\href {https://doi.org/10.1088/1475-7516/2014/05/015}
  {\bibfield  {journal} {\bibinfo  {journal} {JCAP}\ }\textbf {\bibinfo
  {volume} {05}},\ \bibinfo {pages} {015}},\ \Eprint
  {https://arxiv.org/abs/1402.7026} {arXiv:1402.7026 [hep-th]} \BibitemShut
  {NoStop}%
\bibitem [{\citenamefont {Heisenberg}\ \emph {et~al.}(2016)\citenamefont
  {Heisenberg}, \citenamefont {Kase},\ and\ \citenamefont
  {Tsujikawa}}]{Heisenberg:2016eld}%
  \BibitemOpen
  \bibfield  {author} {\bibinfo {author} {\bibfnamefont {L.}~\bibnamefont
  {Heisenberg}}, \bibinfo {author} {\bibfnamefont {R.}~\bibnamefont {Kase}},\
  and\ \bibinfo {author} {\bibfnamefont {S.}~\bibnamefont {Tsujikawa}},\ }\href
  {https://doi.org/10.1016/j.physletb.2016.07.052} {\bibfield  {journal}
  {\bibinfo  {journal} {Phys. Lett. B}\ }\textbf {\bibinfo {volume} {760}},\
  \bibinfo {pages} {617} (\bibinfo {year} {2016})},\ \Eprint
  {https://arxiv.org/abs/1605.05565} {arXiv:1605.05565 [hep-th]} \BibitemShut
  {NoStop}%
\bibitem [{\citenamefont {Kimura}\ \emph {et~al.}(2017)\citenamefont {Kimura},
  \citenamefont {Naruko},\ and\ \citenamefont {Yoshida}}]{Kimura:2016rzw}%
  \BibitemOpen
  \bibfield  {author} {\bibinfo {author} {\bibfnamefont {R.}~\bibnamefont
  {Kimura}}, \bibinfo {author} {\bibfnamefont {A.}~\bibnamefont {Naruko}},\
  and\ \bibinfo {author} {\bibfnamefont {D.}~\bibnamefont {Yoshida}},\ }\href
  {https://doi.org/10.1088/1475-7516/2017/01/002} {\bibfield  {journal}
  {\bibinfo  {journal} {JCAP}\ }\textbf {\bibinfo {volume} {01}},\ \bibinfo
  {pages} {002}},\ \Eprint {https://arxiv.org/abs/1608.07066} {arXiv:1608.07066
  [gr-qc]} \BibitemShut {NoStop}%
\bibitem [{\citenamefont {de~Rham}\ and\ \citenamefont
  {Pozsgay}(2020)}]{deRham:2020yet}%
  \BibitemOpen
  \bibfield  {author} {\bibinfo {author} {\bibfnamefont {C.}~\bibnamefont
  {de~Rham}}\ and\ \bibinfo {author} {\bibfnamefont {V.}~\bibnamefont
  {Pozsgay}},\ }\href {https://doi.org/10.1103/PhysRevD.102.083508} {\bibfield
  {journal} {\bibinfo  {journal} {Phys. Rev. D}\ }\textbf {\bibinfo {volume}
  {102}},\ \bibinfo {pages} {083508} (\bibinfo {year} {2020})},\ \Eprint
  {https://arxiv.org/abs/2003.13773} {arXiv:2003.13773 [hep-th]} \BibitemShut
  {NoStop}%
\bibitem [{\citenamefont {Horndeski}(1974)}]{Horndeski:1974wa}%
  \BibitemOpen
  \bibfield  {author} {\bibinfo {author} {\bibfnamefont {G.~W.}\ \bibnamefont
  {Horndeski}},\ }\href {https://doi.org/10.1007/BF01807638} {\bibfield
  {journal} {\bibinfo  {journal} {Int. J. Theor. Phys.}\ }\textbf {\bibinfo
  {volume} {10}},\ \bibinfo {pages} {363} (\bibinfo {year} {1974})}\BibitemShut
  {NoStop}%
\bibitem [{\citenamefont {Charmousis}\ \emph {et~al.}(2012)\citenamefont
  {Charmousis}, \citenamefont {Copeland}, \citenamefont {Padilla},\ and\
  \citenamefont {Saffin}}]{Charmousis:2011bf}%
  \BibitemOpen
  \bibfield  {author} {\bibinfo {author} {\bibfnamefont {C.}~\bibnamefont
  {Charmousis}}, \bibinfo {author} {\bibfnamefont {E.~J.}\ \bibnamefont
  {Copeland}}, \bibinfo {author} {\bibfnamefont {A.}~\bibnamefont {Padilla}},\
  and\ \bibinfo {author} {\bibfnamefont {P.~M.}\ \bibnamefont {Saffin}},\
  }\href {https://doi.org/10.1103/PhysRevLett.108.051101} {\bibfield  {journal}
  {\bibinfo  {journal} {Phys. Rev. Lett.}\ }\textbf {\bibinfo {volume} {108}},\
  \bibinfo {pages} {051101} (\bibinfo {year} {2012})},\ \Eprint
  {https://arxiv.org/abs/1106.2000} {arXiv:1106.2000 [hep-th]} \BibitemShut
  {NoStop}%
\bibitem [{\citenamefont {Deffayet}\ \emph {et~al.}(2011)\citenamefont
  {Deffayet}, \citenamefont {Gao}, \citenamefont {Steer},\ and\ \citenamefont
  {Zahariade}}]{Deffayet:2011gz}%
  \BibitemOpen
  \bibfield  {author} {\bibinfo {author} {\bibfnamefont {C.}~\bibnamefont
  {Deffayet}}, \bibinfo {author} {\bibfnamefont {X.}~\bibnamefont {Gao}},
  \bibinfo {author} {\bibfnamefont {D.~A.}\ \bibnamefont {Steer}},\ and\
  \bibinfo {author} {\bibfnamefont {G.}~\bibnamefont {Zahariade}},\ }\href
  {https://doi.org/10.1103/PhysRevD.84.064039} {\bibfield  {journal} {\bibinfo
  {journal} {Phys. Rev. D}\ }\textbf {\bibinfo {volume} {84}},\ \bibinfo
  {pages} {064039} (\bibinfo {year} {2011})},\ \Eprint
  {https://arxiv.org/abs/1103.3260} {arXiv:1103.3260 [hep-th]} \BibitemShut
  {NoStop}%
\bibitem [{\citenamefont {Kobayashi}\ \emph {et~al.}(2011)\citenamefont
  {Kobayashi}, \citenamefont {Yamaguchi},\ and\ \citenamefont
  {Yokoyama}}]{Kobayashi:2011nu}%
  \BibitemOpen
  \bibfield  {author} {\bibinfo {author} {\bibfnamefont {T.}~\bibnamefont
  {Kobayashi}}, \bibinfo {author} {\bibfnamefont {M.}~\bibnamefont
  {Yamaguchi}},\ and\ \bibinfo {author} {\bibfnamefont {J.}~\bibnamefont
  {Yokoyama}},\ }\href {https://doi.org/10.1143/PTP.126.511} {\bibfield
  {journal} {\bibinfo  {journal} {Prog. Theor. Phys.}\ }\textbf {\bibinfo
  {volume} {126}},\ \bibinfo {pages} {511} (\bibinfo {year} {2011})},\ \Eprint
  {https://arxiv.org/abs/1105.5723} {arXiv:1105.5723 [hep-th]} \BibitemShut
  {NoStop}%
\bibitem [{\citenamefont {Gleyzes}\ \emph
  {et~al.}(2015{\natexlab{b}})\citenamefont {Gleyzes}, \citenamefont
  {Langlois}, \citenamefont {Piazza},\ and\ \citenamefont
  {Vernizzi}}]{Gleyzes:2014qga}%
  \BibitemOpen
  \bibfield  {author} {\bibinfo {author} {\bibfnamefont {J.}~\bibnamefont
  {Gleyzes}}, \bibinfo {author} {\bibfnamefont {D.}~\bibnamefont {Langlois}},
  \bibinfo {author} {\bibfnamefont {F.}~\bibnamefont {Piazza}},\ and\ \bibinfo
  {author} {\bibfnamefont {F.}~\bibnamefont {Vernizzi}},\ }\href
  {https://doi.org/10.1088/1475-7516/2015/02/018} {\bibfield  {journal}
  {\bibinfo  {journal} {JCAP}\ }\textbf {\bibinfo {volume} {02}},\ \bibinfo
  {pages} {018}},\ \Eprint {https://arxiv.org/abs/1408.1952} {arXiv:1408.1952
  [astro-ph.CO]} \BibitemShut {NoStop}%
\bibitem [{\citenamefont {Langlois}\ and\ \citenamefont
  {Noui}(2016{\natexlab{b}})}]{Langlois:2015skt}%
  \BibitemOpen
  \bibfield  {author} {\bibinfo {author} {\bibfnamefont {D.}~\bibnamefont
  {Langlois}}\ and\ \bibinfo {author} {\bibfnamefont {K.}~\bibnamefont
  {Noui}},\ }\href {https://doi.org/10.1088/1475-7516/2016/07/016} {\bibfield
  {journal} {\bibinfo  {journal} {JCAP}\ }\textbf {\bibinfo {volume} {07}},\
  \bibinfo {pages} {016}},\ \Eprint {https://arxiv.org/abs/1512.06820}
  {arXiv:1512.06820 [gr-qc]} \BibitemShut {NoStop}%
\bibitem [{\citenamefont {Crisostomi}\ \emph
  {et~al.}(2016{\natexlab{a}})\citenamefont {Crisostomi}, \citenamefont
  {Koyama},\ and\ \citenamefont {Tasinato}}]{Crisostomi:2016czh}%
  \BibitemOpen
  \bibfield  {author} {\bibinfo {author} {\bibfnamefont {M.}~\bibnamefont
  {Crisostomi}}, \bibinfo {author} {\bibfnamefont {K.}~\bibnamefont {Koyama}},\
  and\ \bibinfo {author} {\bibfnamefont {G.}~\bibnamefont {Tasinato}},\ }\href
  {https://doi.org/10.1088/1475-7516/2016/04/044} {\bibfield  {journal}
  {\bibinfo  {journal} {JCAP}\ }\textbf {\bibinfo {volume} {04}},\ \bibinfo
  {pages} {044}},\ \Eprint {https://arxiv.org/abs/1602.03119} {arXiv:1602.03119
  [hep-th]} \BibitemShut {NoStop}%
\bibitem [{\citenamefont {Ben~Achour}\ \emph
  {et~al.}(2016{\natexlab{a}})\citenamefont {Ben~Achour}, \citenamefont
  {Crisostomi}, \citenamefont {Koyama}, \citenamefont {Langlois}, \citenamefont
  {Noui},\ and\ \citenamefont {Tasinato}}]{BenAchour:2016fzp}%
  \BibitemOpen
  \bibfield  {author} {\bibinfo {author} {\bibfnamefont {J.}~\bibnamefont
  {Ben~Achour}}, \bibinfo {author} {\bibfnamefont {M.}~\bibnamefont
  {Crisostomi}}, \bibinfo {author} {\bibfnamefont {K.}~\bibnamefont {Koyama}},
  \bibinfo {author} {\bibfnamefont {D.}~\bibnamefont {Langlois}}, \bibinfo
  {author} {\bibfnamefont {K.}~\bibnamefont {Noui}},\ and\ \bibinfo {author}
  {\bibfnamefont {G.}~\bibnamefont {Tasinato}},\ }\href
  {https://doi.org/10.1007/JHEP12(2016)100} {\bibfield  {journal} {\bibinfo
  {journal} {JHEP}\ }\textbf {\bibinfo {volume} {12}},\ \bibinfo {pages}
  {100}},\ \Eprint {https://arxiv.org/abs/1608.08135} {arXiv:1608.08135
  [hep-th]} \BibitemShut {NoStop}%
\bibitem [{\citenamefont {Bekenstein}(1993)}]{Bekenstein:1992pj}%
  \BibitemOpen
  \bibfield  {author} {\bibinfo {author} {\bibfnamefont {J.~D.}\ \bibnamefont
  {Bekenstein}},\ }\href {https://doi.org/10.1103/PhysRevD.48.3641} {\bibfield
  {journal} {\bibinfo  {journal} {Phys. Rev. D}\ }\textbf {\bibinfo {volume}
  {48}},\ \bibinfo {pages} {3641} (\bibinfo {year} {1993})},\ \Eprint
  {https://arxiv.org/abs/gr-qc/9211017} {arXiv:gr-qc/9211017} \BibitemShut
  {NoStop}%
\bibitem [{\citenamefont {Bettoni}\ and\ \citenamefont
  {Liberati}(2013)}]{Bettoni:2013diz}%
  \BibitemOpen
  \bibfield  {author} {\bibinfo {author} {\bibfnamefont {D.}~\bibnamefont
  {Bettoni}}\ and\ \bibinfo {author} {\bibfnamefont {S.}~\bibnamefont
  {Liberati}},\ }\href {https://doi.org/10.1103/PhysRevD.88.084020} {\bibfield
  {journal} {\bibinfo  {journal} {Phys. Rev. D}\ }\textbf {\bibinfo {volume}
  {88}},\ \bibinfo {pages} {084020} (\bibinfo {year} {2013})},\ \Eprint
  {https://arxiv.org/abs/1306.6724} {arXiv:1306.6724 [gr-qc]} \BibitemShut
  {NoStop}%
\bibitem [{\citenamefont {Zumalac{\'a}rregui}\ and\ \citenamefont
  {Garc{\'\i}a-Bellido}(2014)}]{Zumalacarregui:2013pma}%
  \BibitemOpen
  \bibfield  {author} {\bibinfo {author} {\bibfnamefont {M.}~\bibnamefont
  {Zumalac{\'a}rregui}}\ and\ \bibinfo {author} {\bibfnamefont
  {J.}~\bibnamefont {Garc{\'\i}a-Bellido}},\ }\href
  {https://doi.org/10.1103/PhysRevD.89.064046} {\bibfield  {journal} {\bibinfo
  {journal} {Phys. Rev. D}\ }\textbf {\bibinfo {volume} {89}},\ \bibinfo
  {pages} {064046} (\bibinfo {year} {2014})},\ \Eprint
  {https://arxiv.org/abs/1308.4685} {arXiv:1308.4685 [gr-qc]} \BibitemShut
  {NoStop}%
\bibitem [{\citenamefont {Ezquiaga}\ \emph {et~al.}(2017)\citenamefont
  {Ezquiaga}, \citenamefont {Garc{\'\i}a-Bellido},\ and\ \citenamefont
  {Zumalac{\'a}rregui}}]{Ezquiaga:2017ner}%
  \BibitemOpen
  \bibfield  {author} {\bibinfo {author} {\bibfnamefont {J.~M.}\ \bibnamefont
  {Ezquiaga}}, \bibinfo {author} {\bibfnamefont {J.}~\bibnamefont
  {Garc{\'\i}a-Bellido}},\ and\ \bibinfo {author} {\bibfnamefont
  {M.}~\bibnamefont {Zumalac{\'a}rregui}},\ }\href
  {https://doi.org/10.1103/PhysRevD.95.084039} {\bibfield  {journal} {\bibinfo
  {journal} {Phys. Rev. D}\ }\textbf {\bibinfo {volume} {95}},\ \bibinfo
  {pages} {084039} (\bibinfo {year} {2017})},\ \Eprint
  {https://arxiv.org/abs/1701.05476} {arXiv:1701.05476 [hep-th]} \BibitemShut
  {NoStop}%
\bibitem [{\citenamefont {Ben~Achour}\ \emph
  {et~al.}(2016{\natexlab{b}})\citenamefont {Ben~Achour}, \citenamefont
  {Langlois},\ and\ \citenamefont {Noui}}]{BenAchour:2016cay}%
  \BibitemOpen
  \bibfield  {author} {\bibinfo {author} {\bibfnamefont {J.}~\bibnamefont
  {Ben~Achour}}, \bibinfo {author} {\bibfnamefont {D.}~\bibnamefont
  {Langlois}},\ and\ \bibinfo {author} {\bibfnamefont {K.}~\bibnamefont
  {Noui}},\ }\href {https://doi.org/10.1103/PhysRevD.93.124005} {\bibfield
  {journal} {\bibinfo  {journal} {Phys. Rev. D}\ }\textbf {\bibinfo {volume}
  {93}},\ \bibinfo {pages} {124005} (\bibinfo {year} {2016}{\natexlab{b}})},\
  \Eprint {https://arxiv.org/abs/1602.08398} {arXiv:1602.08398 [gr-qc]}
  \BibitemShut {NoStop}%
\bibitem [{\citenamefont {de~Rham}\ and\ \citenamefont
  {Matas}(2016)}]{deRham:2016wji}%
  \BibitemOpen
  \bibfield  {author} {\bibinfo {author} {\bibfnamefont {C.}~\bibnamefont
  {de~Rham}}\ and\ \bibinfo {author} {\bibfnamefont {A.}~\bibnamefont
  {Matas}},\ }\href {https://doi.org/10.1088/1475-7516/2016/06/041} {\bibfield
  {journal} {\bibinfo  {journal} {JCAP}\ }\textbf {\bibinfo {volume} {06}},\
  \bibinfo {pages} {041}},\ \Eprint {https://arxiv.org/abs/1604.08638}
  {arXiv:1604.08638 [hep-th]} \BibitemShut {NoStop}%
\bibitem [{\citenamefont {Crisostomi}\ \emph
  {et~al.}(2016{\natexlab{b}})\citenamefont {Crisostomi}, \citenamefont {Hull},
  \citenamefont {Koyama},\ and\ \citenamefont {Tasinato}}]{Crisostomi:2016tcp}%
  \BibitemOpen
  \bibfield  {author} {\bibinfo {author} {\bibfnamefont {M.}~\bibnamefont
  {Crisostomi}}, \bibinfo {author} {\bibfnamefont {M.}~\bibnamefont {Hull}},
  \bibinfo {author} {\bibfnamefont {K.}~\bibnamefont {Koyama}},\ and\ \bibinfo
  {author} {\bibfnamefont {G.}~\bibnamefont {Tasinato}},\ }\href
  {https://doi.org/10.1088/1475-7516/2016/03/038} {\bibfield  {journal}
  {\bibinfo  {journal} {JCAP}\ }\textbf {\bibinfo {volume} {03}},\ \bibinfo
  {pages} {038}},\ \Eprint {https://arxiv.org/abs/1601.04658} {arXiv:1601.04658
  [hep-th]} \BibitemShut {NoStop}%
\bibitem [{\citenamefont {Renaux-Petel}(2012)}]{Renaux-Petel:2011zgy}%
  \BibitemOpen
  \bibfield  {author} {\bibinfo {author} {\bibfnamefont {S.}~\bibnamefont
  {Renaux-Petel}},\ }\href {https://doi.org/10.1088/1475-7516/2012/02/020}
  {\bibfield  {journal} {\bibinfo  {journal} {JCAP}\ }\textbf {\bibinfo
  {volume} {02}},\ \bibinfo {pages} {020}},\ \Eprint
  {https://arxiv.org/abs/1107.5020} {arXiv:1107.5020 [astro-ph.CO]}
  \BibitemShut {NoStop}%
\bibitem [{\citenamefont {Fasiello}\ and\ \citenamefont
  {Renaux-Petel}(2014)}]{Fasiello:2014aqa}%
  \BibitemOpen
  \bibfield  {author} {\bibinfo {author} {\bibfnamefont {M.}~\bibnamefont
  {Fasiello}}\ and\ \bibinfo {author} {\bibfnamefont {S.}~\bibnamefont
  {Renaux-Petel}},\ }\href {https://doi.org/10.1088/1475-7516/2014/10/037}
  {\bibfield  {journal} {\bibinfo  {journal} {JCAP}\ }\textbf {\bibinfo
  {volume} {10}},\ \bibinfo {pages} {037}},\ \Eprint
  {https://arxiv.org/abs/1407.7280} {arXiv:1407.7280 [astro-ph.CO]}
  \BibitemShut {NoStop}%
\bibitem [{\citenamefont {Akita}\ and\ \citenamefont
  {Kobayashi}(2016)}]{Akita:2015mho}%
  \BibitemOpen
  \bibfield  {author} {\bibinfo {author} {\bibfnamefont {Y.}~\bibnamefont
  {Akita}}\ and\ \bibinfo {author} {\bibfnamefont {T.}~\bibnamefont
  {Kobayashi}},\ }\href {https://doi.org/10.1103/PhysRevD.93.043519} {\bibfield
   {journal} {\bibinfo  {journal} {Phys. Rev. D}\ }\textbf {\bibinfo {volume}
  {93}},\ \bibinfo {pages} {043519} (\bibinfo {year} {2016})},\ \Eprint
  {https://arxiv.org/abs/1512.01380} {arXiv:1512.01380 [hep-th]} \BibitemShut
  {NoStop}%
\bibitem [{\citenamefont {Passaglia}\ and\ \citenamefont
  {Hu}(2018)}]{Passaglia:2018afq}%
  \BibitemOpen
  \bibfield  {author} {\bibinfo {author} {\bibfnamefont {S.}~\bibnamefont
  {Passaglia}}\ and\ \bibinfo {author} {\bibfnamefont {W.}~\bibnamefont {Hu}},\
  }\href {https://doi.org/10.1103/PhysRevD.98.023526} {\bibfield  {journal}
  {\bibinfo  {journal} {Phys. Rev. D}\ }\textbf {\bibinfo {volume} {98}},\
  \bibinfo {pages} {023526} (\bibinfo {year} {2018})},\ \Eprint
  {https://arxiv.org/abs/1804.07741} {arXiv:1804.07741 [astro-ph.CO]}
  \BibitemShut {NoStop}%
\bibitem [{\citenamefont {Gao}\ \emph {et~al.}(2013)\citenamefont {Gao},
  \citenamefont {Kobayashi}, \citenamefont {Shiraishi}, \citenamefont
  {Yamaguchi}, \citenamefont {Yokoyama},\ and\ \citenamefont
  {Yokoyama}}]{Gao:2012ib}%
  \BibitemOpen
  \bibfield  {author} {\bibinfo {author} {\bibfnamefont {X.}~\bibnamefont
  {Gao}}, \bibinfo {author} {\bibfnamefont {T.}~\bibnamefont {Kobayashi}},
  \bibinfo {author} {\bibfnamefont {M.}~\bibnamefont {Shiraishi}}, \bibinfo
  {author} {\bibfnamefont {M.}~\bibnamefont {Yamaguchi}}, \bibinfo {author}
  {\bibfnamefont {J.}~\bibnamefont {Yokoyama}},\ and\ \bibinfo {author}
  {\bibfnamefont {S.}~\bibnamefont {Yokoyama}},\ }\href
  {https://doi.org/10.1093/ptep/ptt031} {\bibfield  {journal} {\bibinfo
  {journal} {PTEP}\ }\textbf {\bibinfo {volume} {2013}},\ \bibinfo {pages}
  {053E03} (\bibinfo {year} {2013})},\ \Eprint
  {https://arxiv.org/abs/1207.0588} {arXiv:1207.0588 [astro-ph.CO]}
  \BibitemShut {NoStop}%
\bibitem [{\citenamefont {Ezquiaga}\ and\ \citenamefont
  {Zumalac{\'a}rregui}(2017)}]{Ezquiaga:2017ekz}%
  \BibitemOpen
  \bibfield  {author} {\bibinfo {author} {\bibfnamefont {J.~M.}\ \bibnamefont
  {Ezquiaga}}\ and\ \bibinfo {author} {\bibfnamefont {M.}~\bibnamefont
  {Zumalac{\'a}rregui}},\ }\href
  {https://doi.org/10.1103/PhysRevLett.119.251304} {\bibfield  {journal}
  {\bibinfo  {journal} {Phys. Rev. Lett.}\ }\textbf {\bibinfo {volume} {119}},\
  \bibinfo {pages} {251304} (\bibinfo {year} {2017})},\ \Eprint
  {https://arxiv.org/abs/1710.05901} {arXiv:1710.05901 [astro-ph.CO]}
  \BibitemShut {NoStop}%
\bibitem [{\citenamefont {Langlois}\ \emph {et~al.}(2018)\citenamefont
  {Langlois}, \citenamefont {Saito}, \citenamefont {Yamauchi},\ and\
  \citenamefont {Noui}}]{Langlois:2017dyl}%
  \BibitemOpen
  \bibfield  {author} {\bibinfo {author} {\bibfnamefont {D.}~\bibnamefont
  {Langlois}}, \bibinfo {author} {\bibfnamefont {R.}~\bibnamefont {Saito}},
  \bibinfo {author} {\bibfnamefont {D.}~\bibnamefont {Yamauchi}},\ and\
  \bibinfo {author} {\bibfnamefont {K.}~\bibnamefont {Noui}},\ }\href
  {https://doi.org/10.1103/PhysRevD.97.061501} {\bibfield  {journal} {\bibinfo
  {journal} {Phys. Rev. D}\ }\textbf {\bibinfo {volume} {97}},\ \bibinfo
  {pages} {061501} (\bibinfo {year} {2018})},\ \Eprint
  {https://arxiv.org/abs/1711.07403} {arXiv:1711.07403 [gr-qc]} \BibitemShut
  {NoStop}%
\bibitem [{\citenamefont {Sakstein}\ and\ \citenamefont
  {Jain}(2017)}]{Sakstein:2017xjx}%
  \BibitemOpen
  \bibfield  {author} {\bibinfo {author} {\bibfnamefont {J.}~\bibnamefont
  {Sakstein}}\ and\ \bibinfo {author} {\bibfnamefont {B.}~\bibnamefont
  {Jain}},\ }\href {https://doi.org/10.1103/PhysRevLett.119.251303} {\bibfield
  {journal} {\bibinfo  {journal} {Phys. Rev. Lett.}\ }\textbf {\bibinfo
  {volume} {119}},\ \bibinfo {pages} {251303} (\bibinfo {year} {2017})},\
  \Eprint {https://arxiv.org/abs/1710.05893} {arXiv:1710.05893 [astro-ph.CO]}
  \BibitemShut {NoStop}%
\bibitem [{\citenamefont {Crisostomi}\ and\ \citenamefont
  {Koyama}(2018)}]{Crisostomi:2017lbg}%
  \BibitemOpen
  \bibfield  {author} {\bibinfo {author} {\bibfnamefont {M.}~\bibnamefont
  {Crisostomi}}\ and\ \bibinfo {author} {\bibfnamefont {K.}~\bibnamefont
  {Koyama}},\ }\href {https://doi.org/10.1103/PhysRevD.97.021301} {\bibfield
  {journal} {\bibinfo  {journal} {Phys. Rev. D}\ }\textbf {\bibinfo {volume}
  {97}},\ \bibinfo {pages} {021301} (\bibinfo {year} {2018})},\ \Eprint
  {https://arxiv.org/abs/1711.06661} {arXiv:1711.06661 [astro-ph.CO]}
  \BibitemShut {NoStop}%
\bibitem [{\citenamefont {Creminelli}\ and\ \citenamefont
  {Vernizzi}(2017)}]{Creminelli:2017sry}%
  \BibitemOpen
  \bibfield  {author} {\bibinfo {author} {\bibfnamefont {P.}~\bibnamefont
  {Creminelli}}\ and\ \bibinfo {author} {\bibfnamefont {F.}~\bibnamefont
  {Vernizzi}},\ }\href {https://doi.org/10.1103/PhysRevLett.119.251302}
  {\bibfield  {journal} {\bibinfo  {journal} {Phys. Rev. Lett.}\ }\textbf
  {\bibinfo {volume} {119}},\ \bibinfo {pages} {251302} (\bibinfo {year}
  {2017})},\ \Eprint {https://arxiv.org/abs/1710.05877} {arXiv:1710.05877
  [astro-ph.CO]} \BibitemShut {NoStop}%
\bibitem [{\citenamefont {Amendola}\ \emph {et~al.}(2018)\citenamefont
  {Amendola}, \citenamefont {Bettoni}, \citenamefont {Dom{\`e}nech},\ and\
  \citenamefont {Gomes}}]{Amendola:2018ltt}%
  \BibitemOpen
  \bibfield  {author} {\bibinfo {author} {\bibfnamefont {L.}~\bibnamefont
  {Amendola}}, \bibinfo {author} {\bibfnamefont {D.}~\bibnamefont {Bettoni}},
  \bibinfo {author} {\bibfnamefont {G.}~\bibnamefont {Dom{\`e}nech}},\ and\
  \bibinfo {author} {\bibfnamefont {A.~R.}\ \bibnamefont {Gomes}},\ }\href
  {https://doi.org/10.1088/1475-7516/2018/06/029} {\bibfield  {journal}
  {\bibinfo  {journal} {JCAP}\ }\textbf {\bibinfo {volume} {06}},\ \bibinfo
  {pages} {029}},\ \Eprint {https://arxiv.org/abs/1803.06368} {arXiv:1803.06368
  [gr-qc]} \BibitemShut {NoStop}%
\bibitem [{\citenamefont {Bordin}\ \emph {et~al.}(2020)\citenamefont {Bordin},
  \citenamefont {Copeland},\ and\ \citenamefont {Padilla}}]{Bordin:2020fww}%
  \BibitemOpen
  \bibfield  {author} {\bibinfo {author} {\bibfnamefont {L.}~\bibnamefont
  {Bordin}}, \bibinfo {author} {\bibfnamefont {E.~J.}\ \bibnamefont
  {Copeland}},\ and\ \bibinfo {author} {\bibfnamefont {A.}~\bibnamefont
  {Padilla}},\ }\href {https://doi.org/10.1088/1475-7516/2020/11/063}
  {\bibfield  {journal} {\bibinfo  {journal} {JCAP}\ }\textbf {\bibinfo
  {volume} {11}},\ \bibinfo {pages} {063}},\ \Eprint
  {https://arxiv.org/abs/2006.06652} {arXiv:2006.06652 [astro-ph.CO]}
  \BibitemShut {NoStop}%
\bibitem [{\citenamefont {de~Rham}\ and\ \citenamefont
  {Melville}(2018)}]{deRham:2018red}%
  \BibitemOpen
  \bibfield  {author} {\bibinfo {author} {\bibfnamefont {C.}~\bibnamefont
  {de~Rham}}\ and\ \bibinfo {author} {\bibfnamefont {S.}~\bibnamefont
  {Melville}},\ }\href {https://doi.org/10.1103/PhysRevLett.121.221101}
  {\bibfield  {journal} {\bibinfo  {journal} {Phys. Rev. Lett.}\ }\textbf
  {\bibinfo {volume} {121}},\ \bibinfo {pages} {221101} (\bibinfo {year}
  {2018})},\ \Eprint {https://arxiv.org/abs/1806.09417} {arXiv:1806.09417
  [hep-th]} \BibitemShut {NoStop}%
\bibitem [{\citenamefont {Tomita}(1967)}]{Tomita}%
  \BibitemOpen
  \bibfield  {author} {\bibinfo {author} {\bibfnamefont {K.}~\bibnamefont
  {Tomita}},\ }\href {https://doi.org/10.1143/PTP.37.831} {\bibfield  {journal}
  {\bibinfo  {journal} {Progress of Theoretical Physics}\ }\textbf {\bibinfo
  {volume} {37}},\ \bibinfo {pages} {831} (\bibinfo {year} {1967})},\ \Eprint
  {https://arxiv.org/abs/https://academic.oup.com/ptp/article-pdf/37/5/831/5234391/37-5-831.pdf}
  {https://academic.oup.com/ptp/article-pdf/37/5/831/5234391/37-5-831.pdf}
  \BibitemShut {NoStop}%
\bibitem [{\citenamefont {Matarrese}\ \emph {et~al.}(1993)\citenamefont
  {Matarrese}, \citenamefont {Pantano},\ and\ \citenamefont
  {Saez}}]{Matarrese:1992rp}%
  \BibitemOpen
  \bibfield  {author} {\bibinfo {author} {\bibfnamefont {S.}~\bibnamefont
  {Matarrese}}, \bibinfo {author} {\bibfnamefont {O.}~\bibnamefont {Pantano}},\
  and\ \bibinfo {author} {\bibfnamefont {D.}~\bibnamefont {Saez}},\ }\href
  {https://doi.org/10.1103/PhysRevD.47.1311} {\bibfield  {journal} {\bibinfo
  {journal} {Phys. Rev. D}\ }\textbf {\bibinfo {volume} {47}},\ \bibinfo
  {pages} {1311} (\bibinfo {year} {1993})}\BibitemShut {NoStop}%
\bibitem [{\citenamefont {Matarrese}\ \emph {et~al.}(1994)\citenamefont
  {Matarrese}, \citenamefont {Pantano},\ and\ \citenamefont
  {Saez}}]{Matarrese:1993zf}%
  \BibitemOpen
  \bibfield  {author} {\bibinfo {author} {\bibfnamefont {S.}~\bibnamefont
  {Matarrese}}, \bibinfo {author} {\bibfnamefont {O.}~\bibnamefont {Pantano}},\
  and\ \bibinfo {author} {\bibfnamefont {D.}~\bibnamefont {Saez}},\ }\href
  {https://doi.org/10.1103/PhysRevLett.72.320} {\bibfield  {journal} {\bibinfo
  {journal} {Phys. Rev. Lett.}\ }\textbf {\bibinfo {volume} {72}},\ \bibinfo
  {pages} {320} (\bibinfo {year} {1994})},\ \Eprint
  {https://arxiv.org/abs/astro-ph/9310036} {arXiv:astro-ph/9310036}
  \BibitemShut {NoStop}%
\bibitem [{\citenamefont {Matarrese}\ \emph {et~al.}(1998)\citenamefont
  {Matarrese}, \citenamefont {Mollerach},\ and\ \citenamefont
  {Bruni}}]{Matarrese:1997ay}%
  \BibitemOpen
  \bibfield  {author} {\bibinfo {author} {\bibfnamefont {S.}~\bibnamefont
  {Matarrese}}, \bibinfo {author} {\bibfnamefont {S.}~\bibnamefont
  {Mollerach}},\ and\ \bibinfo {author} {\bibfnamefont {M.}~\bibnamefont
  {Bruni}},\ }\href {https://doi.org/10.1103/PhysRevD.58.043504} {\bibfield
  {journal} {\bibinfo  {journal} {Phys. Rev. D}\ }\textbf {\bibinfo {volume}
  {58}},\ \bibinfo {pages} {043504} (\bibinfo {year} {1998})},\ \Eprint
  {https://arxiv.org/abs/astro-ph/9707278} {arXiv:astro-ph/9707278}
  \BibitemShut {NoStop}%
\bibitem [{\citenamefont {Ananda}\ \emph {et~al.}(2007)\citenamefont {Ananda},
  \citenamefont {Clarkson},\ and\ \citenamefont {Wands}}]{Ananda:2006af}%
  \BibitemOpen
  \bibfield  {author} {\bibinfo {author} {\bibfnamefont {K.~N.}\ \bibnamefont
  {Ananda}}, \bibinfo {author} {\bibfnamefont {C.}~\bibnamefont {Clarkson}},\
  and\ \bibinfo {author} {\bibfnamefont {D.}~\bibnamefont {Wands}},\ }\href
  {https://doi.org/10.1103/PhysRevD.75.123518} {\bibfield  {journal} {\bibinfo
  {journal} {Phys. Rev. D}\ }\textbf {\bibinfo {volume} {75}},\ \bibinfo
  {pages} {123518} (\bibinfo {year} {2007})},\ \Eprint
  {https://arxiv.org/abs/gr-qc/0612013} {arXiv:gr-qc/0612013} \BibitemShut
  {NoStop}%
\bibitem [{\citenamefont {Baumann}\ \emph {et~al.}(2007)\citenamefont
  {Baumann}, \citenamefont {Steinhardt}, \citenamefont {Takahashi},\ and\
  \citenamefont {Ichiki}}]{Baumann:2007zm}%
  \BibitemOpen
  \bibfield  {author} {\bibinfo {author} {\bibfnamefont {D.}~\bibnamefont
  {Baumann}}, \bibinfo {author} {\bibfnamefont {P.~J.}\ \bibnamefont
  {Steinhardt}}, \bibinfo {author} {\bibfnamefont {K.}~\bibnamefont
  {Takahashi}},\ and\ \bibinfo {author} {\bibfnamefont {K.}~\bibnamefont
  {Ichiki}},\ }\href {https://doi.org/10.1103/PhysRevD.76.084019} {\bibfield
  {journal} {\bibinfo  {journal} {Phys. Rev. D}\ }\textbf {\bibinfo {volume}
  {76}},\ \bibinfo {pages} {084019} (\bibinfo {year} {2007})},\ \Eprint
  {https://arxiv.org/abs/hep-th/0703290} {arXiv:hep-th/0703290} \BibitemShut
  {NoStop}%
\bibitem [{\citenamefont {Assadullahi}\ and\ \citenamefont
  {Wands}(2009)}]{Assadullahi:2009nf}%
  \BibitemOpen
  \bibfield  {author} {\bibinfo {author} {\bibfnamefont {H.}~\bibnamefont
  {Assadullahi}}\ and\ \bibinfo {author} {\bibfnamefont {D.}~\bibnamefont
  {Wands}},\ }\href {https://doi.org/10.1103/PhysRevD.79.083511} {\bibfield
  {journal} {\bibinfo  {journal} {Phys. Rev. D}\ }\textbf {\bibinfo {volume}
  {79}},\ \bibinfo {pages} {083511} (\bibinfo {year} {2009})},\ \Eprint
  {https://arxiv.org/abs/0901.0989} {arXiv:0901.0989 [astro-ph.CO]}
  \BibitemShut {NoStop}%
\bibitem [{\citenamefont {Alabidi}\ \emph {et~al.}(2013)\citenamefont
  {Alabidi}, \citenamefont {Kohri}, \citenamefont {Sasaki},\ and\ \citenamefont
  {Sendouda}}]{Alabidi:2013lya}%
  \BibitemOpen
  \bibfield  {author} {\bibinfo {author} {\bibfnamefont {L.}~\bibnamefont
  {Alabidi}}, \bibinfo {author} {\bibfnamefont {K.}~\bibnamefont {Kohri}},
  \bibinfo {author} {\bibfnamefont {M.}~\bibnamefont {Sasaki}},\ and\ \bibinfo
  {author} {\bibfnamefont {Y.}~\bibnamefont {Sendouda}},\ }\href
  {https://doi.org/10.1088/1475-7516/2013/05/033} {\bibfield  {journal}
  {\bibinfo  {journal} {JCAP}\ }\textbf {\bibinfo {volume} {05}},\ \bibinfo
  {pages} {033}},\ \Eprint {https://arxiv.org/abs/1303.4519} {arXiv:1303.4519
  [astro-ph.CO]} \BibitemShut {NoStop}%
\bibitem [{\citenamefont {Yuan}\ and\ \citenamefont
  {Huang}(2021)}]{Yuan:2021qgz}%
  \BibitemOpen
  \bibfield  {author} {\bibinfo {author} {\bibfnamefont {C.}~\bibnamefont
  {Yuan}}\ and\ \bibinfo {author} {\bibfnamefont {Q.-G.}\ \bibnamefont
  {Huang}},\ }\href {https://doi.org/10.1016/j.isci.2021.102860} {\bibfield
  {journal} {\bibinfo  {journal} {iScience}\ }\textbf {\bibinfo {volume}
  {24}},\ \bibinfo {pages} {102860} (\bibinfo {year} {2021})},\ \Eprint
  {https://arxiv.org/abs/2103.04739} {arXiv:2103.04739 [astro-ph.GA]}
  \BibitemShut {NoStop}%
\bibitem [{\citenamefont {Dom\`enech}(2021)}]{Domenech:2021ztg}%
  \BibitemOpen
  \bibfield  {author} {\bibinfo {author} {\bibfnamefont {G.}~\bibnamefont
  {Dom\`enech}},\ }\href {https://doi.org/10.3390/universe7110398} {\bibfield
  {journal} {\bibinfo  {journal} {Universe}\ }\textbf {\bibinfo {volume} {7}},\
  \bibinfo {pages} {398} (\bibinfo {year} {2021})},\ \Eprint
  {https://arxiv.org/abs/2109.01398} {arXiv:2109.01398 [gr-qc]} \BibitemShut
  {NoStop}%
\bibitem [{\citenamefont {Dom\`enech}(2024)}]{Domenech:2024kmh}%
  \BibitemOpen
  \bibfield  {author} {\bibinfo {author} {\bibfnamefont {G.}~\bibnamefont
  {Dom\`enech}},\ }\href@noop {} {\  (\bibinfo {year} {2024})},\ \Eprint
  {https://arxiv.org/abs/2402.17388} {arXiv:2402.17388 [gr-qc]} \BibitemShut
  {NoStop}%
\bibitem [{\citenamefont {Roshan}\ and\ \citenamefont
  {White}(2025)}]{Roshan:2024qnv}%
  \BibitemOpen
  \bibfield  {author} {\bibinfo {author} {\bibfnamefont {R.}~\bibnamefont
  {Roshan}}\ and\ \bibinfo {author} {\bibfnamefont {G.}~\bibnamefont {White}},\
  }\href {https://doi.org/10.1103/RevModPhys.97.015001} {\bibfield  {journal}
  {\bibinfo  {journal} {Rev. Mod. Phys.}\ }\textbf {\bibinfo {volume} {97}},\
  \bibinfo {pages} {015001} (\bibinfo {year} {2025})},\ \Eprint
  {https://arxiv.org/abs/2401.04388} {arXiv:2401.04388 [hep-ph]} \BibitemShut
  {NoStop}%
\bibitem [{\citenamefont {Dom\`enech}\ and\ \citenamefont
  {Ganz}(2025)}]{Domenech:2024drm}%
  \BibitemOpen
  \bibfield  {author} {\bibinfo {author} {\bibfnamefont {G.}~\bibnamefont
  {Dom\`enech}}\ and\ \bibinfo {author} {\bibfnamefont {A.}~\bibnamefont
  {Ganz}},\ }\href {https://doi.org/10.1088/1475-7516/2025/01/020} {\bibfield
  {journal} {\bibinfo  {journal} {JCAP}\ }\textbf {\bibinfo {volume} {01}},\
  \bibinfo {pages} {020}},\ \Eprint {https://arxiv.org/abs/2406.19950}
  {arXiv:2406.19950 [gr-qc]} \BibitemShut {NoStop}%
\bibitem [{\citenamefont {Cai}\ \emph {et~al.}(2020)\citenamefont {Cai},
  \citenamefont {Pi},\ and\ \citenamefont {Sasaki}}]{Cai:2019cdl}%
  \BibitemOpen
  \bibfield  {author} {\bibinfo {author} {\bibfnamefont {R.-G.}\ \bibnamefont
  {Cai}}, \bibinfo {author} {\bibfnamefont {S.}~\bibnamefont {Pi}},\ and\
  \bibinfo {author} {\bibfnamefont {M.}~\bibnamefont {Sasaki}},\ }\href
  {https://doi.org/10.1103/PhysRevD.102.083528} {\bibfield  {journal} {\bibinfo
   {journal} {Phys. Rev. D}\ }\textbf {\bibinfo {volume} {102}},\ \bibinfo
  {pages} {083528} (\bibinfo {year} {2020})},\ \Eprint
  {https://arxiv.org/abs/1909.13728} {arXiv:1909.13728 [astro-ph.CO]}
  \BibitemShut {NoStop}%
\bibitem [{\citenamefont {Kugarajh}\ \emph {et~al.}(2025)\citenamefont
  {Kugarajh}, \citenamefont {Traforetti}, \citenamefont {Maselli},
  \citenamefont {Matarrese},\ and\ \citenamefont
  {Ricciardone}}]{Kugarajh:2025rbt}%
  \BibitemOpen
  \bibfield  {author} {\bibinfo {author} {\bibfnamefont {A.~A.}\ \bibnamefont
  {Kugarajh}}, \bibinfo {author} {\bibfnamefont {M.}~\bibnamefont
  {Traforetti}}, \bibinfo {author} {\bibfnamefont {A.}~\bibnamefont {Maselli}},
  \bibinfo {author} {\bibfnamefont {S.}~\bibnamefont {Matarrese}},\ and\
  \bibinfo {author} {\bibfnamefont {A.}~\bibnamefont {Ricciardone}},\
  }\href@noop {} {\  (\bibinfo {year} {2025})},\ \Eprint
  {https://arxiv.org/abs/2502.20137} {arXiv:2502.20137 [gr-qc]} \BibitemShut
  {NoStop}%
\bibitem [{\citenamefont {Maeda}(1989)}]{Maeda:1988ab}%
  \BibitemOpen
  \bibfield  {author} {\bibinfo {author} {\bibfnamefont {K.-i.}\ \bibnamefont
  {Maeda}},\ }\href {https://doi.org/10.1103/PhysRevD.39.3159} {\bibfield
  {journal} {\bibinfo  {journal} {Phys. Rev. D}\ }\textbf {\bibinfo {volume}
  {39}},\ \bibinfo {pages} {3159} (\bibinfo {year} {1989})}\BibitemShut
  {NoStop}%
\bibitem [{\citenamefont {L{\'o}pez}\ and\ \citenamefont
  {Terente~D{\'\i}az}(2025)}]{Lopez:2025gfu}%
  \BibitemOpen
  \bibfield  {author} {\bibinfo {author} {\bibfnamefont {S.~S.}\ \bibnamefont
  {L{\'o}pez}}\ and\ \bibinfo {author} {\bibfnamefont {J.~J.}\ \bibnamefont
  {Terente~D{\'\i}az}},\ }\href@noop {} {\  (\bibinfo {year} {2025})},\ \Eprint
  {https://arxiv.org/abs/2505.13420} {arXiv:2505.13420 [astro-ph.CO]}
  \BibitemShut {NoStop}%
\bibitem [{\citenamefont {Hu}\ and\ \citenamefont {Gao}(2024)}]{Hu:2024hzo}%
  \BibitemOpen
  \bibfield  {author} {\bibinfo {author} {\bibfnamefont {Y.-M.}\ \bibnamefont
  {Hu}}\ and\ \bibinfo {author} {\bibfnamefont {X.}~\bibnamefont {Gao}},\
  }\href {https://doi.org/10.1103/PhysRevD.110.064038} {\bibfield  {journal}
  {\bibinfo  {journal} {Phys. Rev. D}\ }\textbf {\bibinfo {volume} {110}},\
  \bibinfo {pages} {064038} (\bibinfo {year} {2024})},\ \Eprint
  {https://arxiv.org/abs/2405.20158} {arXiv:2405.20158 [hep-th]} \BibitemShut
  {NoStop}%
\bibitem [{\citenamefont {Feng}\ \emph {et~al.}(2024)\citenamefont {Feng},
  \citenamefont {Zhang},\ and\ \citenamefont {Gao}}]{Feng:2024yic}%
  \BibitemOpen
  \bibfield  {author} {\bibinfo {author} {\bibfnamefont {J.-X.}\ \bibnamefont
  {Feng}}, \bibinfo {author} {\bibfnamefont {F.}~\bibnamefont {Zhang}},\ and\
  \bibinfo {author} {\bibfnamefont {X.}~\bibnamefont {Gao}},\ }\href
  {https://doi.org/10.1140/epjc/s10052-024-13097-7} {\bibfield  {journal}
  {\bibinfo  {journal} {Eur. Phys. J. C}\ }\textbf {\bibinfo {volume} {84}},\
  \bibinfo {pages} {736} (\bibinfo {year} {2024})},\ \Eprint
  {https://arxiv.org/abs/2404.05289} {arXiv:2404.05289 [gr-qc]} \BibitemShut
  {NoStop}%
\bibitem [{\citenamefont {Zhang}\ \emph {et~al.}(2024)\citenamefont {Zhang},
  \citenamefont {Feng},\ and\ \citenamefont {Gao}}]{Zhang:2024vfw}%
  \BibitemOpen
  \bibfield  {author} {\bibinfo {author} {\bibfnamefont {F.}~\bibnamefont
  {Zhang}}, \bibinfo {author} {\bibfnamefont {J.-X.}\ \bibnamefont {Feng}},\
  and\ \bibinfo {author} {\bibfnamefont {X.}~\bibnamefont {Gao}},\ }\href
  {https://doi.org/10.1103/PhysRevD.110.023537} {\bibfield  {journal} {\bibinfo
   {journal} {Phys. Rev. D}\ }\textbf {\bibinfo {volume} {110}},\ \bibinfo
  {pages} {023537} (\bibinfo {year} {2024})},\ \Eprint
  {https://arxiv.org/abs/2404.02922} {arXiv:2404.02922 [gr-qc]} \BibitemShut
  {NoStop}%
\bibitem [{\citenamefont {Creminelli}\ \emph {et~al.}(2018)\citenamefont
  {Creminelli}, \citenamefont {Lewandowski}, \citenamefont {Tambalo},\ and\
  \citenamefont {Vernizzi}}]{Creminelli:2018xsv}%
  \BibitemOpen
  \bibfield  {author} {\bibinfo {author} {\bibfnamefont {P.}~\bibnamefont
  {Creminelli}}, \bibinfo {author} {\bibfnamefont {M.}~\bibnamefont
  {Lewandowski}}, \bibinfo {author} {\bibfnamefont {G.}~\bibnamefont
  {Tambalo}},\ and\ \bibinfo {author} {\bibfnamefont {F.}~\bibnamefont
  {Vernizzi}},\ }\href {https://doi.org/10.1088/1475-7516/2018/12/025}
  {\bibfield  {journal} {\bibinfo  {journal} {JCAP}\ }\textbf {\bibinfo
  {volume} {12}},\ \bibinfo {pages} {025}},\ \Eprint
  {https://arxiv.org/abs/1809.03484} {arXiv:1809.03484 [astro-ph.CO]}
  \BibitemShut {NoStop}%
\bibitem [{\citenamefont {De~Felice}\ \emph {et~al.}(2021)\citenamefont
  {De~Felice}, \citenamefont {Mukohyama},\ and\ \citenamefont
  {Takahashi}}]{DeFelice:2021hps}%
  \BibitemOpen
  \bibfield  {author} {\bibinfo {author} {\bibfnamefont {A.}~\bibnamefont
  {De~Felice}}, \bibinfo {author} {\bibfnamefont {S.}~\bibnamefont
  {Mukohyama}},\ and\ \bibinfo {author} {\bibfnamefont {K.}~\bibnamefont
  {Takahashi}},\ }\href {https://doi.org/10.1088/1475-7516/2021/12/020}
  {\bibfield  {journal} {\bibinfo  {journal} {JCAP}\ }\textbf {\bibinfo
  {volume} {12}}\bibfield  {number} {\bibinfo  {number} { (12)},\ \bibinfo
  {pages} {020}},\ }\Eprint {https://arxiv.org/abs/2110.03194}
  {arXiv:2110.03194 [gr-qc]} \BibitemShut {NoStop}%
\bibitem [{\citenamefont {De~Felice}\ \emph {et~al.}(2022)\citenamefont
  {De~Felice}, \citenamefont {Mukohyama},\ and\ \citenamefont
  {Takahashi}}]{DeFelice:2022xvq}%
  \BibitemOpen
  \bibfield  {author} {\bibinfo {author} {\bibfnamefont {A.}~\bibnamefont
  {De~Felice}}, \bibinfo {author} {\bibfnamefont {S.}~\bibnamefont
  {Mukohyama}},\ and\ \bibinfo {author} {\bibfnamefont {K.}~\bibnamefont
  {Takahashi}},\ }\href {https://doi.org/10.1103/PhysRevLett.129.031103}
  {\bibfield  {journal} {\bibinfo  {journal} {Phys. Rev. Lett.}\ }\textbf
  {\bibinfo {volume} {129}},\ \bibinfo {pages} {031103} (\bibinfo {year}
  {2022})},\ \Eprint {https://arxiv.org/abs/2204.02032} {arXiv:2204.02032
  [gr-qc]} \BibitemShut {NoStop}%
\bibitem [{\citenamefont {Motohashi}\ and\ \citenamefont
  {Mukohyama}(2020)}]{Motohashi:2019ymr}%
  \BibitemOpen
  \bibfield  {author} {\bibinfo {author} {\bibfnamefont {H.}~\bibnamefont
  {Motohashi}}\ and\ \bibinfo {author} {\bibfnamefont {S.}~\bibnamefont
  {Mukohyama}},\ }\href {https://doi.org/10.1088/1475-7516/2020/01/030}
  {\bibfield  {journal} {\bibinfo  {journal} {JCAP}\ }\textbf {\bibinfo
  {volume} {01}},\ \bibinfo {pages} {030}},\ \Eprint
  {https://arxiv.org/abs/1912.00378} {arXiv:1912.00378 [gr-qc]} \BibitemShut
  {NoStop}%
\bibitem [{\citenamefont {Gorji}\ \emph {et~al.}(2021)\citenamefont {Gorji},
  \citenamefont {Motohashi},\ and\ \citenamefont {Mukohyama}}]{Gorji:2020bfl}%
  \BibitemOpen
  \bibfield  {author} {\bibinfo {author} {\bibfnamefont {M.~A.}\ \bibnamefont
  {Gorji}}, \bibinfo {author} {\bibfnamefont {H.}~\bibnamefont {Motohashi}},\
  and\ \bibinfo {author} {\bibfnamefont {S.}~\bibnamefont {Mukohyama}},\ }\href
  {https://doi.org/10.1088/1475-7516/2021/03/081} {\bibfield  {journal}
  {\bibinfo  {journal} {JCAP}\ }\textbf {\bibinfo {volume} {03}},\ \bibinfo
  {pages} {081}},\ \Eprint {https://arxiv.org/abs/2009.11606} {arXiv:2009.11606
  [gr-qc]} \BibitemShut {NoStop}%
\bibitem [{\citenamefont {Gorji}\ \emph {et~al.}(2022)\citenamefont {Gorji},
  \citenamefont {Motohashi},\ and\ \citenamefont {Mukohyama}}]{Gorji:2021isn}%
  \BibitemOpen
  \bibfield  {author} {\bibinfo {author} {\bibfnamefont {M.~A.}\ \bibnamefont
  {Gorji}}, \bibinfo {author} {\bibfnamefont {H.}~\bibnamefont {Motohashi}},\
  and\ \bibinfo {author} {\bibfnamefont {S.}~\bibnamefont {Mukohyama}},\ }\href
  {https://doi.org/10.1088/1475-7516/2022/02/030} {\bibfield  {journal}
  {\bibinfo  {journal} {JCAP}\ }\textbf {\bibinfo {volume} {02}}\bibfield
  {number} {\bibinfo  {number} { (02)},\ \bibinfo {pages} {030}},\ }\Eprint
  {https://arxiv.org/abs/2110.10731} {arXiv:2110.10731 [hep-th]} \BibitemShut
  {NoStop}%
\bibitem [{\citenamefont {Ben~Achour}(2025)}]{BenAchour:2024hbg}%
  \BibitemOpen
  \bibfield  {author} {\bibinfo {author} {\bibfnamefont {J.}~\bibnamefont
  {Ben~Achour}},\ }\href {https://doi.org/10.1140/epjc/s10052-025-13983-8}
  {\bibfield  {journal} {\bibinfo  {journal} {Eur. Phys. J. C}\ }\textbf
  {\bibinfo {volume} {85}},\ \bibinfo {pages} {424} (\bibinfo {year} {2025})},\
  \Eprint {https://arxiv.org/abs/2412.04135} {arXiv:2412.04135 [gr-qc]}
  \BibitemShut {NoStop}%
\bibitem [{\citenamefont {Ben~Achour}\ \emph {et~al.}(2025)\citenamefont
  {Ben~Achour}, \citenamefont {Gorji},\ and\ \citenamefont
  {Roussille}}]{BenAchour:2024tqt}%
  \BibitemOpen
  \bibfield  {author} {\bibinfo {author} {\bibfnamefont {J.}~\bibnamefont
  {Ben~Achour}}, \bibinfo {author} {\bibfnamefont {M.~A.}\ \bibnamefont
  {Gorji}},\ and\ \bibinfo {author} {\bibfnamefont {H.}~\bibnamefont
  {Roussille}},\ }\href {https://doi.org/10.1088/1475-7516/2025/02/015}
  {\bibfield  {journal} {\bibinfo  {journal} {JCAP}\ }\textbf {\bibinfo
  {volume} {02}},\ \bibinfo {pages} {015}},\ \Eprint
  {https://arxiv.org/abs/2402.01487} {arXiv:2402.01487 [gr-qc]} \BibitemShut
  {NoStop}%
\bibitem [{\citenamefont {Armendariz-Picon}\ \emph {et~al.}(1999)\citenamefont
  {Armendariz-Picon}, \citenamefont {Damour},\ and\ \citenamefont
  {Mukhanov}}]{Armendariz-Picon:1999hyi}%
  \BibitemOpen
  \bibfield  {author} {\bibinfo {author} {\bibfnamefont {C.}~\bibnamefont
  {Armendariz-Picon}}, \bibinfo {author} {\bibfnamefont {T.}~\bibnamefont
  {Damour}},\ and\ \bibinfo {author} {\bibfnamefont {V.~F.}\ \bibnamefont
  {Mukhanov}},\ }\href {https://doi.org/10.1016/S0370-2693(99)00603-6}
  {\bibfield  {journal} {\bibinfo  {journal} {Phys. Lett. B}\ }\textbf
  {\bibinfo {volume} {458}},\ \bibinfo {pages} {209} (\bibinfo {year}
  {1999})},\ \Eprint {https://arxiv.org/abs/hep-th/9904075}
  {arXiv:hep-th/9904075} \BibitemShut {NoStop}%
\bibitem [{\citenamefont {Hwang}\ \emph {et~al.}(2017)\citenamefont {Hwang},
  \citenamefont {Jeong},\ and\ \citenamefont {Noh}}]{Hwang:2017oxa}%
  \BibitemOpen
  \bibfield  {author} {\bibinfo {author} {\bibfnamefont {J.-C.}\ \bibnamefont
  {Hwang}}, \bibinfo {author} {\bibfnamefont {D.}~\bibnamefont {Jeong}},\ and\
  \bibinfo {author} {\bibfnamefont {H.}~\bibnamefont {Noh}},\ }\href
  {https://doi.org/10.3847/1538-4357/aa74be} {\bibfield  {journal} {\bibinfo
  {journal} {Astrophys. J.}\ }\textbf {\bibinfo {volume} {842}},\ \bibinfo
  {pages} {46} (\bibinfo {year} {2017})},\ \Eprint
  {https://arxiv.org/abs/1704.03500} {arXiv:1704.03500 [astro-ph.CO]}
  \BibitemShut {NoStop}%
\bibitem [{\citenamefont {Gong}(2022)}]{Gong:2019mui}%
  \BibitemOpen
  \bibfield  {author} {\bibinfo {author} {\bibfnamefont {J.-O.}\ \bibnamefont
  {Gong}},\ }\href {https://doi.org/10.3847/1538-4357/ac3a6c} {\bibfield
  {journal} {\bibinfo  {journal} {Astrophys. J.}\ }\textbf {\bibinfo {volume}
  {925}},\ \bibinfo {pages} {102} (\bibinfo {year} {2022})},\ \Eprint
  {https://arxiv.org/abs/1909.12708} {arXiv:1909.12708 [gr-qc]} \BibitemShut
  {NoStop}%
\bibitem [{\citenamefont {Tomikawa}\ and\ \citenamefont
  {Kobayashi}(2020)}]{Tomikawa:2019tvi}%
  \BibitemOpen
  \bibfield  {author} {\bibinfo {author} {\bibfnamefont {K.}~\bibnamefont
  {Tomikawa}}\ and\ \bibinfo {author} {\bibfnamefont {T.}~\bibnamefont
  {Kobayashi}},\ }\href {https://doi.org/10.1103/PhysRevD.101.083529}
  {\bibfield  {journal} {\bibinfo  {journal} {Phys. Rev. D}\ }\textbf {\bibinfo
  {volume} {101}},\ \bibinfo {pages} {083529} (\bibinfo {year} {2020})},\
  \Eprint {https://arxiv.org/abs/1910.01880} {arXiv:1910.01880 [gr-qc]}
  \BibitemShut {NoStop}%
\bibitem [{\citenamefont {Dom\`enech}\ and\ \citenamefont
  {Sasaki}(2021)}]{Domenech:2020xin}%
  \BibitemOpen
  \bibfield  {author} {\bibinfo {author} {\bibfnamefont {G.}~\bibnamefont
  {Dom\`enech}}\ and\ \bibinfo {author} {\bibfnamefont {M.}~\bibnamefont
  {Sasaki}},\ }\href {https://doi.org/10.1103/PhysRevD.103.063531} {\bibfield
  {journal} {\bibinfo  {journal} {Phys. Rev. D}\ }\textbf {\bibinfo {volume}
  {103}},\ \bibinfo {pages} {063531} (\bibinfo {year} {2021})},\ \Eprint
  {https://arxiv.org/abs/2012.14016} {arXiv:2012.14016 [gr-qc]} \BibitemShut
  {NoStop}%
\bibitem [{\citenamefont {De~Luca}\ \emph {et~al.}(2020)\citenamefont
  {De~Luca}, \citenamefont {Franciolini}, \citenamefont {Kehagias},\ and\
  \citenamefont {Riotto}}]{DeLuca:2019ufz}%
  \BibitemOpen
  \bibfield  {author} {\bibinfo {author} {\bibfnamefont {V.}~\bibnamefont
  {De~Luca}}, \bibinfo {author} {\bibfnamefont {G.}~\bibnamefont
  {Franciolini}}, \bibinfo {author} {\bibfnamefont {A.}~\bibnamefont
  {Kehagias}},\ and\ \bibinfo {author} {\bibfnamefont {A.}~\bibnamefont
  {Riotto}},\ }\href {https://doi.org/10.1088/1475-7516/2020/03/014} {\bibfield
   {journal} {\bibinfo  {journal} {JCAP}\ }\textbf {\bibinfo {volume} {03}},\
  \bibinfo {pages} {014}},\ \Eprint {https://arxiv.org/abs/1911.09689}
  {arXiv:1911.09689 [gr-qc]} \BibitemShut {NoStop}%
\bibitem [{\citenamefont {Inomata}\ and\ \citenamefont
  {Terada}(2020)}]{Inomata:2019yww}%
  \BibitemOpen
  \bibfield  {author} {\bibinfo {author} {\bibfnamefont {K.}~\bibnamefont
  {Inomata}}\ and\ \bibinfo {author} {\bibfnamefont {T.}~\bibnamefont
  {Terada}},\ }\href {https://doi.org/10.1103/PhysRevD.101.023523} {\bibfield
  {journal} {\bibinfo  {journal} {Phys. Rev. D}\ }\textbf {\bibinfo {volume}
  {101}},\ \bibinfo {pages} {023523} (\bibinfo {year} {2020})},\ \Eprint
  {https://arxiv.org/abs/1912.00785} {arXiv:1912.00785 [gr-qc]} \BibitemShut
  {NoStop}%
\bibitem [{\citenamefont {Yuan}\ \emph
  {et~al.}(2020{\natexlab{a}})\citenamefont {Yuan}, \citenamefont {Chen},\ and\
  \citenamefont {Huang}}]{Yuan:2019fwv}%
  \BibitemOpen
  \bibfield  {author} {\bibinfo {author} {\bibfnamefont {C.}~\bibnamefont
  {Yuan}}, \bibinfo {author} {\bibfnamefont {Z.-C.}\ \bibnamefont {Chen}},\
  and\ \bibinfo {author} {\bibfnamefont {Q.-G.}\ \bibnamefont {Huang}},\ }\href
  {https://doi.org/10.1103/PhysRevD.101.063018} {\bibfield  {journal} {\bibinfo
   {journal} {Phys. Rev. D}\ }\textbf {\bibinfo {volume} {101}},\ \bibinfo
  {pages} {6} (\bibinfo {year} {2020}{\natexlab{a}})},\ \Eprint
  {https://arxiv.org/abs/1912.00885} {arXiv:1912.00885 [astro-ph.CO]}
  \BibitemShut {NoStop}%
\bibitem [{\citenamefont {Lu}\ \emph {et~al.}(2020)\citenamefont {Lu},
  \citenamefont {Ali}, \citenamefont {Gong}, \citenamefont {Lin},\ and\
  \citenamefont {Zhang}}]{Lu:2020diy}%
  \BibitemOpen
  \bibfield  {author} {\bibinfo {author} {\bibfnamefont {Y.}~\bibnamefont
  {Lu}}, \bibinfo {author} {\bibfnamefont {A.}~\bibnamefont {Ali}}, \bibinfo
  {author} {\bibfnamefont {Y.}~\bibnamefont {Gong}}, \bibinfo {author}
  {\bibfnamefont {J.}~\bibnamefont {Lin}},\ and\ \bibinfo {author}
  {\bibfnamefont {F.}~\bibnamefont {Zhang}},\ }\href
  {https://doi.org/10.1103/PhysRevD.102.083503(2020)} {\bibfield  {journal}
  {\bibinfo  {journal} {Phys. Rev. D}\ }\textbf {\bibinfo {volume} {102}},\
  \bibinfo {pages} {083503} (\bibinfo {year} {2020})},\ \Eprint
  {https://arxiv.org/abs/2006.03450} {arXiv:2006.03450 [gr-qc]} \BibitemShut
  {NoStop}%
\bibitem [{\citenamefont {Yuan}\ \emph {et~al.}(2025)\citenamefont {Yuan},
  \citenamefont {Chen},\ and\ \citenamefont {Liu}}]{Yuan:2024qfz}%
  \BibitemOpen
  \bibfield  {author} {\bibinfo {author} {\bibfnamefont {C.}~\bibnamefont
  {Yuan}}, \bibinfo {author} {\bibfnamefont {Z.-C.}\ \bibnamefont {Chen}},\
  and\ \bibinfo {author} {\bibfnamefont {L.}~\bibnamefont {Liu}},\ }\href
  {https://doi.org/10.1103/PhysRevD.111.103528} {\bibfield  {journal} {\bibinfo
   {journal} {Phys. Rev. D}\ }\textbf {\bibinfo {volume} {111}},\ \bibinfo
  {pages} {103528} (\bibinfo {year} {2025})},\ \Eprint
  {https://arxiv.org/abs/2410.18996} {arXiv:2410.18996 [gr-qc]} \BibitemShut
  {NoStop}%
\bibitem [{\citenamefont {Dom\`enech}\ and\ \citenamefont
  {Sasaki}(2018)}]{Domenech:2017ems}%
  \BibitemOpen
  \bibfield  {author} {\bibinfo {author} {\bibfnamefont {G.}~\bibnamefont
  {Dom\`enech}}\ and\ \bibinfo {author} {\bibfnamefont {M.}~\bibnamefont
  {Sasaki}},\ }\href {https://doi.org/10.1103/PhysRevD.97.023521} {\bibfield
  {journal} {\bibinfo  {journal} {Phys. Rev. D}\ }\textbf {\bibinfo {volume}
  {97}},\ \bibinfo {pages} {023521} (\bibinfo {year} {2018})},\ \Eprint
  {https://arxiv.org/abs/1709.09804} {arXiv:1709.09804 [gr-qc]} \BibitemShut
  {NoStop}%
\bibitem [{\citenamefont {Kohri}\ and\ \citenamefont
  {Terada}(2018)}]{Kohri:2018awv}%
  \BibitemOpen
  \bibfield  {author} {\bibinfo {author} {\bibfnamefont {K.}~\bibnamefont
  {Kohri}}\ and\ \bibinfo {author} {\bibfnamefont {T.}~\bibnamefont {Terada}},\
  }\href {https://doi.org/10.1103/PhysRevD.97.123532} {\bibfield  {journal}
  {\bibinfo  {journal} {Phys. Rev. D}\ }\textbf {\bibinfo {volume} {97}},\
  \bibinfo {pages} {123532} (\bibinfo {year} {2018})},\ \Eprint
  {https://arxiv.org/abs/1804.08577} {arXiv:1804.08577 [gr-qc]} \BibitemShut
  {NoStop}%
\bibitem [{\citenamefont {Dom{\`e}nech}\ \emph {et~al.}(2020)\citenamefont
  {Dom{\`e}nech}, \citenamefont {Pi},\ and\ \citenamefont
  {Sasaki}}]{Domenech:2020kqm}%
  \BibitemOpen
  \bibfield  {author} {\bibinfo {author} {\bibfnamefont {G.}~\bibnamefont
  {Dom{\`e}nech}}, \bibinfo {author} {\bibfnamefont {S.}~\bibnamefont {Pi}},\
  and\ \bibinfo {author} {\bibfnamefont {M.}~\bibnamefont {Sasaki}},\ }\href
  {https://doi.org/10.1088/1475-7516/2020/08/017} {\bibfield  {journal}
  {\bibinfo  {journal} {JCAP}\ }\textbf {\bibinfo {volume} {08}},\ \bibinfo
  {pages} {017}},\ \Eprint {https://arxiv.org/abs/2005.12314} {arXiv:2005.12314
  [gr-qc]} \BibitemShut {NoStop}%
\bibitem [{\citenamefont {Cai}\ \emph {et~al.}(2019)\citenamefont {Cai},
  \citenamefont {Pi},\ and\ \citenamefont {Sasaki}}]{Cai:2018dig}%
  \BibitemOpen
  \bibfield  {author} {\bibinfo {author} {\bibfnamefont {R.-g.}\ \bibnamefont
  {Cai}}, \bibinfo {author} {\bibfnamefont {S.}~\bibnamefont {Pi}},\ and\
  \bibinfo {author} {\bibfnamefont {M.}~\bibnamefont {Sasaki}},\ }\href
  {https://doi.org/10.1103/PhysRevLett.122.201101} {\bibfield  {journal}
  {\bibinfo  {journal} {Phys. Rev. Lett.}\ }\textbf {\bibinfo {volume} {122}},\
  \bibinfo {pages} {201101} (\bibinfo {year} {2019})},\ \Eprint
  {https://arxiv.org/abs/1810.11000} {arXiv:1810.11000 [astro-ph.CO]}
  \BibitemShut {NoStop}%
\bibitem [{\citenamefont {Yuan}\ \emph
  {et~al.}(2020{\natexlab{b}})\citenamefont {Yuan}, \citenamefont {Chen},\ and\
  \citenamefont {Huang}}]{Yuan:2019wwo}%
  \BibitemOpen
  \bibfield  {author} {\bibinfo {author} {\bibfnamefont {C.}~\bibnamefont
  {Yuan}}, \bibinfo {author} {\bibfnamefont {Z.-C.}\ \bibnamefont {Chen}},\
  and\ \bibinfo {author} {\bibfnamefont {Q.-G.}\ \bibnamefont {Huang}},\ }\href
  {https://doi.org/10.1103/PhysRevD.101.043019} {\bibfield  {journal} {\bibinfo
   {journal} {Phys. Rev. D}\ }\textbf {\bibinfo {volume} {101}},\ \bibinfo
  {pages} {4} (\bibinfo {year} {2020}{\natexlab{b}})},\ \Eprint
  {https://arxiv.org/abs/1910.09099} {arXiv:1910.09099 [astro-ph.CO]}
  \BibitemShut {NoStop}%
\bibitem [{\citenamefont {Dom{\`e}nech}\ and\ \citenamefont
  {Chluba}(2025)}]{Domenech:2025bvr}%
  \BibitemOpen
  \bibfield  {author} {\bibinfo {author} {\bibfnamefont {G.}~\bibnamefont
  {Dom{\`e}nech}}\ and\ \bibinfo {author} {\bibfnamefont {J.}~\bibnamefont
  {Chluba}},\ }\href {https://doi.org/10.1088/1475-7516/2025/07/034} {\bibfield
   {journal} {\bibinfo  {journal} {JCAP}\ }\textbf {\bibinfo {volume} {07}},\
  \bibinfo {pages} {034}},\ \Eprint {https://arxiv.org/abs/2503.13670}
  {arXiv:2503.13670 [gr-qc]} \BibitemShut {NoStop}%
\bibitem [{\citenamefont {Thrane}\ and\ \citenamefont
  {Romano}(2013)}]{Thrane:2013oya}%
  \BibitemOpen
  \bibfield  {author} {\bibinfo {author} {\bibfnamefont {E.}~\bibnamefont
  {Thrane}}\ and\ \bibinfo {author} {\bibfnamefont {J.~D.}\ \bibnamefont
  {Romano}},\ }\href {https://doi.org/10.1103/PhysRevD.88.124032} {\bibfield
  {journal} {\bibinfo  {journal} {Phys. Rev.}\ }\textbf {\bibinfo {volume}
  {D88}},\ \bibinfo {pages} {124032} (\bibinfo {year} {2013})},\ \Eprint
  {https://arxiv.org/abs/1310.5300} {arXiv:1310.5300 [astro-ph.IM]}
  \BibitemShut {NoStop}%
\bibitem [{\citenamefont {Schmitz}(2021)}]{Schmitz:2020syl}%
  \BibitemOpen
  \bibfield  {author} {\bibinfo {author} {\bibfnamefont {K.}~\bibnamefont
  {Schmitz}},\ }\href {https://doi.org/10.1007/JHEP01(2021)097} {\bibfield
  {journal} {\bibinfo  {journal} {JHEP}\ }\textbf {\bibinfo {volume} {01}},\
  \bibinfo {pages} {097}},\ \Eprint {https://arxiv.org/abs/2002.04615}
  {arXiv:2002.04615 [hep-ph]} \BibitemShut {NoStop}%
\bibitem [{A+()}]{A+}%
  \BibitemOpen
  \href@noop {} {\bibinfo {title} {The {A+} design curve}},\ \bibinfo
  {howpublished} {\url{https://dcc.ligo.org/LIGO-T1800042/public}},\ \bibinfo
  {note} {[Online; accessed 05-May-2023]}\BibitemShut {NoStop}%
\bibitem [{\citenamefont {Branchesi}\ \emph {et~al.}(2023)\citenamefont
  {Branchesi} \emph {et~al.}}]{Branchesi:2023mws}%
  \BibitemOpen
  \bibfield  {author} {\bibinfo {author} {\bibfnamefont {M.}~\bibnamefont
  {Branchesi}} \emph {et~al.},\ }\href
  {https://doi.org/10.1088/1475-7516/2023/07/068} {\bibfield  {journal}
  {\bibinfo  {journal} {JCAP}\ }\textbf {\bibinfo {volume} {07}},\ \bibinfo
  {pages} {068}},\ \Eprint {https://arxiv.org/abs/2303.15923} {arXiv:2303.15923
  [gr-qc]} \BibitemShut {NoStop}%
\bibitem [{ce()}]{ce}%
  \BibitemOpen
  \href@noop {} {\bibinfo {title} {Cosmic explorer sensitivity curve}},\
  \bibinfo {howpublished} {\url{https://cosmicexplorer.org/sensitivity.html}},\
  \bibinfo {note} {[Online; accessed 05-May-2023]}\BibitemShut {NoStop}%
\bibitem [{\citenamefont {Yagi}\ and\ \citenamefont
  {Seto}(2011)}]{Yagi:2011wg}%
  \BibitemOpen
  \bibfield  {author} {\bibinfo {author} {\bibfnamefont {K.}~\bibnamefont
  {Yagi}}\ and\ \bibinfo {author} {\bibfnamefont {N.}~\bibnamefont {Seto}},\
  }\href {https://doi.org/10.1103/PhysRevD.95.109901,
  10.1103/PhysRevD.83.044011} {\bibfield  {journal} {\bibinfo  {journal} {Phys.
  Rev.}\ }\textbf {\bibinfo {volume} {D83}},\ \bibinfo {pages} {044011}
  (\bibinfo {year} {2011})},\ \bibinfo {note} {[Erratum: Phys.
  Rev.D95,no.10,109901(2017)]},\ \Eprint {https://arxiv.org/abs/1101.3940}
  {arXiv:1101.3940 [astro-ph.CO]} \BibitemShut {NoStop}%
\bibitem [{\citenamefont {Kawamura}\ \emph {et~al.}(2020)\citenamefont
  {Kawamura} \emph {et~al.}}]{Kawamura:2020pcg}%
  \BibitemOpen
  \bibfield  {author} {\bibinfo {author} {\bibfnamefont {S.}~\bibnamefont
  {Kawamura}} \emph {et~al.},\ }\href@noop {} {\  (\bibinfo {year} {2020})},\
  \Eprint {https://arxiv.org/abs/2006.13545} {arXiv:2006.13545 [gr-qc]}
  \BibitemShut {NoStop}%
\bibitem [{\citenamefont {Barke}\ \emph {et~al.}(2015)\citenamefont {Barke},
  \citenamefont {Wang}, \citenamefont {Esteban~Delgado}, \citenamefont
  {Tr\"obs}, \citenamefont {Heinzel},\ and\ \citenamefont
  {Danzmann}}]{Barke:2014lsa}%
  \BibitemOpen
  \bibfield  {author} {\bibinfo {author} {\bibfnamefont {S.}~\bibnamefont
  {Barke}}, \bibinfo {author} {\bibfnamefont {Y.}~\bibnamefont {Wang}},
  \bibinfo {author} {\bibfnamefont {J.~J.}\ \bibnamefont {Esteban~Delgado}},
  \bibinfo {author} {\bibfnamefont {M.}~\bibnamefont {Tr\"obs}}, \bibinfo
  {author} {\bibfnamefont {G.}~\bibnamefont {Heinzel}},\ and\ \bibinfo {author}
  {\bibfnamefont {K.}~\bibnamefont {Danzmann}},\ }\href
  {https://doi.org/10.1088/0264-9381/32/9/095004} {\bibfield  {journal}
  {\bibinfo  {journal} {Class. Quant. Grav.}\ }\textbf {\bibinfo {volume}
  {32}},\ \bibinfo {pages} {095004} (\bibinfo {year} {2015})},\ \Eprint
  {https://arxiv.org/abs/1411.1260} {arXiv:1411.1260 [physics.ins-det]}
  \BibitemShut {NoStop}%
\bibitem [{\citenamefont {Sesana}\ \emph {et~al.}(2021)\citenamefont {Sesana}
  \emph {et~al.}}]{Sesana:2019vho}%
  \BibitemOpen
  \bibfield  {author} {\bibinfo {author} {\bibfnamefont {A.}~\bibnamefont
  {Sesana}} \emph {et~al.},\ }\href
  {https://doi.org/10.1007/s10686-021-09709-9} {\bibfield  {journal} {\bibinfo
  {journal} {Exper. Astron.}\ }\textbf {\bibinfo {volume} {51}},\ \bibinfo
  {pages} {1333} (\bibinfo {year} {2021})},\ \Eprint
  {https://arxiv.org/abs/1908.11391} {arXiv:1908.11391 [astro-ph.IM]}
  \BibitemShut {NoStop}%
\bibitem [{\citenamefont {Abbott}\ \emph {et~al.}(2021)\citenamefont {Abbott}
  \emph {et~al.}}]{KAGRA:2021kbb}%
  \BibitemOpen
  \bibfield  {author} {\bibinfo {author} {\bibfnamefont {R.}~\bibnamefont
  {Abbott}} \emph {et~al.} (\bibinfo {collaboration} {KAGRA, Virgo, LIGO
  Scientific}),\ }\href {https://doi.org/10.1103/PhysRevD.104.022004}
  {\bibfield  {journal} {\bibinfo  {journal} {Phys. Rev. D}\ }\textbf {\bibinfo
  {volume} {104}},\ \bibinfo {pages} {022004} (\bibinfo {year} {2021})},\
  \Eprint {https://arxiv.org/abs/2101.12130} {arXiv:2101.12130 [gr-qc]}
  \BibitemShut {NoStop}%
\bibitem [{\citenamefont {Vainshtein}(1972)}]{Vainshtein:1972sx}%
  \BibitemOpen
  \bibfield  {author} {\bibinfo {author} {\bibfnamefont {A.~I.}\ \bibnamefont
  {Vainshtein}},\ }\href {https://doi.org/10.1016/0370-2693(72)90147-5}
  {\bibfield  {journal} {\bibinfo  {journal} {Phys. Lett. B}\ }\textbf
  {\bibinfo {volume} {39}},\ \bibinfo {pages} {393} (\bibinfo {year}
  {1972})}\BibitemShut {NoStop}%
\bibitem [{\citenamefont {Kimura}\ \emph {et~al.}(2012)\citenamefont {Kimura},
  \citenamefont {Kobayashi},\ and\ \citenamefont {Yamamoto}}]{Kimura:2011dc}%
  \BibitemOpen
  \bibfield  {author} {\bibinfo {author} {\bibfnamefont {R.}~\bibnamefont
  {Kimura}}, \bibinfo {author} {\bibfnamefont {T.}~\bibnamefont {Kobayashi}},\
  and\ \bibinfo {author} {\bibfnamefont {K.}~\bibnamefont {Yamamoto}},\ }\href
  {https://doi.org/10.1103/PhysRevD.85.024023} {\bibfield  {journal} {\bibinfo
  {journal} {Phys. Rev. D}\ }\textbf {\bibinfo {volume} {85}},\ \bibinfo
  {pages} {024023} (\bibinfo {year} {2012})},\ \Eprint
  {https://arxiv.org/abs/1111.6749} {arXiv:1111.6749 [astro-ph.CO]}
  \BibitemShut {NoStop}%
\bibitem [{\citenamefont {Koyama}\ \emph {et~al.}(2013)\citenamefont {Koyama},
  \citenamefont {Niz},\ and\ \citenamefont {Tasinato}}]{Koyama:2013paa}%
  \BibitemOpen
  \bibfield  {author} {\bibinfo {author} {\bibfnamefont {K.}~\bibnamefont
  {Koyama}}, \bibinfo {author} {\bibfnamefont {G.}~\bibnamefont {Niz}},\ and\
  \bibinfo {author} {\bibfnamefont {G.}~\bibnamefont {Tasinato}},\ }\href
  {https://doi.org/10.1103/PhysRevD.88.021502} {\bibfield  {journal} {\bibinfo
  {journal} {Phys. Rev. D}\ }\textbf {\bibinfo {volume} {88}},\ \bibinfo
  {pages} {021502} (\bibinfo {year} {2013})},\ \Eprint
  {https://arxiv.org/abs/1305.0279} {arXiv:1305.0279 [hep-th]} \BibitemShut
  {NoStop}%
\bibitem [{\citenamefont {Babichev}\ and\ \citenamefont
  {Deffayet}(2013)}]{Babichev:2013usa}%
  \BibitemOpen
  \bibfield  {author} {\bibinfo {author} {\bibfnamefont {E.}~\bibnamefont
  {Babichev}}\ and\ \bibinfo {author} {\bibfnamefont {C.}~\bibnamefont
  {Deffayet}},\ }\href {https://doi.org/10.1088/0264-9381/30/18/184001}
  {\bibfield  {journal} {\bibinfo  {journal} {Class. Quant. Grav.}\ }\textbf
  {\bibinfo {volume} {30}},\ \bibinfo {pages} {184001} (\bibinfo {year}
  {2013})},\ \Eprint {https://arxiv.org/abs/1304.7240} {arXiv:1304.7240
  [gr-qc]} \BibitemShut {NoStop}%
\bibitem [{\citenamefont {Gubitosi}\ \emph {et~al.}(2013)\citenamefont
  {Gubitosi}, \citenamefont {Piazza},\ and\ \citenamefont
  {Vernizzi}}]{Gubitosi:2012hu}%
  \BibitemOpen
  \bibfield  {author} {\bibinfo {author} {\bibfnamefont {G.}~\bibnamefont
  {Gubitosi}}, \bibinfo {author} {\bibfnamefont {F.}~\bibnamefont {Piazza}},\
  and\ \bibinfo {author} {\bibfnamefont {F.}~\bibnamefont {Vernizzi}},\ }\href
  {https://doi.org/10.1088/1475-7516/2013/02/032} {\bibfield  {journal}
  {\bibinfo  {journal} {JCAP}\ }\textbf {\bibinfo {volume} {02}},\ \bibinfo
  {pages} {032}},\ \Eprint {https://arxiv.org/abs/1210.0201} {arXiv:1210.0201
  [hep-th]} \BibitemShut {NoStop}%
\bibitem [{\citenamefont {Inomata}\ \emph {et~al.}(2020)\citenamefont
  {Inomata}, \citenamefont {Kawasaki}, \citenamefont {Mukaida}, \citenamefont
  {Terada},\ and\ \citenamefont {Yanagida}}]{Inomata:2020lmk}%
  \BibitemOpen
  \bibfield  {author} {\bibinfo {author} {\bibfnamefont {K.}~\bibnamefont
  {Inomata}}, \bibinfo {author} {\bibfnamefont {M.}~\bibnamefont {Kawasaki}},
  \bibinfo {author} {\bibfnamefont {K.}~\bibnamefont {Mukaida}}, \bibinfo
  {author} {\bibfnamefont {T.}~\bibnamefont {Terada}},\ and\ \bibinfo {author}
  {\bibfnamefont {T.~T.}\ \bibnamefont {Yanagida}},\ }\href
  {https://doi.org/10.1103/PhysRevD.101.123533} {\bibfield  {journal} {\bibinfo
   {journal} {Phys. Rev. D}\ }\textbf {\bibinfo {volume} {101}},\ \bibinfo
  {pages} {123533} (\bibinfo {year} {2020})},\ \Eprint
  {https://arxiv.org/abs/2003.10455} {arXiv:2003.10455 [astro-ph.CO]}
  \BibitemShut {NoStop}%
\bibitem [{\citenamefont {Papanikolaou}\ \emph {et~al.}(2021)\citenamefont
  {Papanikolaou}, \citenamefont {Vennin},\ and\ \citenamefont
  {Langlois}}]{Papanikolaou:2020qtd}%
  \BibitemOpen
  \bibfield  {author} {\bibinfo {author} {\bibfnamefont {T.}~\bibnamefont
  {Papanikolaou}}, \bibinfo {author} {\bibfnamefont {V.}~\bibnamefont
  {Vennin}},\ and\ \bibinfo {author} {\bibfnamefont {D.}~\bibnamefont
  {Langlois}},\ }\href {https://doi.org/10.1088/1475-7516/2021/03/053}
  {\bibfield  {journal} {\bibinfo  {journal} {JCAP}\ }\textbf {\bibinfo
  {volume} {03}},\ \bibinfo {pages} {053}},\ \Eprint
  {https://arxiv.org/abs/2010.11573} {arXiv:2010.11573 [astro-ph.CO]}
  \BibitemShut {NoStop}%
\bibitem [{\citenamefont {Dom{\`e}nech}\ \emph {et~al.}(2021)\citenamefont
  {Dom{\`e}nech}, \citenamefont {Lin},\ and\ \citenamefont
  {Sasaki}}]{Domenech:2020ssp}%
  \BibitemOpen
  \bibfield  {author} {\bibinfo {author} {\bibfnamefont {G.}~\bibnamefont
  {Dom{\`e}nech}}, \bibinfo {author} {\bibfnamefont {C.}~\bibnamefont {Lin}},\
  and\ \bibinfo {author} {\bibfnamefont {M.}~\bibnamefont {Sasaki}},\ }\href
  {https://doi.org/10.1088/1475-7516/2021/11/E01} {\bibfield  {journal}
  {\bibinfo  {journal} {JCAP}\ }\textbf {\bibinfo {volume} {04}},\ \bibinfo
  {pages} {062}},\ \bibinfo {note} {[Erratum: JCAP 11, E01 (2021)]},\ \Eprint
  {https://arxiv.org/abs/2012.08151} {arXiv:2012.08151 [gr-qc]} \BibitemShut
  {NoStop}%
\bibitem [{\citenamefont {Jiang}\ \emph {et~al.}(2025)\citenamefont {Jiang},
  \citenamefont {Lin},\ and\ \citenamefont {Gao}}]{Jiang:2025ysb}%
  \BibitemOpen
  \bibfield  {author} {\bibinfo {author} {\bibfnamefont {J.}~\bibnamefont
  {Jiang}}, \bibinfo {author} {\bibfnamefont {J.}~\bibnamefont {Lin}},\ and\
  \bibinfo {author} {\bibfnamefont {X.}~\bibnamefont {Gao}},\ }\href@noop {} {\
   (\bibinfo {year} {2025})},\ \Eprint {https://arxiv.org/abs/2508.20000}
  {arXiv:2508.20000 [gr-qc]} \BibitemShut {NoStop}%
\end{thebibliography}%

\end{document}